\documentclass[11pt]{article}
\usepackage[utf8]{inputenc}
\usepackage{standalone}
\usepackage{float}
\usepackage{tikz}
\usepackage{safecolours}
\usepackage{xcolor,soul}
\usetikzlibrary{decorations.pathreplacing,shapes.multipart, arrows.meta, calc, quotes, tikzmark, positioning}
\pdfoutput=1
\usepackage[margin=1.1in,letterpaper]{geometry}
\usepackage{physics}
\usepackage{amsmath,amssymb,amsthm}
\usepackage{enumitem}
\usepackage{graphicx}
\usepackage[numbers, square, sort&compress]{natbib}
\DeclareMathOperator\sign{sgn}
\usepackage{subcaption}
\usepackage[colorlinks=true, urlcolor=blue, citecolor=black, linkcolor=black]{hyperref}
\usepackage{authblk}

\theoremstyle{definition}

\title{A minimal model of cognition based on oscillatory and current-based reinforcement processes}
\author[1,2]{Linnéa Gyllingberg$^*$}
\author[3]{Yu Tian$^*$}
\author[2,4]{David J. T. Sumpter}
\affil[1]{\textit{Department of Mathematics, University of California, Los Angeles, CA, USA}}
\affil[2]{{\textit{Department of Mathematics, Uppsala University, Uppsala, Sweden}}}
\affil[3]{{\textit{Nordita, Stockholm University and KTH Royal Institute of Technology, Stockholm, Sweden }}}
\affil[4]{{\textit{Department of Information Technology, Uppsala University, Uppsala, Sweden}}}
\date{}

\begin{document}

\maketitle

\def\thefootnote{*}\footnotetext{These authors contributed equally to this work}\def\thefootnote{\arabic{footnote}}

% \tableofcontents

\begin{abstract}
Building mathematical models of brains is difficult because of the sheer complexity of the problem. One potential starting point is through basal cognition, which gives abstract representation of a range of organisms without central nervous systems, including fungi, slime moulds and bacteria. We propose one such model, demonstrating how a combination of oscillatory and current-based reinforcement processes can be used to couple resources in an efficient manner, mimicking the way these organisms function. A key ingredient in our model, not found in previous basal cognition models, is that we explicitly model oscillations in the number of particles (i.e. the nutrients, chemical signals or similar, which make up the biological system) and the flow of these particles within the modelled organisms. Using this approach, we find that our model builds efficient solutions, provided the environmental oscillations are sufficiently out of phase. We further demonstrate that amplitude differences can promote efficient solutions and that the system is robust to frequency differences. In the context of these findings, we discuss connections between our model and basal cognition in biological systems and slime moulds, in particular, how oscillations might contribute to self-organised problem-solving by these organisms.
\end{abstract}

\section{Introduction}
Understanding cognitive abilities, such as perception, problem solving, and learning, is a central and longstanding question in neurobiology.  One crucial aspect of cognition is the role of {\it oscillatory processes} \cite{bacsar2000brain}. In the human brain, for example, theta oscillations are linked to active learning in infants \cite{begus2020rhythm} and play an important role in memory formation \cite{herweg2020theta}; beta oscillations have shown pivotal for multisensory learning \cite{gnaedinger2019multisensory};  alpha oscillations are central in temporal attention \cite{hanslmayr2011role}; and cross-frequency coupling, i.e. coupling between neurons of different frequencies, has been proposed as a mechanism for working memory \cite{lisman2013theta, jensen2014temporal}. Another crucial aspect of cognitive abilities is synaptic plasticity, i.e. changing the strength of the connections between neurons based on previous activity  \cite{martin2000synaptic, holscher1999synaptic}. Such  adaptive processes,  often referred to as {\it reinforcement}, have been studied in both experimental and computational settings. For example, a study on rats shows that reinforcement determines the timing dependence of corticostriatal synaptic plasticity in vivo \cite{fisher2017reinforcement}, and dopamine-dependent synaptic plasticity has been studied in mice \cite{shindou2019silent}. 

While the human brain serves as the primary model system for cognitive research, the field of basal cognition (also referred to as minimal cognition) aims to articulate the fundamental requirements necessary for the generation of cognitive phenomena \cite{van2006principles,lyon2021reframing}. Cognitive-like abilities are not limited to organisms with central nervous systems, but are also found in aneural organisms \cite{boussard2021adaptive}. All living organisms employ sensory and information-processing mechanisms to assess and engage with both their internal environment and the world around them, and  oscillations often play a crucial role in their function \cite{boussard2021adaptive}. For example, self-sustained oscillations that encode various types of information have been observed in cells \cite{behar2010understanding, purvis2013encoding}. Oscillations are also found in the root apex of many plants \cite{baluvska2013root}, and have been suggested to be a driving mechanism for deciding where the plant should branch its roots \cite{traas2010oscillating}.

\textit{Physarum polycephalum} is a popular model organism for studying problem-solving in aneural organisms and better understanding basal cognition \cite{oettmeier2017physarum, vallverdu2018slime, smith2020needs}. In its vegetative state, known as the plasmodium, this organism takes the form of a colossal, mobile cell. The plasmodium's fundamental structure encompasses a syncytium of nuclei and an intricate intracellular cytoskeleton, forming a complex network of cytoplasmic veins. It can expand to cover extensive areas, reaching dimensions of hundreds of square centimetres, and can divide into smaller, viable subunits. When these subunits come into contact, they have the remarkable ability to fuse and share information, leading to the reformation of a giant plasmodium. These slime moulds demonstrate habituation \cite{boisseau2016habituation} as well as anticipation \cite{saigusa2008amoebae}. Early experiments on these single cellular organisms show that they can find the shortest paths between food sources through a labyrinth \cite{nakagaki2000maze,beekman2015brainless}. Further experiments have shown that it can construct efficient and robust transport networks with the same topology as the Tokyo rail network \cite{tero2010rules}, avoid already exploited patches \cite{reid2013amoeboid}, and solve a Towers of Hanoi maze \cite{reid2013solving}. 

Experiments show that partial bodies in the plasmodium of the slime mould can also be viewed as nonlinear oscillators, where the interactions between the oscillators are strongly affected by the geometry of the tube network \cite{takamatsu2000controlling,takamatsu2001spatiotemporal}. Thus both oscillatory and reinforcement processes are involved in \textit{Physarum polycephalum}'s cognitive abilities \cite{boussard2021adaptive}. Oscillations in slime mould take place in different parts of the organism and on multiple time scales; from short-period oscillations in the contraction cycle, to long period oscillations in the cell cycle \cite{dussutour2021learning}. 

In this article we look at how oscillators, coupled through current-based reinforcement, can exhibit both problem solving (i.e. shortest path finding) abilities and long-range synchronisation observed in many cognitive phenomena.  Following Boussard et al., who emphasize the slime mould as an ideal model system for relating basal cognitive functions to biological mechanisms \cite{boussard2021adaptive}, it is such a coupling of feedback with oscillations which we refer to as minimal cognition. This definition is also in line with discussions around properties of minimal cognition for other biological systems \cite{beer1992evolving, beer1996toward, brancazio2022easy}. We can also see our work as contributing a model of \textit{basal cognition}, a term first introduced in a special double issue of Philosophical Transactions of the Royal Society B in March 2021 \cite{lyon2021reframing}. This emerging field, as elucidated by Lyon et al., underscores the presence of cognitive abilities predating the evolution of nervous systems. Indeed, Figure 1 in \cite{lyon2021reframing} illustrates various organisms that have been studied to comprehend basal cognition, including \textit{Bacillus subtilis} \cite{liu2017coupling}, the aneural placozoan \textit{Trichoplax adhaerens} \cite{moroz2021neural}, and the planarian flatworm \textit{Platyhelminthes} \cite{pezzulo2021bistability}. 

Aligning with the broader scope of basal cognition, we aim here to define a mathematical model that elucidates the role of oscillations and reinforcement mechanisms for cognitive phenomona in aneural settings. We apply Beer's four steps for theory creation around cognitive phenomena \cite{beer2020lost} to basal cognition. In this introduction, as a first step, we have presented a conceptual framework, i.e. the idea that reinforcement and oscillations are sufficient to produce some basal cognitive abilities. Beer's second step, which we perform in section \ref{sec:model}, is to present a mathematical model based on oscillators and reinforcement on networks. As a third step, in section \ref{sec:results}, we study the mathematical properties of this model with a particular focus on what types of oscillations produce basal cognitive phenomena in aneural organisms. Finally, the fourth step discusses how our model might be further developed in light of the results and the degree to which they help us understand biological systems.

From a mathematical point of view, in this paper, we start by studying the dynamics of a small network, a cycle graph with two oscillators. We show that this model can generate shortest path networks, localised oscillatory behaviour and global oscillatory dynamics. Then, addressing questions raised by Boussard et al. \cite{boussard2021adaptive}, we look at how phase, frequency and amplitude differences can be used to induce dynamic problem solving. Finally, we look at how larger networks of coupled oscillators can create complex networks of connections that mimic some notion of basal cognition.
%\david{I wonder if this last paragraph can be omitted, depending on the journal?}

%One sentence somewhere here about how oscillations is found in microbes cognition \cite{wan2023active}

\section{Model}
\label{sec:model}
Our aim is to define a model which has a combination of reinforcement and oscillatory processes, which can both `solve' the problem of finding the shortest path between two or more points and exhibit long-range oscillations. We abstract away from the specific biological systems, and find a model of basal (or minimal) cognition \cite{lyon2021reframing}. To this end, we describe the motion of particles which, depending on the system, might be nutrients, chemical signals (e.g. in slime moulds \cite{boussard2021adaptive} and many fungi \cite{schmieder2019bidirectional}), or information within a decentralised system (e.g. macromolecular networks in microorganisms \cite{westerhoff2014macromolecular}). The system itself is modelled as a network or a graph over which the particles move. The particles travel on the edges of the graph and in doing so produce current reinforcement. The coupling between the system and its environment is modelled through an in- and out-flow of particles at some of the points on the network. In order to capture oscillations, which are also widespread in systems of microbes \cite{wan2023active}, slime moulds \cite{boussard2021adaptive}, and fungi \cite{schmieder2019bidirectional}, we will make this in- and out-flow of particles change over time. 

\subsection{Previous models}

We build primarily on the work of Tero et al. in modelling slime mould \cite{tero2006physarum}.
They model the movement of particles, which  are nutrient and chemical signals, on nodes on a graph, which can be thought of as the cytoskeleton of the plasmodium. The dynamic weights on the edges of the graph are the thickness of the tubes within that form the cytoskeleton. Similar analogies can be made for other basal cognitive systems. For example, for fungi, the graph is the hyphae, and the particles are nutrients \cite{schmieder2019bidirectional}.

 Tero et al.'s model uses an analogy to electrical networks, to describe a flux of particles, $Q_{ij}$, between the nodes. Equations for evolution of the conductivity of the edges, $D_{ij}(t)$,  (i.e. the thickness of the edges) are then proposed as follows: 
\begin{equation}
    \dv{D_{ij}(t)}{t}=f(|Q_{ij}|)-\lambda D_{ij},
\end{equation} 
where $f$ is a non-decreasing function.
The equation above states that edges in the tubular network are thickened if there is a sufficient volume flow, and diminished otherwise. In the case where $f(q)=q$, this model has been proven to eventually converge to the shortest path between food sources, independently of the network structure \cite{ito2011convergence,bonifaci2012physarum}.  In this work, the flow of particles is assumed to be in steady state, i.e., $Q_{ij}=N_i-N_j=0$, where $N_i$ is the number of particles. In their work, Tero et al., assume that all flow between all nodes, except for the input and output nodes, is zero. They also assume that the total flux of the system is constant, meaning that the sum of the input and output is always zero. Another adaptation of the Tero et al.-model has been presented by Ma et al., who introduced a stochastic version of the current reinforcement model, where the flow of the particles between the nodes is not assumed to be in steady state \cite{ma2013current}, and (in simulations) the model also finds the shortest paths between food sources. 

Mathematical models of coupled oscillators, such as the Kuramoto model \cite{kuramoto1975international}, have played a pivotal role in investigating the dynamics of neuronal processes within the brain \cite{breakspear2010generative,cabral2011role}. And, while it has been adapted to include synaptic plasticity, by evolving coupling strength based on phase differences and Hebbian learning principles \cite{maistrenko2007multistability,timms2014synchronization,ruangkriengsin2022low}, these models do not incorporate the current reinforcement mechanism, which is central to problem solving by slime moulds outlined in the previous paragraph. Indeed, while the Kuramoto model has been used to model the anticipation of periodic events in slime moulds \cite{saigusa2008amoebae}, there is a lack of models using coupled oscillators to study other problem solving abilities of slime moulds, such as finding shortest paths. 
Alim et al. have made an interesting starting point into this, by studying the mechanism of signalling propagation in slime moulds \cite{alim2017mechanism}. Using the Stokes equations to model the cytoplasmic flow velocity with the active tension of the walls of the slime mould oscillating, they argue that their model will obey similar dynamics to the Tero et al. model \cite{tero2006physarum}, and thus lead to shortest paths solutions. 

Watanabe et al. have also studied an algorithm based on the current reinforcement model, but with oscillatory inputs and outputs \cite{watanabe2014transportation}. In this model, the nodes switch between being inputs and outputs explicitly. In terms of other applications, Ben-Ami et al. have recently studied a model of flow of blood through a series of cylindrical capillaries that form a three-node network, which gives rise to self-sustained oscillations \cite{ben2022structural}.  What is missing from these models, however, is an explicit representation of the number of particles on a node in combination with oscillations. It is such a model we now define.

\subsection{Model definition}
Our model consists of an undirected network, $G(V, E)$, where $V = \{1, 2, \dots, n\}$ is the node set, and an edge $(i, j)\in E$ is an unordered pair of two distinct nodes in the set $V$, where each edge $(i, j)$ is associated with a length $l_{ij}$ ($l_{ij} = \infty$ if there is no edge). We denote the number of particles at each node $i$ by $N_i$. For each edge $(i,j)$, there is a current of particles inversely proportional to the length of the connection, $\abs{N_i-N_j}/l_{ij}$. The edge is also characterised by its conductivity (corresponding to thickness of the tubes in slime mould), $D_{ij}$, which will also affect the current. The number of particles at a node $i$, will change based on the flow of the particles from the neighbouring nodes of $i$, $j \in \Gamma(i)$ according to the following equation:  

\begin{equation}
    \dv{N_i(t)}{t} = \sum_{j\in\Gamma(i)}\frac{N_j(t) - N_i(t)}{l_{ij}}D_{ij}(t),
\end{equation}
(ignoring input and output nodes for now). The conductivity of each edge $(i,j)\in E$ changes based on the following equation: 
\begin{align}\label{equ:odes-d}
    \dv{D_{ij}(t)}{t} = q\frac{\abs{N_i(t) - N_j(t)}}{l_{ij}}D_{ij}(t) - \lambda D_{ij}(t).
\end{align}
Here, $q$ is the reinforcement strength and $\lambda$ is the decay rate. Note that, unlike previous deterministic models of current reinforcement \cite{tero2006physarum,tero2007mathematical,tero2008flow}, we give dynamic equations for the number of particles at each node, as we do not assume the flow of particles to be in equilibrium on each edge as in \cite{tero2006physarum}. This approach is similar to that taken by \cite{ma2013current}, who looked at a stochastic model of current reinforcement similar to the one we study here. Another difference to much of the work by Tero and co-workers is that we only consider the case where the increase in conductivity is proportional to (and not a non-linear function of) the current. Although we don't explicitly write down the equations for the particle flow on the edges, the rate at which particles are transported from one node to another can be derived in the same way as done by Tero et al. \cite{tero2006physarum}.

In the above description there is no in- or out-flow of particles in the system and thus no interaction between the system and the environment. We incorporate such interactions in two different cases: (i) the non-oscillatory case, where particles continuously enter the network at sources and leave at sinks with fixed rates, and (ii) the oscillatory case, where the sources and sinks change over time, in the sense that sources can become sinks and vice versa. We now describe each of these in more detail.

\subsection{Non-oscillatory sources and sinks}

We start by studying non-oscillatory sources and sinks. In this setting, particles enter the network at some source(s) $S = \{s_1, \dots\}\subset V$ with a constant rate $a$, and leave the network at some sink(s) $T = \{t_1, \dots\}\subset V$ with a rate $b N_i$ for each sink $i$, where $b$ is a constant. Thus, the equations describing the number of particles at each node $i$ are given by:
\begin{align}\label{equ:particles_non_osc}
    \dv{N_i(t)}{t} = 
    \begin{cases}
        a + \sum_{v_j\in\Gamma(v_i)}\frac{N_j(t) - N_i(t)}{l_{ij}}D_{ij}(t),\ &\text{if } i\in S,\\
        -bN_i(t) + \sum_{v_j\in\Gamma(v_i)}\frac{N_j(t) - N_i(t)}{l_{ij}}D_{ij}(t),\ &\text{if } i\in T,\\
        \sum_{v_j\in\Gamma(v_i)}\frac{N_j(t) - N_i(t)}{l_{ij}}D_{ij}(t),\ &\text{otherwise}. 
    \end{cases}
\end{align}

\begin{figure}
    \centering
\begin{subfigure}[t]{0.4\linewidth}
\begin{tikzpicture}
[vertex/.style={circle, draw}] 
\foreach \Z\X[count=\Y] in {1/wong-colour1,2/wong-colour7,3/wong-colour4,4/wong-colour7,5/wong-colour7}
{\node[vertex, fill=\X] (x-\Y) at ({72*\Y+162}:2){\Z}; }
\foreach \X[count=\Y] in {0,...,4}
\node [font = {\small}, left of = x-1, xshift=0.5cm, yshift=-1.5cm, align=left] (a_t) {$a$};
\node [right of = x-3,  xshift=0.9cm, yshift=0.5cm] (b_t) {$-b\cdot N_3$};
\draw[line width=1.5pt] (x-1) -- (x-5);
\draw[->, line width = 0.7pt] (a_t) -- (x-1);
\draw[->, line width = 0.7pt] (x-3) -- (b_t);
\draw[line width=1.5pt] (x-1) -- (x-2); %node[midway,above] {$l_{ij}$};
\draw[thick, decorate, decoration={brace, ,mirror,raise=1pt,amplitude=10pt}, shorten <= 0.05cm, shorten >= 0.05cm](x-4) -- (x-5) node [midway,above=8pt,xshift=-0.04cm] {$l_{ij}$};
%\draw [decorate,decoration={brace,amplitude=10pt},xshift=-0.5cm,yshift=0pt](x-1) -- (x-2) node [black,midway,xshift=-0.6cm]
%\draw[brace] (x-1) -- (x-2);
\draw[line width=1.5pt] (x-2) -- (x-3);
\draw[line width=1.5pt] (x-3) -- (x-4) node[midway] (d_ij) {};
\node [above right of = d_ij, yshift=0.5cm] (D_ij) {$D_{ij}$};
\draw[line width = 0.7pt, shorten <= -0.1cm, shorten >= -0.1cm] (D_ij) to[bend right=20] (d_ij);
\draw[line width=1.5pt](x-4) -- (x-5);
%{\ifnum\X=0
%\draw (x-\Y) -- (x-5);
%\else
%\draw (x-\Y) -- (x-\X);
%\fi}
\end{tikzpicture}
    \caption{Non-oscillating nodes} \label{fig:non-osc_graph}  
  \end{subfigure}
\begin{subfigure}[t]{0.4\linewidth}
\hspace*{-0.5cm}
\begin{tikzpicture}
[vertex/.style={circle, draw}] 
\foreach \Z\X[count=\Y] in {1/wong-colour1,2/wong-colour7,3/solin-colour3,4/wong-colour7,5/wong-colour7}
{\node[vertex, fill=\X] (x-\Y) at ({72*\Y+162}:2){\Z}; }
\foreach \X[count=\Y] in {0,...,4}
\node [font = {\small}, left of = x-1, xshift=0.5cm, yshift=-1.5cm, align=left] (a_t) {$A_1(1 + \sin(2\pi\theta_1t + \phi_1))- bN_1(t)$};
\node [right of = x-3,  xshift=0.7cm, yshift=1.0cm, align=left] (b_t) {$A_3(1 + \sin(2\pi\theta_3t + \phi_3))$\\ $- bN_3(t)$};
\draw[line width=1.5pt] (x-1) -- (x-5);
\draw[->, line width = 0.7pt] (a_t) -- (x-1);
\draw[->, line width = 0.7pt] (b_t) -- (x-3);
\draw[line width=1.5pt] (x-1) -- (x-2); %node[midway,above] {$l_{ij}$};
\draw[thick, decorate, decoration={brace, ,mirror,raise=1pt,amplitude=10pt}, shorten <= 0.05cm, shorten >= 0.05cm](x-4) -- (x-5) node [midway,above=8pt,xshift=-0.04cm] {$l_{ij}$};
%\draw [decorate,decoration={brace,amplitude=10pt},xshift=-0.5cm,yshift=0pt](x-1) -- (x-2) node [black,midway,xshift=-0.6cm]
%\draw[brace] (x-1) -- (x-2);
\draw[line width=1.5pt] (x-2) -- (x-3);
\draw[line width=1.5pt] (x-3) -- (x-4) node[midway] (d_ij) {};
\node [above right of = d_ij, yshift=0.5cm] (D_ij) {$D_{ij}$};
\draw[line width = 0.7pt, shorten <= -0.1cm, shorten >= -0.1cm] (D_ij) to[bend right=20] (d_ij);
\draw[line width=1.5pt](x-4) -- (x-5);
%{\ifnum\X=0
%\draw (x-\Y) -- (x-5);
%\else
%\draw (x-\Y) -- (x-\X);
%\fi}
\end{tikzpicture}
\caption{Oscillating nodes} \label{fig:osc_graph}  
\end{subfigure}
    \caption{The model on a $C_5$ graph, where $l_{ij}$ is the length of the edge between node $i$ and $j$, $D_{ij}$ is the conductivity (thickness) of the edge between $i$ and $j$. In (a) the nodes are non-oscillating, and node 1 is a source, with an inflow at rate $a$ and node 3 is a sink with an outflow at rate $bN_3$. In (b) the nodes are oscillating as $A_1(1 + \sin(2\pi\theta_1t + \phi_1))- bN_1(t)$ and $A_3(1 + \sin(2\pi\theta_3t + \phi_3))$\\ $- bN_3(t)$ at node 1 and 3 respectively, meaning that the different colours of the oscillating nodes indicates different amplitude, frequencies and phases in the oscillators.}
    \label{fig:5cycle}
\end{figure}
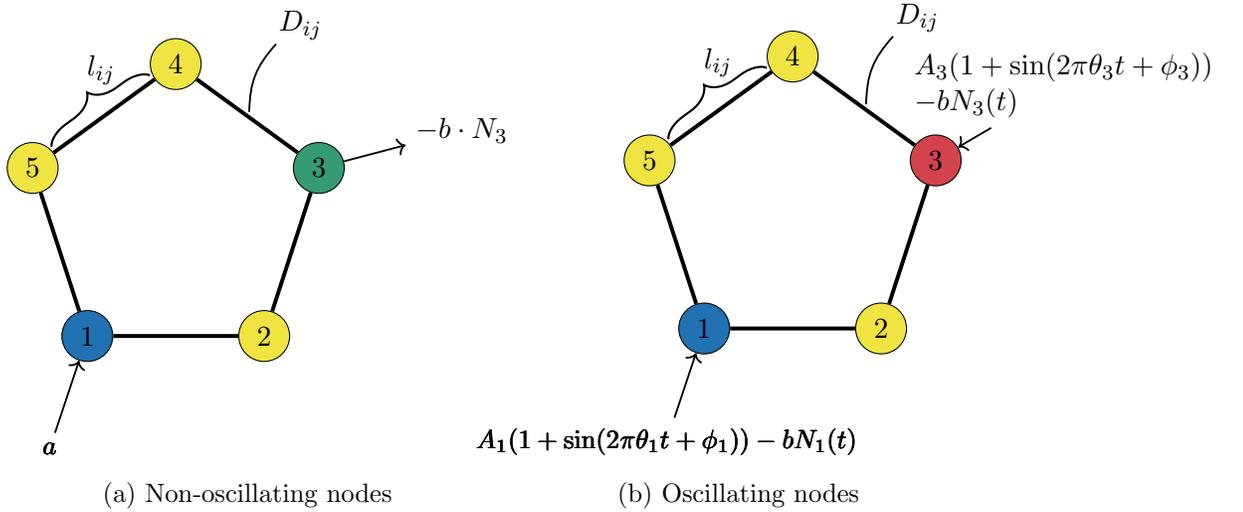

We start by studying these equations in the case of a cyclic graph of five nodes, $C_5$. Here, the nodes are labeled $(1,2,3,4,5)$, and node $1$ is a source, with an inflow $a$, while node $3$ is a sink, with outflow $-b N_3$ at that node. See Figure \ref{fig:5cycle}a. 

%At each node $i$ at time $t$ there are $N_i(t)$ particles. The nodes are connected with edges associated with a conductivity $D$. At each time, the conductivity of the edge between node $i$ and  $j$ is given by $D_{ij}(t)$. 

%\todo{CHANGE POSITION OF NODES TO MATCH SIMULATIONS}
%\todo{LINE FROM A SHOULD BE THICKER + ONE PLOT FOR THE OSCILLATORY CASE}
% \yu{[The following is from previous discussion of the structure, and may not be necessary.]}
% \textbf{Connecting edges}
% \textbf{Input/Output nodes}
% They have an inflow which is determined by the edge conductivity% \textbf{Internal nodes}
% \subsection{Inputs and outputs}
% We will start by studying this system on a 
% \subsubsection{Constant input/output}
%\yu{[The following is under the previous section ``Constant input/output", and may be appropriate here, so I moved it here.]}

\subsection{Oscillatory nodes}\label{sec:model-oscil}
 To incorporate oscillations, we consider nodes which alternate between being sources and sinks. For these nodes, the rate at which particles enter or leave the network through the environment changes over time. To this end, we model the interaction with the environment as the following function, for each node $i\in O$,
\begin{align}
    A_i(1 + \sin(2\pi\theta_i t + \phi_i)) - bN_i(t),
    \label{equ:oscil-func}
\end{align}
where $O$ denotes the set of oscillatory nodes, and $A_i, \theta_i, \phi_i$ characterise the features of the environment around node $i$. Specifically, $A_i$ is the amplitude, $\theta_i$ is the frequency, $\phi_i$ is the phase, and $b$ is the output rate. We choose parameters for the function \eqref{equ:oscil-func}, so it can have both positive and negative values at varying time $t$, to model that particles can both enter and leave the network at node $i$. For $C_5$, we let nodes $1$ and $3$ be the oscillatory nodes, i.e., $O = \{1,3\}$ (see Figure \ref{fig:5cycle}b).
 
We choose this function to ensure that the model reflects the dynamic behaviour of slime moulds, which do not conserve mass but constantly grow and reorganize. By designing $A_i(1 + \sin(2\pi\theta_i t + \phi_i))$ to be non-negative, we ensure a positive in-flow, while $-bN_i(t)$ manages the out-flow, preventing any negative number of particles at any node at any time step. This self-stabilizing mechanism couples a positive in-flow with a negative out-flow, ensuring the system remains balanced. Unlike the model by Watanabe et al., where switching is manually controlled, our model allows for spontaneous, self-organizing, and autonomous switching, reflecting more natural, spontaneous transitions.

% LINNÈA
% Paragraph why, we choose this function.. 
% Slime moulds expand, No conservation of mass, instead we want it to always be positive, some mass. We want $A_i(1 + \sin(2\pi\theta_i t + \phi_i))$ to be non-negative, and that $-bN_i(t)$, can't have negative number of particles... we want it to be self stabilizing... 
% --There should never negative number of particles at a node at any time step
% -- We want to couple a positive in-flow with a negative out flow
% IN CONTRAST TO WATANABE... we want the switching to happen spontaneously, instead turning the switching "on and of"  Self-organising/autonomous switching... 

% In this paper, we will consider exclusively when there are two input/output nodes (i.e., $\abs{\mathcal{A}} = 2$). We will start from the case when there is one initial ``source" $v_i$ where it starts with the maximum value in the sine function with $\phi_i = 0$ and one initial ``sink" $v_j$ where it starts with the minimum value in the sine function with with $\phi_i = \pi$, while other parameters are set to be identical for both nodes, i.e., $A_i=A_j=A$ and $\theta_i=\theta_j=\theta$. Then, we explore the behaviour of the system in more detail numerically. Later, we will further investigate the system by varying the phase change $\Delta\phi = \phi_j - \phi_i$, the amplitude ratio $A_j/A_i$ and the frequency ratio $\theta_j/\theta_i$, and illustrate the rich behaviour the system exhibits. 

When the phase difference between two nodes is $\pi$, the sine waves corresponding to the two nodes have exactly opposite signs. In this case, when one node has particles entering the network, the other node has particles leaving it, and vice versa. To model this, we set $\phi_1 = 0$ and $\phi_3 = \pi$, while we set the other parameters to be identical for both nodes, i.e., $A_1=A_3=A$ and $\theta_1=\theta_3=\theta$. Hence, the evolution of $N_1$ and $N_3$ is
\begin{align}\label{eq:C_5_eq_ph-pi}
    \begin{cases}
        \dv{N_1(t)}{t} &= A(1 + \sin(2\pi\theta t)) - bN_1(t) + \frac{(N_5(t)-N_1(t))}{l_{51}}D_{51} + \frac{(N_2(t)-N_1(t))}{l_{12}}D_{12}\\
        % \vspace*{.5em}
        \dv{N_3(t)}{t} &= A(1 + \sin(2\pi\theta t + \pi)) - bN_3(t) + \frac{(N_2(t)-N_3(t))}{l_{23}}D_{23} + \frac{(N_4(t)-N_3(t))}{l_{34}}D_{34}
    \end{cases},
\end{align}
while others maintain the same as in Eqs.~\eqref{equ:particles_non_osc}.

Building on this special case, we now allow the phase of node $3$ to take values between $-\pi$ and $\pi$, i.e., $\phi_3 = \phi\in (-\pi, \pi]$. Similarly, the amplitude, which corresponds to the number of particles waiting at one node to enter the network and also the capacity of the node for them to enter the network, is  set to $A_1 = A$ and $A_3 = \alpha A$, where $\alpha = A_3/A_1$ encodes the amplitude ratio. Similarly, the frequency, which corresponds to how fast the environment changes around each node, is set to $\theta_1 = \theta$ and $\theta_3 = \gamma\theta$, where $\gamma = \theta_3/\theta_1$ encodes the frequency ratio. The evolution of $N_1$ and $N_3$ are then modified to be
\begin{align}\label{eq:C_5_eq_freq}
    \begin{cases}
        \dv{N_1(t)}{t} &= A(1 + \sin(2\pi\theta t)) - bN_1(t) + \frac{(N_5(t)-N_1(t))}{l_{51}}D_{51} + \frac{(N_2(t)-N_1(t))}{l_{12}}D_{12}\\
        % \vspace*{.5em}
        \dv{N_3(t)}{t} &= \alpha A(1 + \sin(2\pi\gamma\theta t+ \phi)) - bN_3(t) + \frac{(N_2(t)-N_3(t))}{l_{23}}D_{23} + \frac{(N_4(t)-N_3(t))}{l_{34}}D_{34}
    \end{cases},
\end{align}
 while the other equations maintain the same as in Eqs.~\eqref{equ:particles_non_osc}.

\subsection{Large graphs}\label{sec:large_graphs}
\begin{figure}[H]
    \centering
    \includegraphics[width=.6\textwidth]{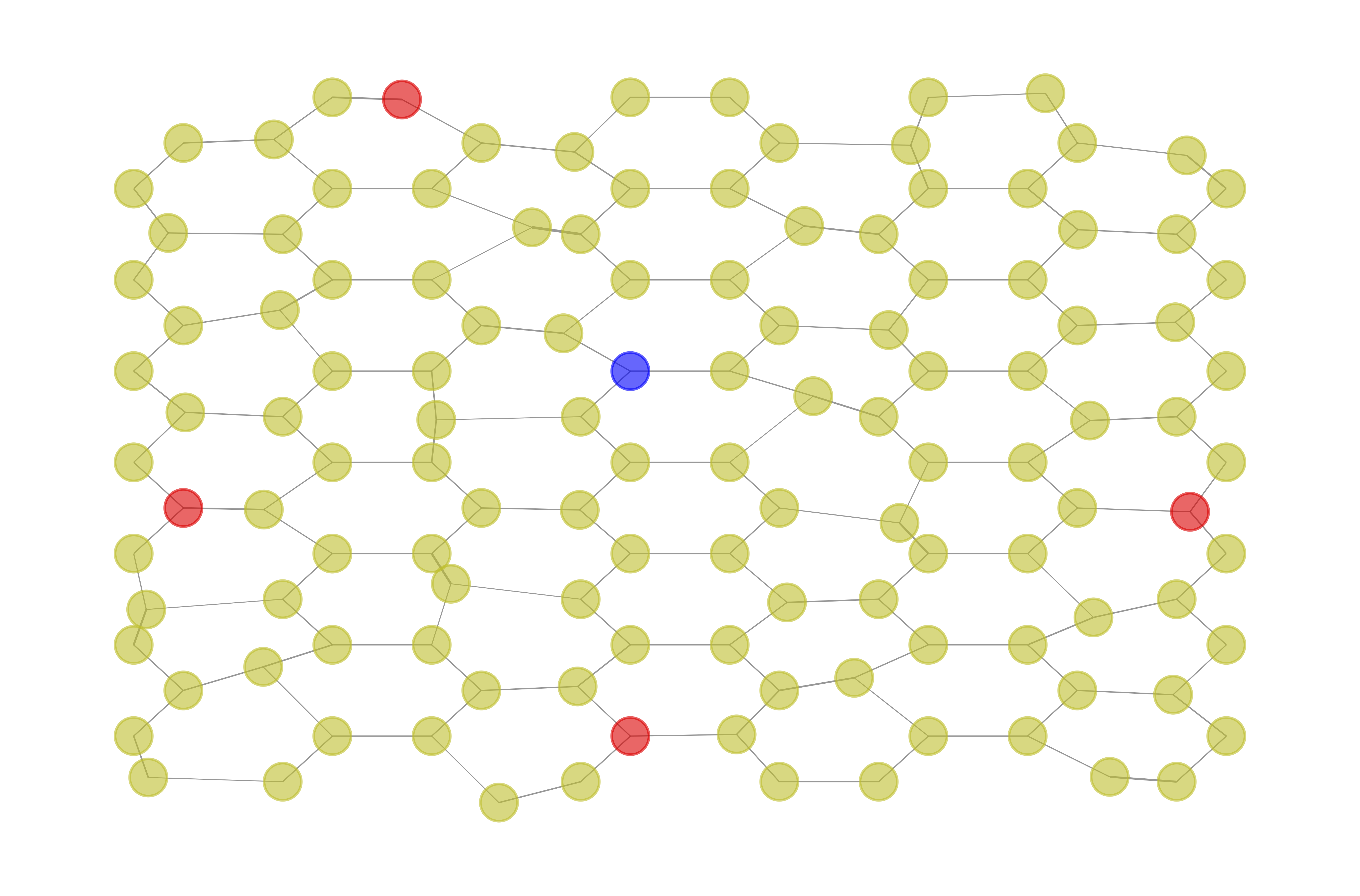}
    \caption{Example of a large regular graph of degree 3, where the node in blue corresponds to an oscillatory node with phase $0$ and the nodes in red correspond to oscillatory nodes with phase $\pi$ in later simulations.}
    \label{fig:large-hexagon-graph}
\end{figure}

% \todo{To be restructured: only explain the structures.}
 The simplest graph for two nodes to have at least two paths connecting them is a cycle graph, and the smallest one to have two paths of different lengths while the two nodes are not directly connected is a cycle graph of length $5$; thus $C_5$ is selected to illustrate our model in the previous sections. 
There are many ways to generalise this to larger graphs, where, e.g., we can consider cycle graphs of arbitrary sizes (larger than $5$), but there are still only two paths connecting the nodes. To systematically incorporate more complexity to the graph, we consider regular graphs of degree $3$, meaning that each node has $3$ neighbours instead of $2$. This choice is inspired by the observation that, as like many other biological networks, the average degree of a node in \textit{Physarum polycephalum} networks is approximately three, with only minor variations due to end branches \cite{dirnberger2017characterizing, dussutour2024flow}. For the graph to have a clear geometric meaning, we choose nodes and edges to be the hexagonal tiling of the plane, leading to the hexagonal lattice graph. We have also included a small level of noise to the position of some nodes, so that paths between the nodes are mostly of different lengths; see Figure \ref{fig:large-hexagon-graph} for example. In this way, we have more geometrically meaningful paths connecting the nodes, which better models biological systems.
% Note that we maintain the same mechanisms of the interaction between the network and the environment, thus the differences lie in the network size, and the neighbourhood structure. 

\section{Results}
\label{sec:results}
We now show that our proposed model can find the shortest path in the cycle graph of size $5$ for a fixed source and sink, in the sense that at steady state conductivity is non-zero only on edges on the shortest path between the source and sink. We then show, in section \ref{sec:res-osc}, that this feature is maintained in simulations in the oscillatory setting when phase difference is maximised, and further investigate the role of phase, amplitude and frequency. Finally, in section \ref{sec:res-large}, we look at larger graphs with multiple oscillating nodes.
Throughout this section, we maintain our assumption that in the change of conductivity, the reinforcement effect, as parameterized in $q$, should be orders of magnitude larger than the decaying effect, as parameterized in $\lambda$, over time. Specifically in the numerical experiments, we choose $q/\lambda=10$. Similar results will be obtained when we change $q$ but maintain a similar ratio, and similar final results will be reached when we choose the ratio to be much larger, but after a longer time. 

\subsection{Non-oscillatory sources and sinks}\label{sec:res-const}

Our model on $C_5$ (shown in Figure \ref{fig:non-osc_graph}) is described by the following set of equations:
\begin{align}\label{eq:C_5_eq} 
\begin{split}
    \dv{N_1(t)}{t} &=   a + \frac{(N_2-N_1)}{l_{12}}D_{12}+\frac{(N_5-N_1)}{l_{51}}D_{51} \\
    \dv{N_2(t)}{t} &=   \frac{(N_1-N_2)}{l_{12}}D_{12}+\frac{(N_3-N_2)}{l_{23}}D_{23} \\
    \dv{N_3(t)}{t} &=  -bN_3+ \frac{(N_2-N_3)}{l_{23}}D_{23}+\frac{(N_4-N_3)}{l_{34}}D_{34} \\
    \dv{N_4(t)}{t} &=   \frac{(N_3-N_4)}{l_{34}}D_{34}+\frac{(N_5-N_4)}{l_{45}}D_{45} \\
    \dv{N_5(t)}{t} &=   \frac{(N_1-N_5)}{l_{51}}D_{51}+\frac{(N_4-N_5)}{l_{45}}D_{45} \\
    \dv{D_{12}(t)}{t} &= q\frac{\abs{N_1 - N_2}}{l_{12}}D_{12} - \lambda D_{12}, \\
    \dv{D_{23}(t)}{t} &= q\frac{\abs{N_2 - N_3}}{l_{23}}D_{23} - \lambda D_{23}, \\
    \dv{D_{34}(t)}{t} &= q\frac{\abs{N_3 - N_4}}{l_{34}}D_{34} - \lambda D_{34}, \\
    \dv{D_{45}(t)}{t} &= q\frac{\abs{N_4 - N_5}}{l_{45}}D_{45} - \lambda D_{45}, \\
    \dv{D_{51}(t)}{t} &= q\frac{\abs{N_5 - N_1}}{l_{51}}D_{51}- \lambda D_{51}. 
    \end{split}
\end{align}
To derive the equilibrium points, we set all time derivatives in the system \eqref{eq:C_5_eq} to zero. We solve the resulting system of algebraic equations by sequentially isolating the variables for concentrations $N_i$ and conductivities $D_{ij}$, recognizing that, at equilibrium, the net flux between any two connected nodes that are neither sources nor sinks must be zero. This approach yields different possible solutions based on which pathways have non-zero conductivities. We derive the first two equilibrium points, $\mathcal{E}_1$ and $\mathcal{E}_2$, by considering cases where flow is confined to distinct paths: 1-2-3 or 1-5-4-3, respectively. The equilibrium points are given by:

\begin{align*}
  \mathcal{E}_{1} 
  &= (N_1^*,N_2^*,N_3^*,N_4^*, N_5^*, D_{12}^*,D_{23}^*,D_{34}^*,D_{45}^*,D_{51}^*) \\
   &= \left(\frac{a}{b}+ \frac{\lambda l_{23}}{q} + \frac{\lambda l_{12}}{q}, \frac{a}{b}+\frac{\lambda l_{23}}{q}, \frac{a}{b}, N_4^*, N_5^*,  \frac{a q}{\lambda}, \frac{a q}{\lambda},0,0,0\right), 
\end{align*}
and
    \begin{align*}
    \mathcal{E}_{2} 
    &= 
    (N_1^*,N_2^*,N_3^*,N_4^*, N_5^*, D_{12}^*,D_{23}^*,D_{34}^*,D_{45}^*,D_{51}^*) \\
    &= \left(\frac{a}{b} + \frac{\lambda l_{34}}{q} + \frac{\lambda l_{45}}{q} +  \frac{\lambda l_{51}}{q}, N_2^*, \frac{a}{b}, \frac{a}{b}+\frac{\lambda l_{34}}{q}, \frac{a}{b} + \frac{\lambda l_{34}}{q} + \frac{\lambda l_{45}}{q}, 0,0, \frac{a q}{\lambda}, \frac{a q}{\lambda}, \frac{a q}{\lambda}\right).
\end{align*} 
When these paths have equal effective lengths, an additional equilibrium, $\mathcal{E}_3$, arises where both paths have balanced flow, given by:
\begin{align*}
    \mathcal{E}_{3} 
    &=(N_1^*,N_2^*,N_3^*,N_4^*, N_5^*, D_{12}^*,D_{23}^*,D_{34}^*,D_{45}^*,D_{51}^*) \\
   &=  \left(\frac{a}{b}+\frac{\lambda l_{23}}{q}+\frac{\lambda l_{12}}{q}, \frac{a}{b}+\frac{\lambda l_{23}}{q}, \frac{a}{b}, \frac{a}{b}+\frac{\lambda l_{34}}{q}, \frac{a}{b} + \frac{\lambda l_{34}}{q} + \frac{\lambda l_{45}}{q}, \frac{aq}{2\lambda},\frac{aq}{2\lambda}, \frac{a q}{2\lambda}, \frac{a q}{2\lambda}, \frac{a q}{2\lambda}\right).
\end{align*}
The derivation requires examining these cases to fully describe all potential steady states. For the detailed step-by-step process, we refer to Appendix \ref{sec:app-non-osc}. The three different combinations of steady states are shown in Figure \ref{fig:C_5-steady}.

\begin{figure}[H]
    \centering
    \includegraphics[width=.96\textwidth]{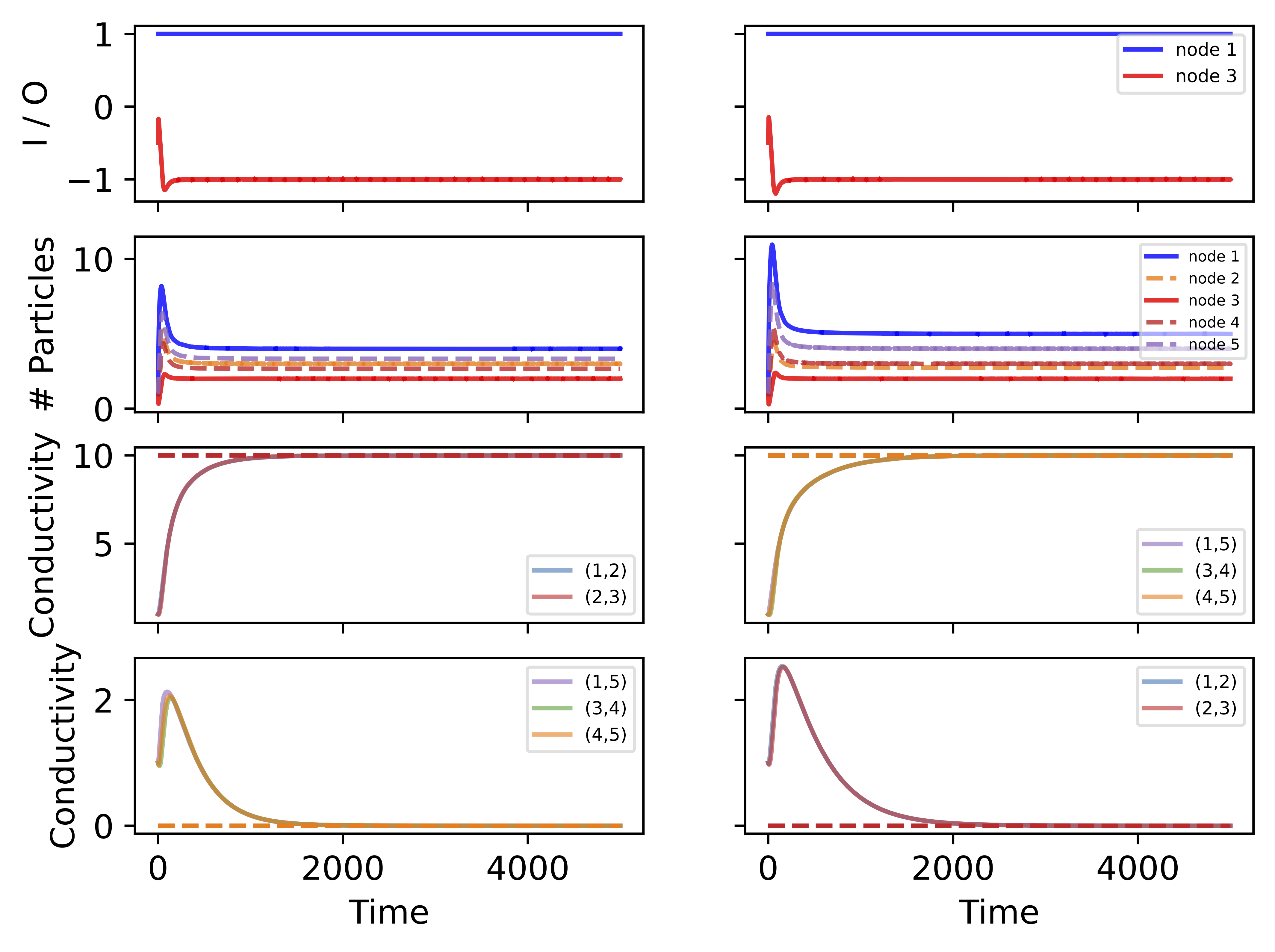}
    \caption{Results from the model on a $C_5$ graph, with node $1$ as a source, node $3$ as a sink, an inflow at rate $a=1$, an outflow at rate $b=0.5$, and the lengths of all edges are $l_{ij} = 10$ apart from $l_{12}$. On the left, we set $l_{12} = 10$, thus the path going through 1-2-3, is the shortest one. On the right, we set $l_{12} = 30$, thus the path going through 1-5-4-3 is the shortest one. The first row shows the input at node $1$ and output at node $3$, the second row plots the number of particles on each node, $N_i(t)$, the third row indicates the conductivity of the edges in the shortest path, and the last row indicates the conductivity of the remaining edges in the graph, where the dashed lines in the corresponding colour show the analytical results of the steady state. }
    \label{fig:res-1s1s}
\end{figure}

In Figure \ref{fig:res-1s1s} we see the temporal dynamics of our model with non-oscillatory sources and sinks on the $C_5$ graph, for two different values of the parameter $l_{12}$. In these simulations, we set the inflow rate $a=1$, the outflow rate $b=0.5$, the parameter of reinforcement $q=0.1$, and the decay parameter $\lambda =0.01$. %as in the bifurcation diagram in Figure \ref{fig:const-bifur}. 
In the left panel $l_{12} = 10$, meaning that 1-2-3 is the shortest path between the source and the sink. Whereas in the right panel $l_{12} = 30$, meaning that 1-5-4-3 is the shortest path between the source and the sink. We observe that the conductivity of the edges in the shortest path increases over time until the steady state is reached, while the conductivity of the edges that are not in the shortest path can also increase at the beginning, but will eventually converge to $0$. For the number of particles, almost all nodes have an initial accumulating period before converging to the steady states. %Furthermore, we observe that the model can find the shortest path on cycle graphs of larger sizes, and also graphs of more general structure, which implies that the feature of finding the shortest path can potentially be generalised to a wider class of graphs; \yu{see more details in Appendix \ref{sec:app-nonoscillatory}}.

The Jacobian matrix of the system \eqref{eq:C_5_eq}, is  given by the following: 
\begin{equation}
 \mathbf{J}= \resizebox{0.85\textwidth}{!}{$
\begin{pmatrix}
-\frac{D_{12}}{l_{12}}-\frac{D_{51}}{l_{51}} & \frac{D_{12}}{l_{12}} & 0 & 0 & \frac{D_{51}}{l_{51}} & \frac{N_2-N_1}{l_{12}} & 0 & 0 & 0& \frac{N_5-N_1}{l_{51}}   \\
\frac{D_{12}}{l_{12}} & -\frac{D_{12}}{l_{12}}-\frac{D_{23}}{l_{23}}  & \frac{D_{23}}{l_{23}} & 0 & 0 & \frac{N_2-N_1}{l_{12}} &\frac{N_3-N_2}{l_{23}} & 0 & 0& 0  \\
0 & \frac{D_{23}}{l_{23}}  & -b-\frac{D_{23}}{l_{23}}-\frac{D_{34}}{l_{34}} & \frac{D_{34}}{l_{34}} & 0 & 0 &\frac{N_3-N_2}{l_{23}} & \frac{N_4-N_3}{l_{34}}& 0& 0  \\
0 & 0  & \frac{D_{34}}{l_{34}} & -\frac{D_{34}}{l_{34}}-\frac{D_{45}}{l_{45}} & \frac{D_{45}}{l_{45}} & 0 & 0 & \frac{N_3-N_4}{l_{34}} & \frac{N_5-N_4}{l_{45}} & 0 \\
\frac{D_{51}}{l_{51}} & 0  & 0  & \frac{D_{45}}{l_{45}} & -\frac{D_{45}}{l_{45}} - \frac{D_{51}}{l_{51}} & 0 & 0 & 0 & \frac{N_4-N_5}{l_{45}} & \frac{N_1-N_5}{l_{51}} \\
\frac{q \sign{(N_1-N_2)}D_{12}}{l_{12}} & \frac{q \sign{(N_1-N_2)}D_{12}}{l_{12}} & 0 & 0 & 0 & \frac{q |N_1-N_2|}{l_{12}}-\lambda & 0 & 0 & 0 & 0 \\
0 & \frac{q \sign{(N_2-N_3)}D_{23}}{l_{23}} & \frac{q \sign{(N_2-N_3)}D_{23}}{l_{23}}  & 0 & 0 &0  &  \frac{q |N_2-N_3|}{l_{23}}-\lambda & 0 & 0 & 0 \\
0 & 0 & \frac{q \sign{(N_3-N_4)}D_{34}}{l_{34}}  & \frac{q \sign{(N_3-N_4)}D_{34}}{l_{34}}  & 0 &0  & 0 & \frac{q |N_3-N_4|}{l_{34}}-\lambda & 0 & 0 \\
0 & 0 & 0  & \frac{q \sign{(N_4-N_5)}D_{45}}{l_{45}}  & \frac{q \sign{(N_4-N_5)}D_{45}}{l_{45}}  &0  & 0 & 0 & \frac{q |N_4-N_5|}{l_{45}}-\lambda & 0 \\
\frac{q \sign{(N_5-N_1)}D_{51}}{l_{51}}  & 0 & 0  & 0 & \frac{q \sign{(N_5-N_1)}D_{51}}{l_{51}}  &0  & 0 & 0 & 0 & \frac{q |N_5-N_1|}{l_{51}}-\lambda 
\end{pmatrix}$}
\end{equation}

In the appendix \ref{sec:app-non-osc}, we show that for all three steady states, this Jacobian matrix will have zero determinant, meaning that at least one eigenvalue of the Jacobian matrix will be 0. Thus, the steady states will be non-hyperbolic, implying that studying the eigenvalues of the Jacobian is not enough to determine stability of the steady states. To do this, a full analysis of the steady states using center manifold theory would be required \cite{wiggins1990}.

Nonetheless, we can use the Jacobian matrix to understand more about what happens to the system. We note that when $D_{12}^*=D_{23}^*=0$ or $D_{34}^*=D_{45}^*=D_{51}^*=0$, at least one row in the Jacobian matrix will contain only zero values, and thus span a slow manifold. For the Jacobian matrix evaluated at $\mathcal{E}_{1}$, the eigenvectors corresponding to the zero eigenvalue will be $(0,0,0,1,0,0,0,0,0,0)$ and $(0,0,0,0,1,0,0,0,0,0)$. The implication of this observation is that there is a line of steady states for both $N_4^*$ and $N_5^*$, which form the slow manifold. For the Jacobian matrix evaluated at $\mathcal{E}_{2}$, the eigenvector corresponding to the zero eigenvalue will be $(0,1,0,0,0,0,0,0,0,0)$, corresponding to a line of steady states for $N_2^*$ which is also a slow manifold. These slow manifolds correspond to the nodes with no edges, meaning that there are particles left behind at those nodes that are stuck because the conductivity converges to 0 for the longer paths.

\begin{figure}[H]
    \centering
    \begin{tabular}{c}
         \includegraphics[width=.465\textwidth]{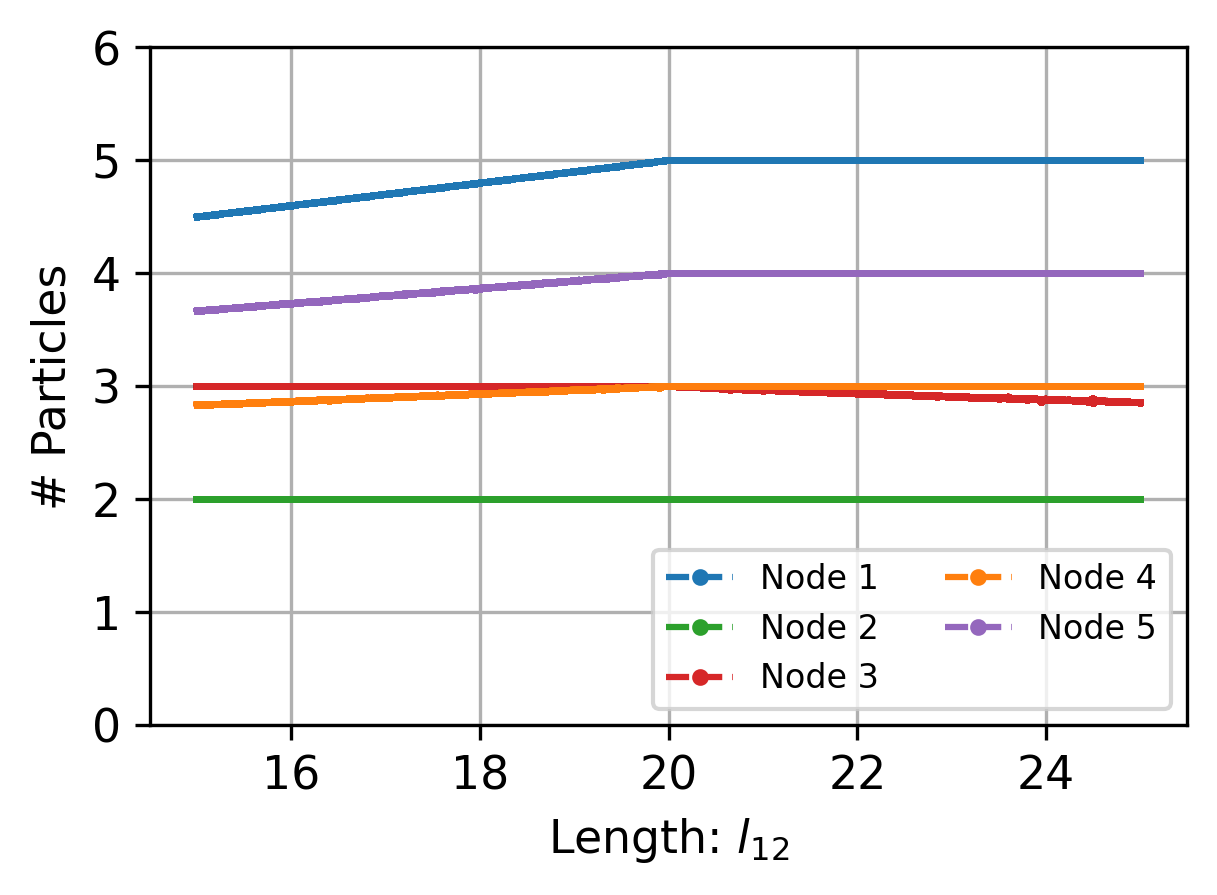} 
        \includegraphics[width=.48\textwidth]{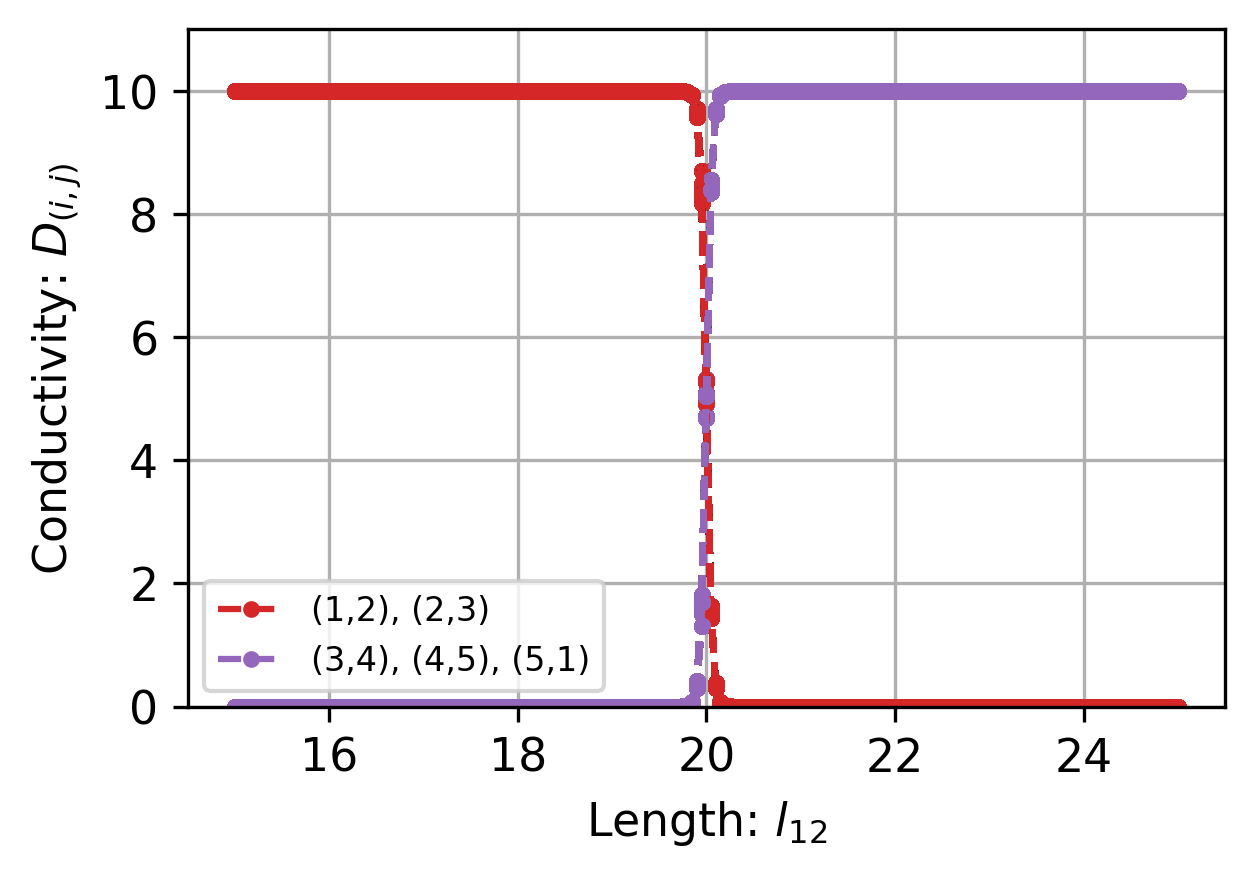}
    \end{tabular}
    \caption{Bifurcation diagrams with $l_{12}$ as bifurcation parameter. Here $l_{23}=l_{34}=l_{45}=l_{51}=10$, $a=1.0$, $b=0.5$, $q=0.1$ and $\lambda=0.01$. The left plot shows the number of particles at each node, while the right plot displays the conductivity $D_{(i,j)}$ of the edges, both as a function of $l_{12}$. The diagrams are generated by simulating the equation system \eqref{eq:C_5_eq_steady} through 100,000 time steps, each time initiated with random initial conditions. This process is repeated 10 times for each value of $l_{12}$, with the bifurcation diagram displaying the results from the last 100 time steps of each simulation}
    \label{fig:const-bifur}
\end{figure}

\begin{figure}[H]
\begin{subfigure}[b]{0.31\textwidth}
   \resizebox{\linewidth}{!}{\begin{tikzpicture}
   [vertex/.style={circle, draw}] 
\foreach \Z\X[count=\Y] in {1/wong-colour1,2/wong-colour7,3/wong-colour4,4/wong-colour7,5/wong-colour7}
{\node[vertex, fill=\X] (x-\Y) at ({72*\Y+162}:2){\Z}; }
\foreach \X[count=\Y] in {0,...,4}
\draw[line width=2.4pt] (x-1) -- (x-2) node[midway] (d_ij) {};
\draw[line width=2.4pt] (x-2) -- (x-3) node[midway] (d_ij2) {};
\node [above of = d_ij] (D_ij) {$\frac{aq}{\lambda}$};
\node [right of = d_ij2] (D_ij2) {$\frac{aq}{\lambda}$};
\draw[line width = 0.7pt, shorten <= -0.1cm, shorten >= -0.1cm] (D_ij) to[bend right=20] (d_ij);
\draw[line width = 0.7pt, shorten <= -0.1cm, shorten >= -0.1cm] (D_ij2) to[bend right=10] (d_ij2);
\node at ([xshift=-1.8cm]x-1.west) (N_1) {$\frac{a}{b}+ \frac{\lambda l_{23}}{q} + \frac{\lambda l_{12}}{q}$};
\draw[line width = 0.7pt, shorten <= -0.1cm, shorten >= -0.2cm] (N_1) to[bend right=5] (x-1);
\node at ([xshift=1.8cm]x-2.south) (N_2) {$\frac{a}{b}+\frac{\lambda l_{23}}{q}$};
\draw[line width = 0.7pt, shorten <= -0.1cm, shorten >= -0.2cm] (N_2) to[bend left=5] (x-2);
\node at ([xshift=1.3cm]x-3.west) (N_3) {$\frac{a}{b}$};
\draw[line width = 0.7pt, shorten <= -0.1cm, shorten >= -0.2cm] (N_3) to[bend right=5] (x-3);
\node at ([yshift=0.7cm]x-4.north) (N_4) {$N_4^*$};
\draw[line width = 0.7pt, shorten <= -0.1cm, shorten >= -0.2cm] (N_4) to[bend right=5] (x-4);
\node at ([xshift=-1.0cm]x-5.west) (N_5) {$N_5^*$};
\draw[line width = 0.7pt, shorten <= -0.1cm, shorten >= -0.2cm] (N_5) to[bend right=5] (x-5);
\end{tikzpicture}
}
\caption{Non-hyperbolic stable equilibrium when
$l_{12}+l_{23}<l_{34}+l_{45}+l_{51}$ ($\mathcal{E}_1$).}
\end{subfigure}
\hfill\begin{subfigure}[b]{0.31\textwidth}
   \resizebox{\linewidth}{!}{
   \begin{tikzpicture}
 [vertex/.style={circle, draw}] 
\foreach \Z\X[count=\Y] in {1/wong-colour1,2/wong-colour7,3/wong-colour4,4/wong-colour7,5/wong-colour7}
{\node[vertex, fill=\X] (x-\Y) at ({72*\Y+162}:2){\Z}; }
\foreach \X[count=\Y] in {0,...,4}
\draw[line width=1.2pt] (x-3) -- (x-4) node[midway] (d_ij3) {};
\draw[line width=1.2pt] (x-4) -- (x-5) node[midway] (d_ij4) {};
\draw[line width=1.2pt] (x-5) -- (x-1) node[midway] (d_ij5) {};
\draw[line width=1.2pt] (x-1) -- (x-2) node[midway] (d_ij) {};
\draw[line width=1.2pt] (x-2) -- (x-3) node[midway] (d_ij2) {};
\node [above right of = d_ij3] (D_ij3) {$\frac{aq}{2\lambda}$};
\node [above left of = d_ij4] (D_ij4) {$\frac{aq}{2\lambda}$};
\node [left of = d_ij5] (D_ij5) {$\frac{aq}{2\lambda}$};
\node [above of = d_ij] (D_ij) {$\frac{aq}{2\lambda}$};
\node [right of = d_ij2] (D_ij2) {$\frac{aq}{2\lambda}$};
\draw[line width = 0.7pt, shorten <= -0.1cm, shorten >= -0.1cm] (D_ij) to[bend right=20] (d_ij);
\draw[line width = 0.7pt, shorten <= -0.1cm, shorten >= -0.1cm] (D_ij2) to[bend right=10] (d_ij2);
\draw[line width = 0.7pt, shorten <= -0.1cm, shorten >= -0.1cm] (D_ij3) to[bend right=20] (d_ij3);
\draw[line width = 0.7pt, shorten <= -0.1cm, shorten >= -0.1cm] (D_ij4) to[bend right=10] (d_ij4);
\draw[line width = 0.7pt, shorten <= -0.1cm, shorten >= -0.1cm] (D_ij5) to[bend right=10] (d_ij5);
\node at ([xshift=-1.8cm]x-1.west) (N_1) {$\frac{a}{b}+\frac{\lambda l_{23}}{q}+\frac{\lambda l_{12}}{q}$};
\draw[line width = 0.7pt, shorten <= -0.1cm, shorten >= -0.2cm] (N_1) to[bend right=5] (x-1);
\node at ([xshift=1.8cm]x-2.south) (N_2) {$\frac{a}{b}+\frac{\lambda l_{23}}{q}$};
\draw[line width = 0.7pt, shorten <= -0.1cm, shorten >= -0.2cm] (N_2) to[bend left=5] (x-2);
\node at ([xshift=1.3cm]x-3.west) (N_3) {$\frac{a}{b}$};
\draw[line width = 0.7pt, shorten <= -0.1cm, shorten >= -0.2cm] (N_3) to[bend right=5] (x-3);
\node at ([yshift=0.7cm]x-4.north) (N_4) {$ \frac{a}{b}+\frac{\lambda l_{34}}{q}$};
\draw[line width = 0.7pt, shorten <= -0.1cm, shorten >= -0.2cm] (N_4) to[bend right=5] (x-4);
\node at ([xshift=-1.9cm]x-5.west) (N_5) {$\frac{a}{b} + \frac{\lambda l_{34}}{q} + \frac{\lambda l_{45}}{q}$};
\draw[line width = 0.7pt, shorten <= -0.1cm, shorten >= -0.2cm] (N_5) to[bend right=5] (x-5);
\end{tikzpicture}}
\caption{Non-hyperbolic stable equilibrium when $l_{12}+l_{23}=l_{34}+l_{45}+l_{51}$ ($\mathcal{E}_3$).}
\end{subfigure}
\hfill
\begin{subfigure}[b]{0.31\textwidth}
   \resizebox{\linewidth}{!}{\begin{tikzpicture}
   [vertex/.style={circle, draw}] 
\foreach \Z\X[count=\Y] in {1/wong-colour1,2/wong-colour7,3/wong-colour4,4/wong-colour7,5/wong-colour7}
{\node[vertex, fill=\X] (x-\Y) at ({72*\Y+162}:2){\Z}; }
\foreach \X[count=\Y] in {0,...,4}
\draw[line width=2.4pt] (x-3) -- (x-4) node[midway] (d_ij3) {};
\draw[line width=2.4pt] (x-4) -- (x-5) node[midway] (d_ij4) {};
\draw[line width=2.4pt] (x-5) -- (x-1) node[midway] (d_ij5) {};
\node [above right of = d_ij3] (D_ij3) {$\frac{aq}{\lambda}$};
\node [above left of = d_ij4] (D_ij4) {$\frac{aq}{\lambda}$};
\node [left of = d_ij5] (D_ij5) {$\frac{aq}{\lambda}$};
\draw[line width = 0.7pt, shorten <= -0.1cm, shorten >= -0.1cm] (D_ij3) to[bend right=20] (d_ij3);
\draw[line width = 0.7pt, shorten <= -0.1cm, shorten >= -0.1cm] (D_ij4) to[bend right=10] (d_ij4);
\draw[line width = 0.7pt, shorten <= -0.1cm, shorten >= -0.1cm] (D_ij5) to[bend right=10] (d_ij5);
\node at ([xshift=-1.8cm]x-1.west) (N_1) {$\frac{a}{b} + \frac{\lambda l_{34}}{q} + \frac{\lambda l_{45}}{q}$};
\draw[line width = 0.7pt, shorten <= -0.1cm, shorten >= -0.2cm] (N_1) to[bend right=5] (x-1);
\node at ([xshift=1.8cm]x-2.south) (N_2) {$N_2^*$};
\draw[line width = 0.7pt, shorten <= -0.1cm, shorten >= -0.2cm] (N_2) to[bend left=5] (x-2);
\node at ([xshift=1.3cm]x-3.west) (N_3) {$\frac{a}{b}$};
\draw[line width = 0.7pt, shorten <= -0.1cm, shorten >= -0.2cm] (N_3) to[bend right=5] (x-3);
\node at ([yshift=0.7cm]x-4.north) (N_4) {$ \frac{a}{b}+\frac{\lambda l_{34}}{q}$};
\draw[line width = 0.7pt, shorten <= -0.1cm, shorten >= -0.2cm] (N_4) to[bend right=5] (x-4);
\node at ([xshift=-1.9cm]x-5.west) (N_5) {$\frac{a}{b} + \frac{\lambda l_{34}}{q} + \frac{\lambda l_{45}}{q}$};
\draw[line width = 0.7pt, shorten <= -0.1cm, shorten >= -0.2cm] (N_5) to[bend right=5] (x-5);
\end{tikzpicture}}
\caption{Non-hyperbolic stable equilibrium when $l_{12}+l_{23}>l_{34}+l_{45}+l_{51}$ ($\mathcal{E}_2$).}
\end{subfigure}
\caption{Different equilibrium points for the model on the $C_5$ graph with non-oscillatory sources and sinks, where node 1 (blue) is a source, and node 3 (green) is a sink. In (a) we see the stable equilibrium when 1-2-3 is the shortest path, in (b) we see the stable equilibrium when both paths have equal length, and in (c) we see the stable equilibrium when 1-5-4-3 is the shortest path. }
  \label{fig:C_5-steady}
\end{figure}
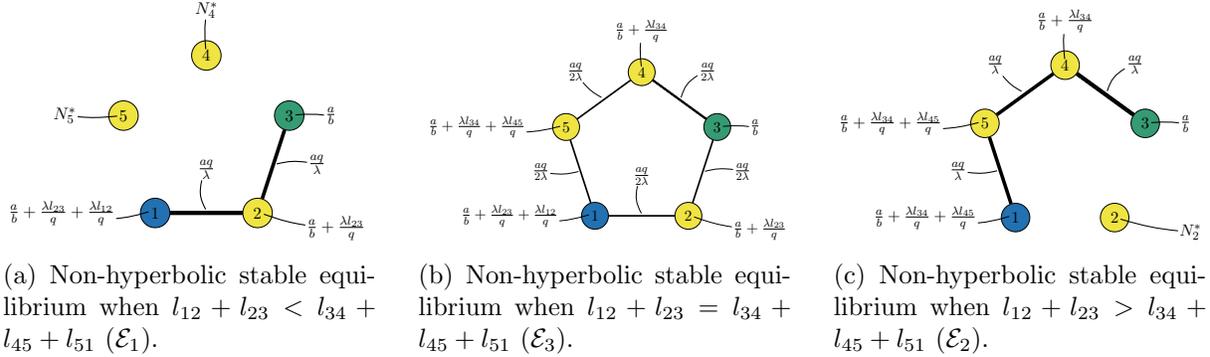

The numerical bifurcation diagram in Figure \ref{fig:const-bifur} looks at the role of $l_{12}$ as a bifurcation parameter. 
In Figure \ref{fig:const-bifur}(left), we see that the number of particles at node 3, $N_3^*$ is always $a/b$, independent on the length $l_{12}$. However, $N_1^*$, $N_4^*$ and $N_5^*$, increase linearly as $l_{12}$ approaches the equilibrium point 20, after which they all are constant. $N_2^*$ is constant ($\frac{a}{b}+\frac{\lambda l_{23}}{q}=3$) until the bifurcation point $l_{12}=20$, where it instead decreases linearly. In Figure \ref{fig:const-bifur}(right), we see that the conductivity of the edges on the shortest path is always $aq/\lambda$, and 0 for the edges that are not the shortest path. When the length of $l_{12}$ approaches the bifurcation point 20, the simulations have not reached steady state after 100,000 time steps, and is thus not 0 and $aq/\lambda$ in the plot. When $l_{12}=20$ the two paths are of equal lengths and $D_{ij}=\frac{aq}{2\lambda}$ for all $(i,j)$. 

Figure \ref{fig:C_5-steady} illustrates the overall picture. In (a) we see that when 1-2-3 is the shortest path, there is a non-hyperbolic equilibrium where the conductivity on the edges of the longest path is zero, but the conductivity of the edges of the shortest path is $\frac{aq}{\lambda}$. The number of particles at the nodes with no connections, $N_4^*$ and $N_5^*$, spans the slow manifold. In (b) the paths 1-2-3 and 3-4-5-1 are of equal lengths and the conductivity of all edges is $\frac{aq}{2\lambda}$. When the path through nodes 1-2-3 is longer than 1-5-4-3 (Figure \ref{fig:C_5-steady} (c)), the non-hyperbolic equilibrium point corresponds to having zero-conductivity on the edges going through node 2, and a line of steady states for the number of particles on node 2, $N_2^*$. The conductivity on the edges on the shortest path is $\frac{aq}{\lambda}$. 

% \yu{Also add the input/output on the top! Left: shortest path 1, Right: shortest path 2.}\todo{Two of these figures. }

\subsection{Oscillatory nodes}
\label{sec:res-osc}
We now use simulations to explore the behaviour of the model with oscillatory nodes on a $C_5$ graph. We start from the case when the two oscillatory nodes have phase difference $\pi$ (i.e. completely out of phase) and then explore the behaviour of the model when we vary the phase difference between $-\pi$ and $\pi$. Finally, we look at how the amplitude and frequency of the oscillatory nodes affect the model. Throughout the section, we set nodes $1$ and $3$ to be the oscillatory nodes, and the length of all edges to be $l_{ij} = 10$; thus the shortest path between the two oscillatory nodes is 1-2-3.

\begin{figure}[H]
    \centering
    \includegraphics[width=.96\textwidth]{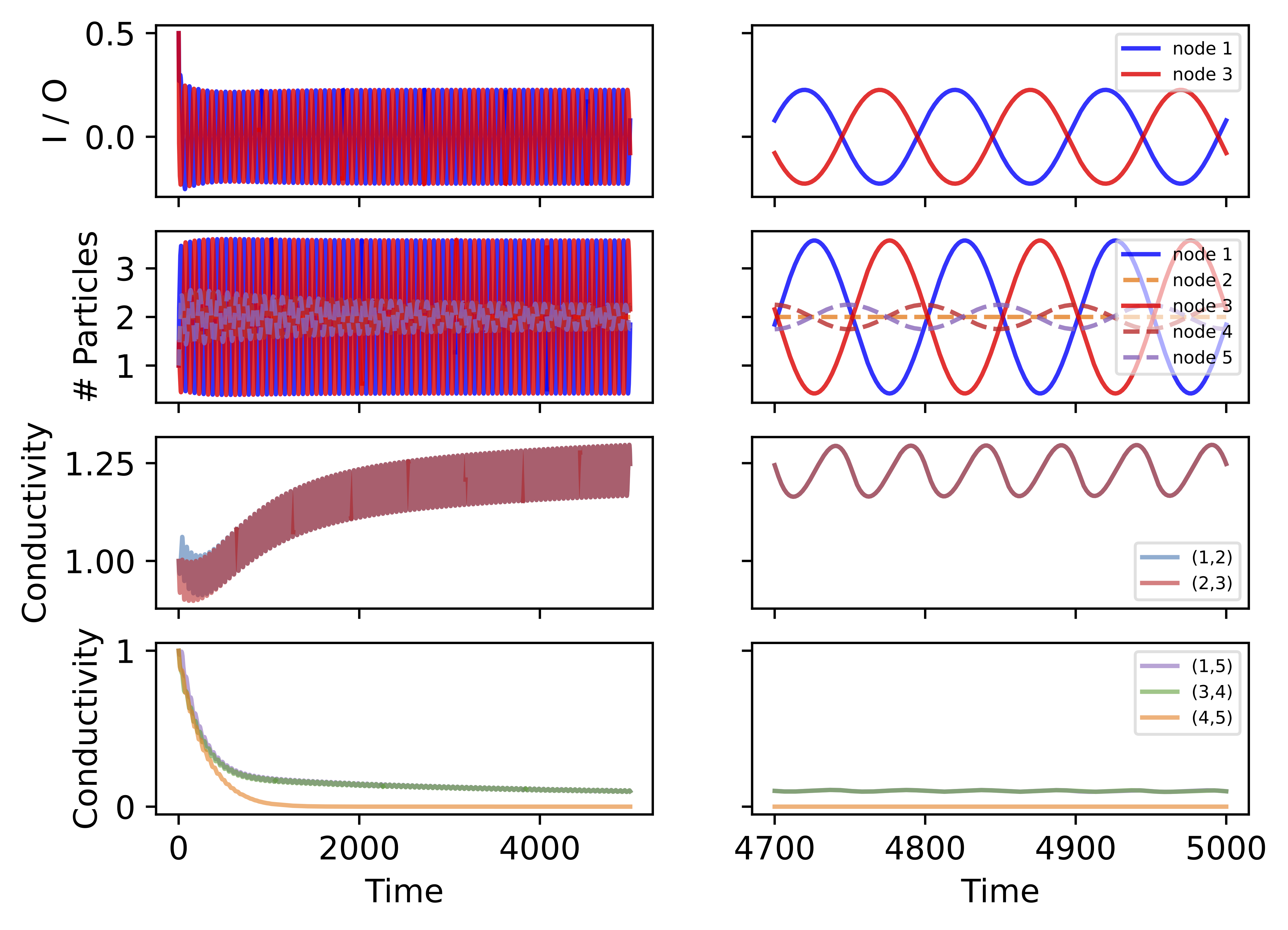}
    \caption{Results from the model on a $C_5$ graph, with nodes 1 and 3 as the oscillating nodes, rate $b=0.5$, amplitude $A_1 = A_3 = A=1$, frequency $\theta_1 = \theta_3 =\theta=0.01$, and phases $\phi_1 = 0$ and $\phi_3 = \pi$, with the same reinforcement parameter, $q=0.1$, and decay parameter, $\lambda=0.01$, as in Figure \ref{fig:res-1s1s}. The first row plots the actual input/output to the two nodes, the second row shows the number of particles in each node, the third row indicates the conductivity of the edges in the shortest path and the last row indicates the conductivity of the remaining edges in the graph, with the full profile on the left and the change in the long term on the right.}
    \label{fig:res-osln-1s1s}
\end{figure}

%In these simulations, the rate $b=0.1$, the amplitude $A_1 = A_3 = A = 1$, frequency $\theta_1 = \theta_3 = \theta = 0.01$, and phases $\phi_1 = 0$ and $\phi_3 = \pi$, with the same reinforcement parameter $q=0.1$ and decay parameter $\lambda=0.01$, as before. 
\textbf{Phase difference.} 
Figure \ref{fig:res-osln-1s1s} shows the temporal dynamics of our model when the two oscillatory nodes have phase difference $\pi$. The top two panels of Figure \ref{fig:res-osln-1s1s} show the oscillating inputs and outputs over two different time windows. The number of particles at node 1 and 3 have the same frequencies as the inputs and outputs, but with a short delay as the particles flow into and out of the nodes.  Oscillations with similar frequencies can be seen at nodes 4 and 5, which are on the longer path, but with much smaller amplitude and a different phase. Node 2, which is on the shortest path, reaches steady state. 

The conductivity of the edges oscillates over time, but they also change over a longer time scale. Specifically, the median conductivity of the edges on the shortest path generally increases, but with a rate of increase that tends very slowly towards $0$.  Interestingly, the number of particles oscillates at a frequency that is almost half that of the conductivity on the edges. This is presumably because the flow shuttles backwards and forwards. Moreover, the amplitude of the conductivity changes is much smaller than that of the oscillations of particles at nodes 1 and 3 (note the scale on the axis for conductivity in Figure \ref{fig:res-osln-1s1s}). The median conductivity of the edges that are not on the shortest path decrease over time and converge to values either $0$ or very close to $0$ (with no or very small oscillations) depending on the initial conditions (see the bottom row of Figure \ref{fig:res-osln-1s1s}).

We observe similar features of the model on cycles of larger sizes and also graphs of more general structure (see more details in Appendix \ref{sec:app-oscil}).
% When the oscillatory nodes have phase difference $\pi$, the conductivity of the edges in the shortest path is significantly larger than that of the edges on the longer path (see Fig.~\ref{fig:res-osln-1s1s}). In the simulations, we set the rate $b=0.5$, the amplitude $A=1$, and the common frequency $\theta=0.01$, with the same reinforcement $q$ and decay parameter $\lambda$ as in Sec.~\ref{sec:res-const}. We observe that the input/output of the two oscillatory nodes do have positive and negative values over time, which indicates that they alternative between effective sources and sinks in the model. The number of particles also oscillate over time, apart from node $2$ which has reached the steady state. We note that node $2$ is exactly in the centre of the shortest path between the two oscillatory nodes. Similarly, the conductivity of the edges also oscillates over time, apart from edge $(4,5)$ which is not incident on any of the two oscillatory nodes. Furthermore, the conductivity oscillate at a higher frequency than the number of particles. We also observe similar features of the model on cycles of larger sizes and also graphs of more general structure; \yu{see more details in Appendix \ref{sec:app-oscil}}.

\begin{figure}[H]
    \centering
    \begin{tabular}{c}
         \includegraphics[width=.96\textwidth]{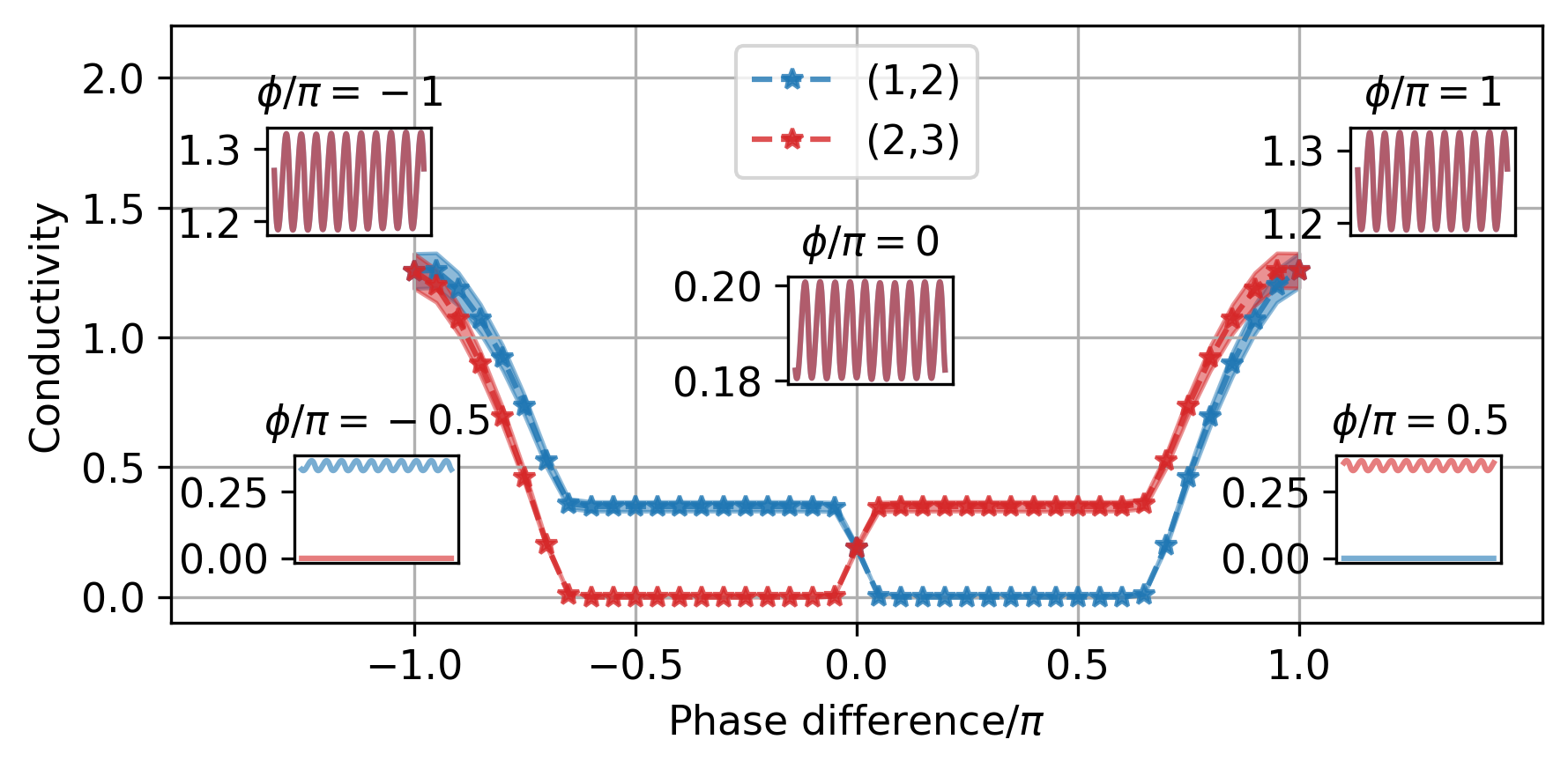} \\
         \includegraphics[width=.96\textwidth]{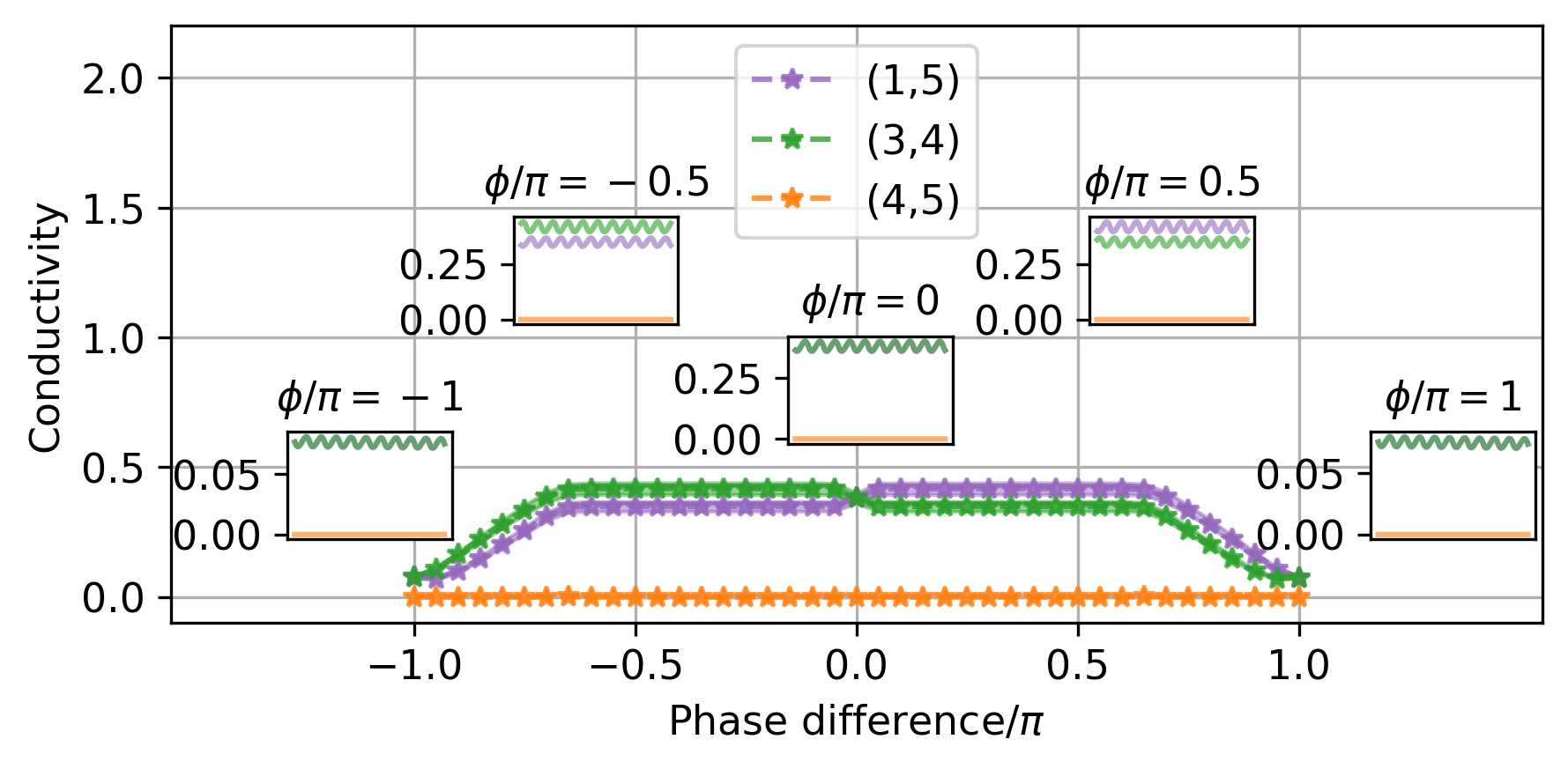}
    \end{tabular}
    \caption{Bifurcation diagrams of the conductivity (thickness) of the edges on a $C_5$ graph, with phase difference between the two oscillators as the bifurcation parameter. The top figure shows the conductivity of the edges on the shortest path, and the bottom figure shows the conductivity of the edges on the longest path. Parameters used are consistent with those in Figure \ref{fig:res-osln-1s1s}. Lines marked with asterisks represent median conductivity values, and shading denotes the amplitude of conductivity. Inserted subfigures show how the conductivity of edges changes as a function of time for different values of the phase difference.}
    \label{fig:osln-cycle5}
\end{figure}

In Figure \ref{fig:osln-cycle5}, we examine the change in conductivity as a function of the phase difference $\phi_3-\phi_1$, ranging from $-\pi$ to $\pi$. When the phase difference is $-\pi$, the conductivity of the edges along the shortest path is at its maximum (see the points corresponding to a phase difference of $-\pi$ at the leftmost part of Figure \ref{fig:osln-cycle5}). As the phase difference approaches one, the conductivity of the edges along the shortest path decreases, with the conductivity of the edge $(2,3)$ approaching $0$. Correspondingly, the conductivity of the edges $(3,4)$ and $(1,5)$ slowly increases. In this scenario, the two oscillating nodes are connected via the shortest path and have additional arms extending along the longer path, although these arms do not form a connection.

For phase differences between approximately $-0.65\pi$ and $0$, the network becomes disconnected: the conductivity of the edges remains constant and equals zero for the edge $(1,2)$, as indicated in the corresponding region of Figure \ref{fig:osln-cycle5}. When the phase difference is $0$, the two edges on the shortest path have identical conductivity values, as do the other two disconnected arms on the longer path. As the phase difference increases from $0$ to about $0.65\pi$, the conductivity of the edges is constant once more, with edge $(2,3)$ now having zero conductivity. This pattern repeats, in the sense that the results are invariant under the relabeling of node $1$ with $3$, and nodes $4$ with $5$.
 
% When the phase difference between the two oscillatory nodes is not necessarily $\pi$, the behaviour of the model has more variance, where when the phase difference is not far from $\pi$ or $-\pi$, it can still separate the edges in the shortest path from the others by higher conductivity values; see Fig.~\ref{fig:osln-cycle5}. We observe symmetric performance of the conductivity between the region when the phase difference is greater than $0$ and the one when the phase difference is smaller than $0$, in the sense that the results are invariant if we change the label of node $1$ with $3$, and also the label of node $4$ with $5$. In the case when there is no phase difference between the two oscillatory nodes, only edges that are incident on these two oscillatory nodes have nonzero conductivity. 

\begin{figure}[H]
    \centering
    \hspace*{-3em}
     \includegraphics[width=1.1\textwidth]{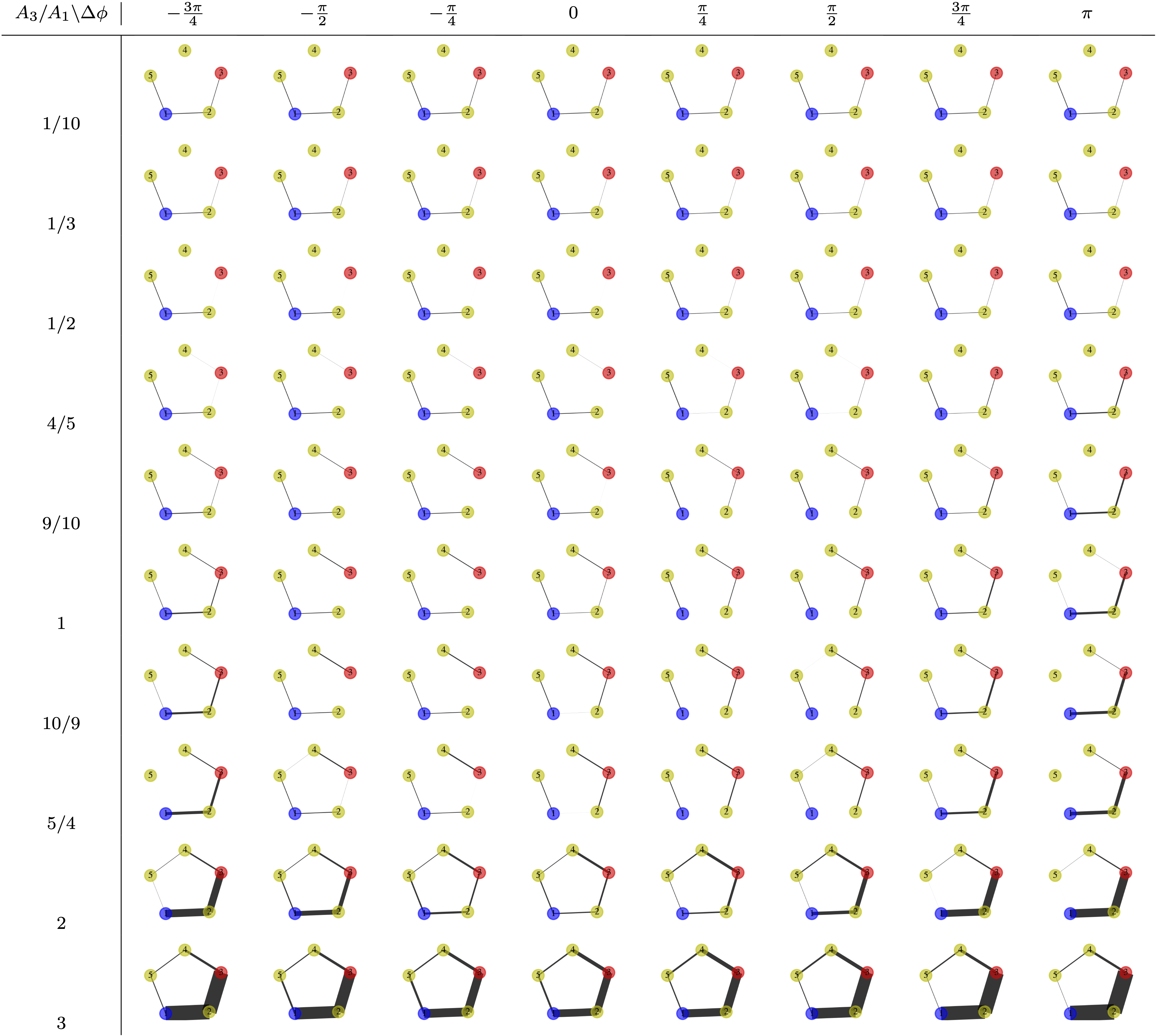}
    \caption{Conductivity results (median) obtained from the model on a $C_5$ graph over a sufficiently long time, subject to a changing amplitude $\alpha = A_3/A_1$ ($A_1 = 1$) and phase difference $\Delta\phi = \phi_3 -\phi_1$ ($\phi_1 = 0$). Other parameters are consistent with those in Figure \ref{fig:res-osln-1s1s}. The thickness of the edges is proportional to the conductivity value.}
    \label{fig:tab-mp}
\end{figure}

\textbf{Amplitude, frequency and phase differences.}
When the amplitude or frequency of the two oscillatory nodes is not exactly the same, the model exhibits a rich span of behaviours. In Figure \ref{fig:tab-mp}, we explore the resulting graphs as we vary both the phase difference $\Delta\phi = \phi_3 - \phi_1$ and the amplitude ratio $A_3/A_1$. In these figures, the thickness of the edges indicates their conductivity after the simulation has run for a sufficient length of time. The middle row, where $A_3/A_1=1$, corresponds to the scenario presented in Figure \ref{fig:osln-cycle5}, and discussed in detail in the previous paragraph. Here, the graph transitions from initially being connected via the shortest path ($\Delta\phi = -3\pi/4$) with additional arms, to becoming disconnected (e.g., $\Delta\phi = -\pi/4$), and then to reconnecting ($\Delta\phi = 3\pi/4$), with the majority of the flow following the shortest path ($\Delta\phi = \pi$).

 When $A_3/A_1$ is close to $1$, the results exhibit similar phase transitions to those seen when $A_3/A_1 = 1$; refer to the rows corresponding to $\Delta\phi = 9/10, 10/9$ in Figure \ref{fig:tab-mp}. As the amplitude ratio $A_3/A_1$ increases, edges generally exhibit higher conductivity values. For instance, when $A_3/A_1 = 2$, all edges display nonzero conductivity values for all possible phase differences $\Delta\phi$. Moreover, if $\Delta\phi$ is close to $-\pi$ or $\pi$, the edges along the shortest path exhibit higher conductivity values than others. With $A_3/A_1=3$ (and for larger ratios), the edges on the shortest path consistently show higher conductivity values than others, regardless of $\Delta\phi$. Thus, differences in amplitude can lead to a greater concentration of flow along the shorter path.
 
When the amplitude ratio $A_3/A_1$ is less than one, edges generally exhibit lower conductivity values. Specifically, when $A_3/A_1=1/2$, no edges incident on oscillatory node $3$ have nonzero conductivity values when the phase difference $\Delta\phi \le 0$. There is also only a weak connection between oscillatory node $3$ and node $2$ along the shortest path when $\Delta\phi > 0$. When $A_3/A_1$ is $1/5$ or $1/10$, this results in a disconnected network for all possible values of $\Delta\phi$.

\begin{figure}[H]
    \centering
    \hspace*{-3em}
    \scriptsize
    \includegraphics[width=1.1\textwidth]{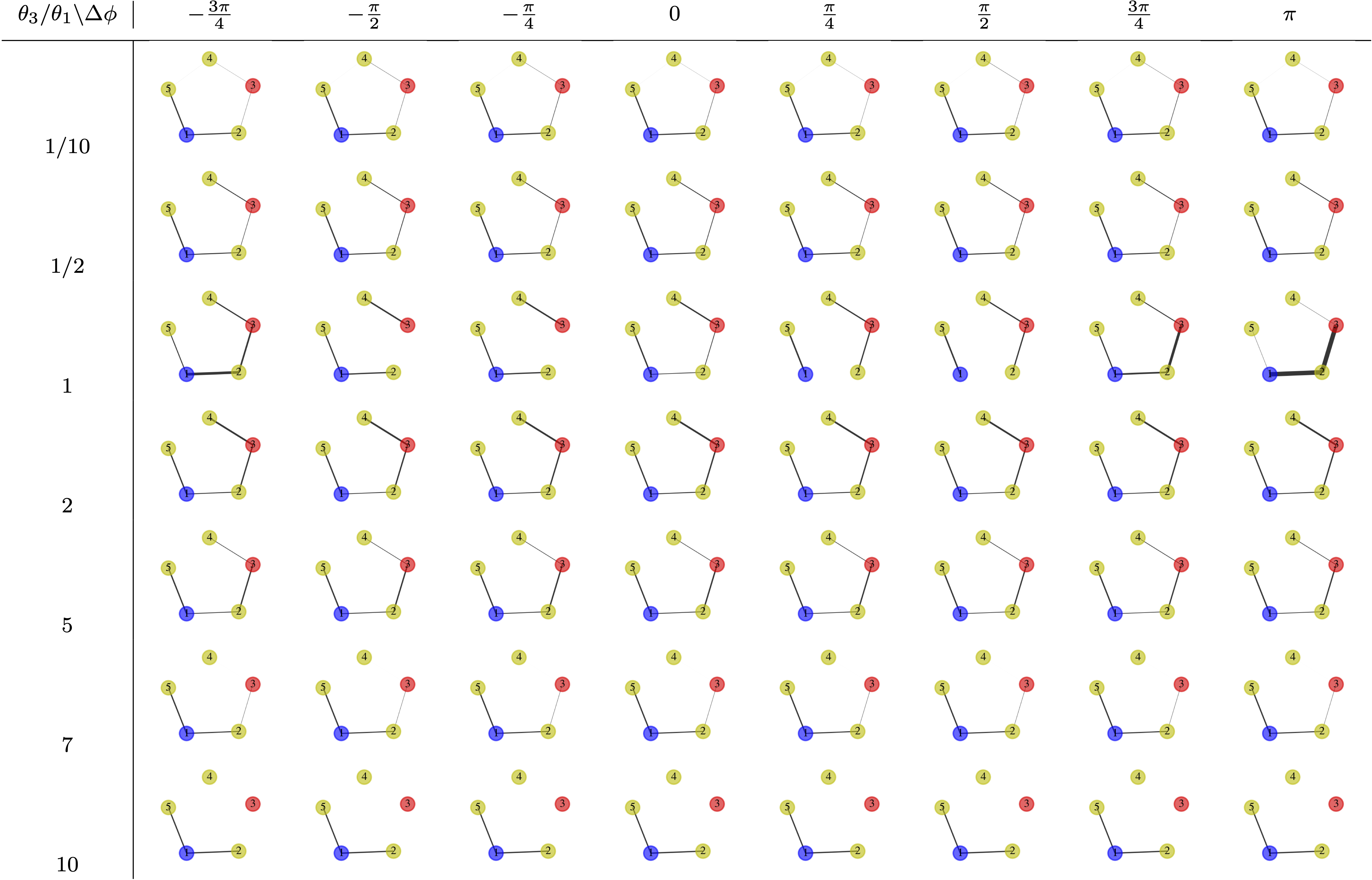}
    \caption{Conductivity results (median) obtained from the model on a $C_5$ graph over a sufficiently long time, subject to a changing frequency ratio $\gamma = \theta_3/\theta_1$ ($\theta_1 = 0.01$) and phase difference $\Delta\phi = \phi_3 -\phi_1$ ($\phi_1 = 0$). Other parameters are consistent with those in Figure \ref{fig:res-osln-1s1s}. The thickness of the edges is proportional to the conductivity value.}
    \label{fig:tab-fp}
\end{figure}

In Figure \ref{fig:tab-fp}, we examine the resulting graphs as we change both phase difference $\Delta\phi = \phi_3 - \phi_1$ and frequency ratio $\theta_3/ \theta_1$. In these figures, the thickness of the edges, again, indicates their conductivity after the simulation has been run for a sufficiently long time. The third row where $\theta_3/ \theta_1 = 1$ corresponds to the case presented in Figure \ref{fig:tab-mp}, exhibiting different features as the phase difference changes, as in the case of $A_3/A_1 = 1$ in the last paragraph. 

However, when the frequency ratio $\theta_3/ \theta_1 \ne 1$, the resulting graphs remain the same at all phase difference values. 
As the frequency ratio $\theta_3/ \theta_1$ increases, less edges have nonzero conductivity in general. For example, when $\theta_3/ \theta_1 = 2$ or $5$, only edges that are incident on either of the two oscillatory nodes have nonzero conductivity values. When $\theta_3/ \theta_1 = 7$, the edge that is incident on oscillatory node $3$ but not in the shortest path now has zero conductivity. In these two cases, the two oscillatory nodes are then connected via the shortest path and have additional arm(s). When $\theta_3/ \theta_1 = 10$, no edges that are incident on oscillatory node $3$ have nonzero conductivity, thus the graph becomes disconnected. Hence, frequency can affect the exploration depth of the flow on the graph.

As the frequency ratio $\theta_3/ \theta_1$ decreases, more edges have nonzero conductivity on the whole. Specifically, when $\theta_3/ \theta_1 = 1/2$, all edges apart from the one that is not incident on either of the oscillatory nodes have nonzero conductivity. When $\theta_3/ \theta_1 = 1/10$, all edges now have nonzero conductivity values, even though the conductivity of the one that has zero conductivity in the previous case is very small. 
% The results imply the important role that frequency play in the effective communication between the oscillatory nodes through the graph, where smaller frequency is preferred for more edges to be explored. 

With all above, we conclude that amplitude, frequency and phase all contain important information for the model to learn interesting patterns in the graph and the change in the environment.

% \subsection{Feedback?}

\subsection{Large graphs}
\label{sec:res-large}
% \yu{@Yu: to-do.}
% \begin{itemize}
%     \item Setting: $n\sim 1000$ nodes RGG, and $10\%$ input/output nodes, and uniformly random phases, amplitude ratios between $1/5$ and $5$, and no frequency differences.
%     \item Output: (i) a snapshot at ``steady" state, and (ii) a video showing the change of behaviour. 
% \end{itemize}

We now proceed to examine the behaviour of the model, featuring oscillatory nodes, on larger graphs (refer to Section \ref{sec:large_graphs} for the setup). The purpose of this investigation is to determine whether the model can reproduce large-scale patterns with shortest path connections between the nodes in the presence of external oscillations.

\begin{figure}[H]
    \centering
    \begin{tabular}{cc}
        \includegraphics[width=.48\textwidth]{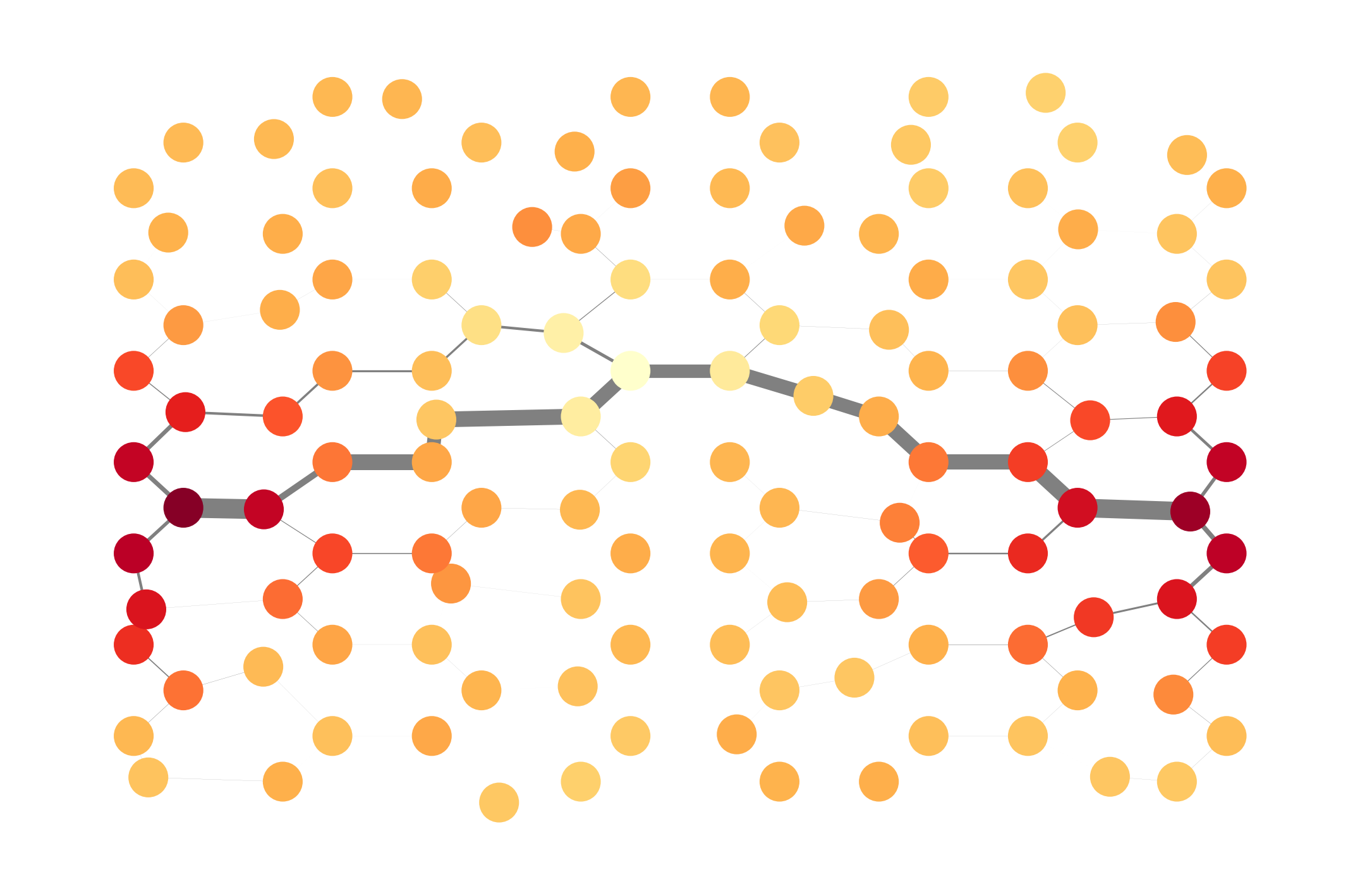} & \includegraphics[width=.48\textwidth]{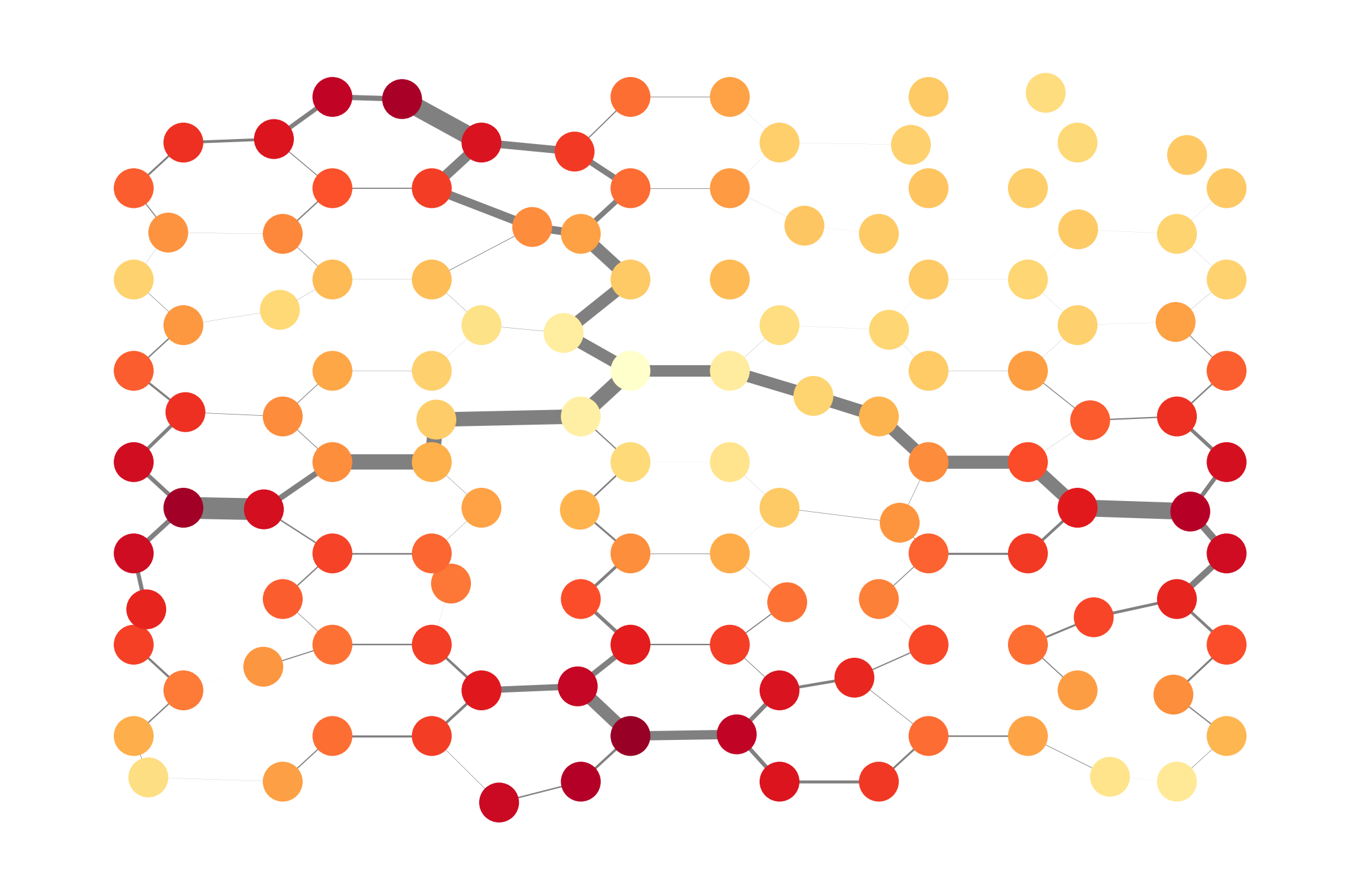} \\
        (a) & (d) \\
        \includegraphics[width=.48\textwidth]{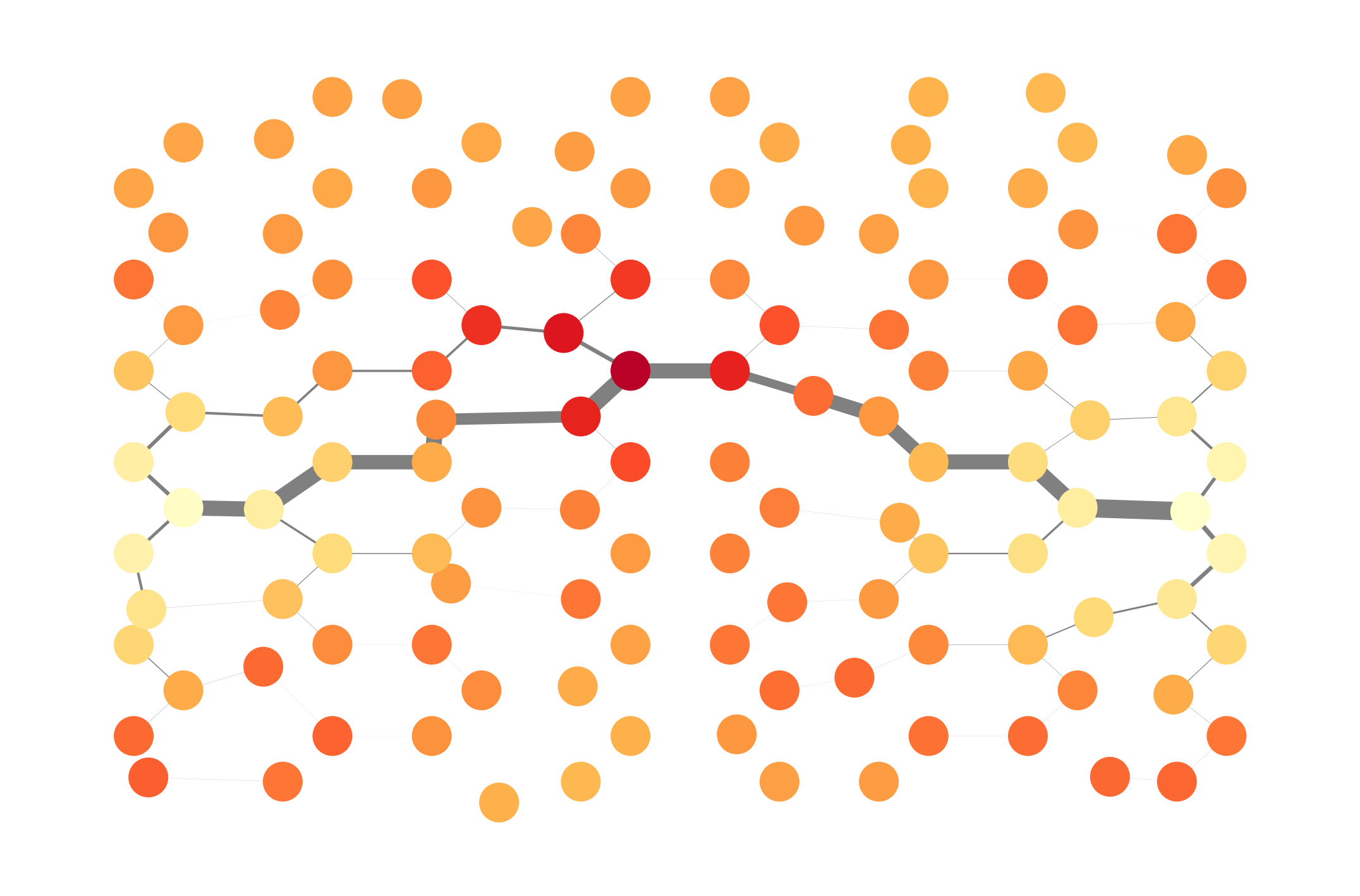} & \includegraphics[width=.48\textwidth]{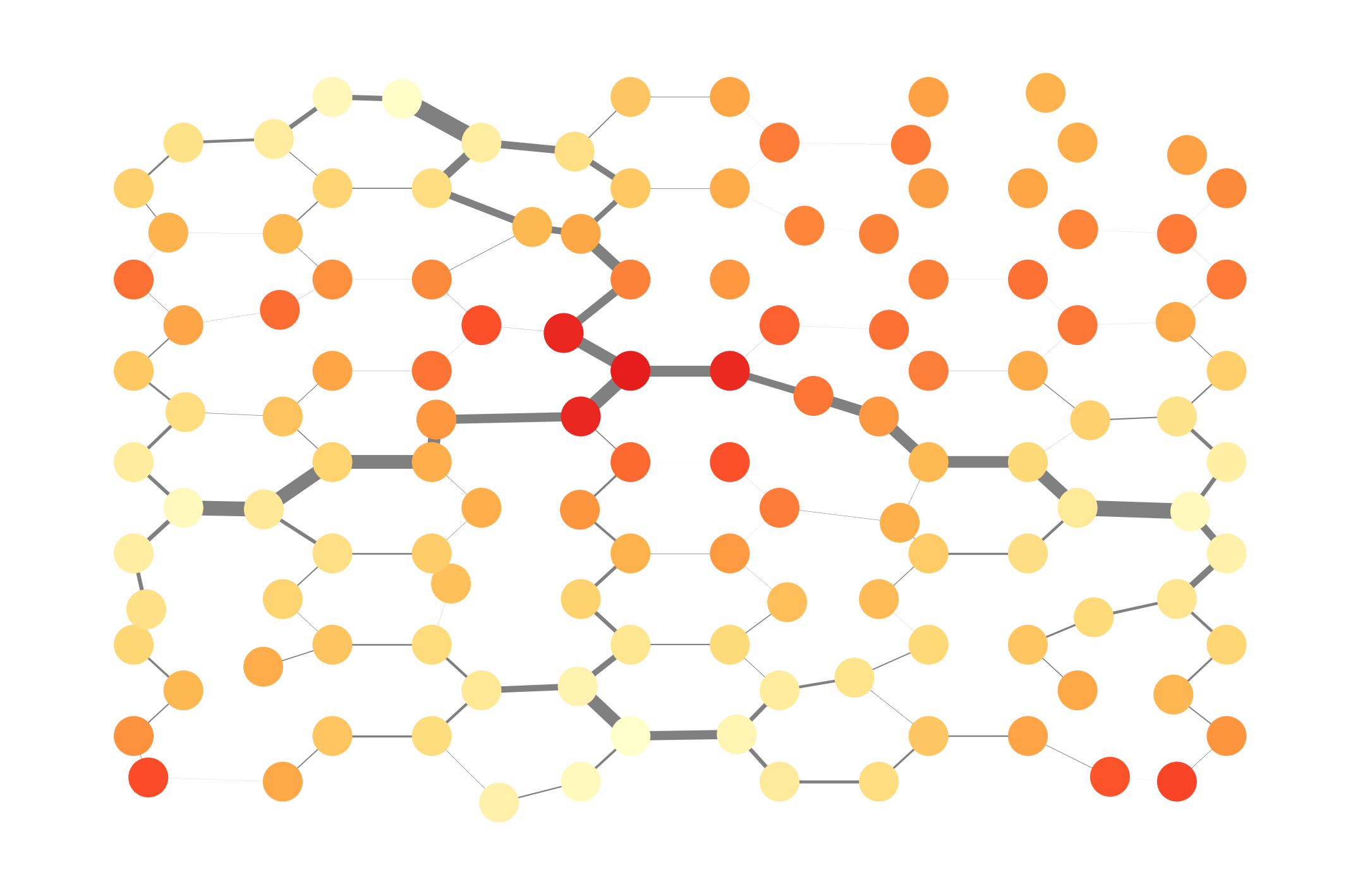}\\
        (b) & (e) \\
        \includegraphics[width=.48\textwidth]{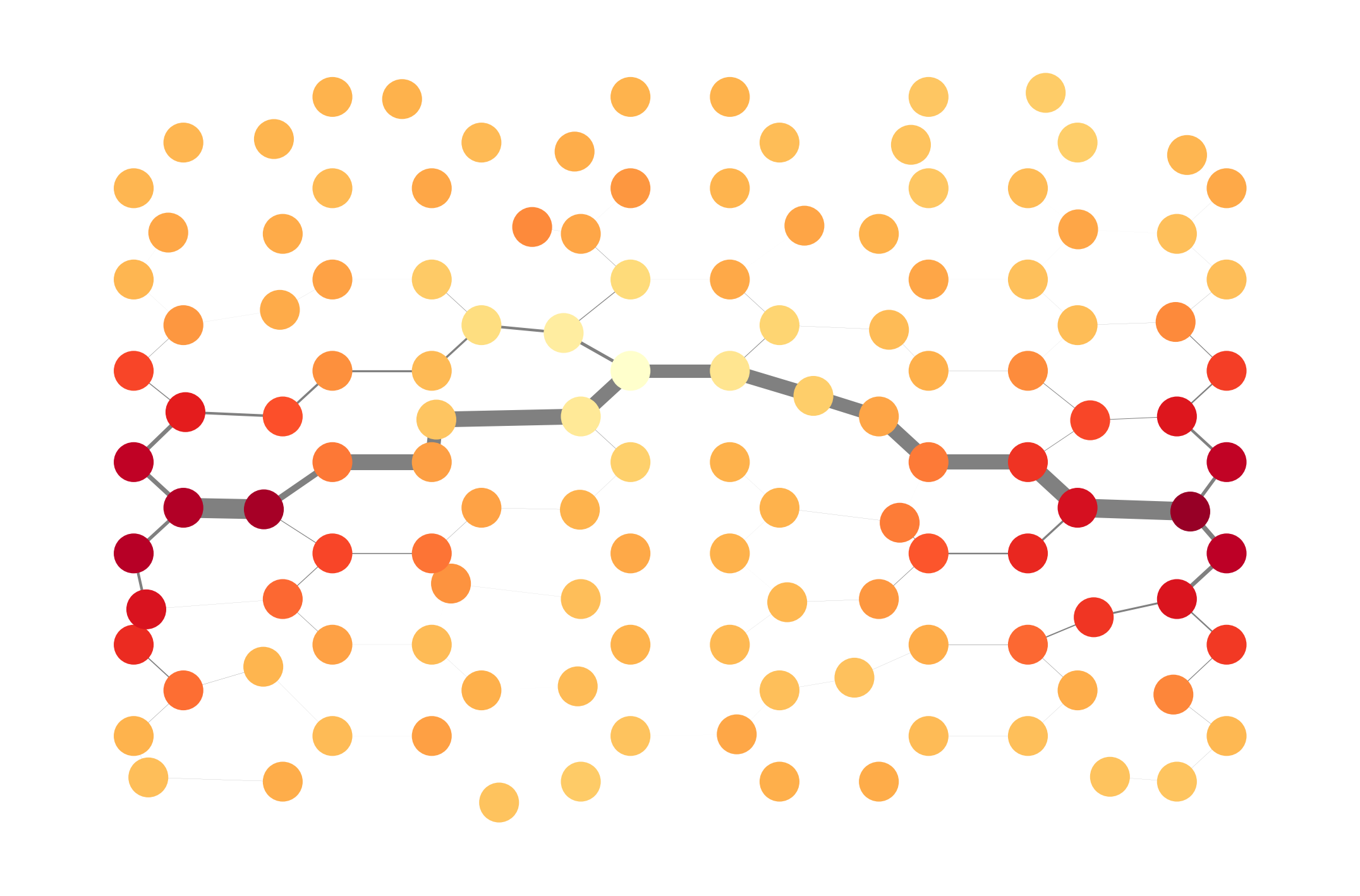} & \includegraphics[width=.48\textwidth]{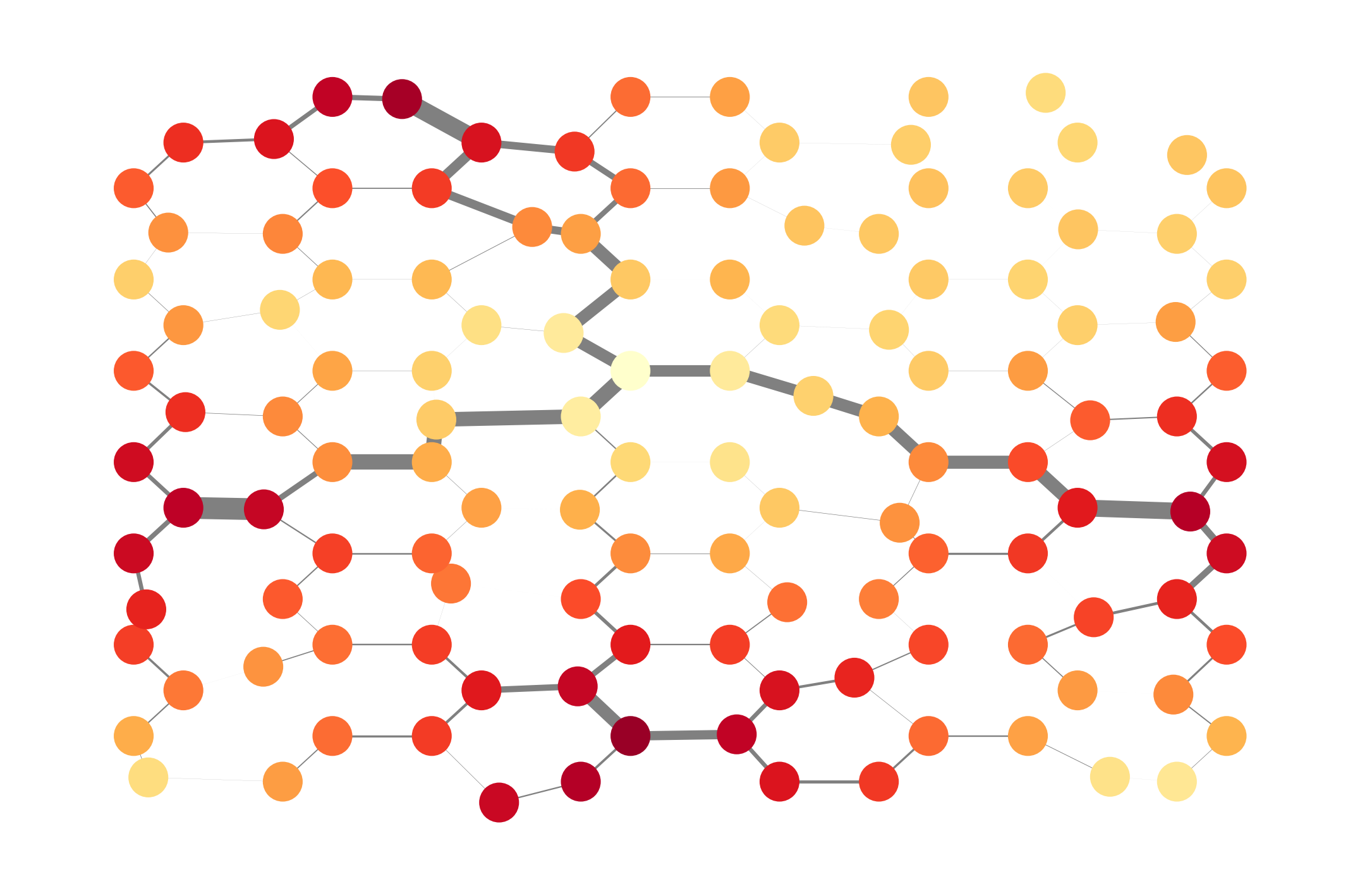}\\
        (c) & (f) \\
    \end{tabular}
    \caption{Snapshots of the graph at three different representative time points after a sufficiently long period, with different numbers of oscillatory nodes. Panels (a), (b), and (c) show snapshots of a graph with $3$ oscillatory nodes, whereas panels (d), (e), and (f) show snapshots of a graph with $5$ oscillatory nodes. The thickness of the edges indicates conductivity, and the colour of the nodes represents the number of particles at each node, with a redder colour indicating a higher number of particles.
    % \yu{The thickness of the edges should be increased?}
    }
    \label{fig:large-hexagon}
\end{figure}

In Figure \ref{fig:large-hexagon}, we explore the behaviour of our model after a sufficiently long period in two different scenarios: one with three oscillatory nodes (the blue node in Figure \ref{fig:large-hexagon-graph} and the two red nodes positioned horizontally in the middle of the figure) and another with five oscillatory nodes (one blue and four red)\footnote{\footnotesize For more details on the simulations, please refer to \url{https://github.com/yuoohmaths/MinimalCognition}.}. In both simulations, the flow primarily focuses on the shortest paths between the oscillatory nodes with the largest phase differences (illustrated by black connecting lines in Figure \ref{fig:large-hexagon}). The oscillations dominate the particle behaviour. Tracing the sequence from Figure \ref{fig:large-hexagon}a through \ref{fig:large-hexagon}b to \ref{fig:large-hexagon}c, we observe that the concentration of particles moves from the periphery to the centre and then back to the periphery again. Figures \ref{fig:large-hexagon}d through \ref{fig:large-hexagon}e to \ref{fig:large-hexagon}f demonstrate the same pattern but with five nodes.

While identifying parameter values that produce this outcome, we note that the frequency of oscillatory nodes plays a more significant role in large graphs than in the smaller graph examples previously examined. The oscillations should be sufficiently slow to allow information transfer from one oscillatory node to another before the particles return along their travelled paths. Simultaneously, the amplitude must be large enough to enable such communication before the conductivity decreases to zero. Therefore, in the simulations on larger graphs, we set the frequency to $\theta = 0.001$ and the amplitude to $A = 5$, while keeping the reinforcement parameter at $q = 0.1$ and the decay parameter at $\lambda = 0.01$.

% \yu{Frequency matters! Now we have changed the frequency to $\theta=0.001$.}

% \section{Methodology and proofs}

% \subsection{Implementation?}

% \subsection{Proofs}

\section{Discussion}

Guided by Beer's four steps \cite{beer1996toward, beer2020lost}, we have provided and investigated a mathematical framework inspired by organisms exhibiting basal cognition \cite{lyon2021reframing}. Under this framework, the formulation of the model is itself a contribution to this research area:  we have built upon previous models of current reinforcement to explicitly include oscillations, demonstrating that such systems often exhibit the greatest flow on the shortest path between the oscillators.  

We can think of our model as a demonstration of how organisms can monitor a dynamic environment, transporting particles (e.g. nutrients or signals) effectively between different locations. These abilities existed well before the development of nervous systems, let alone central nervous systems \cite{lyon2021reframing}. This makes our model a potential step towards a more realistic slime mould model \cite{dussutour2021learning}, as it captures both synchronisation and network construction observed in experiments of these organisms \cite{saigusa2008amoebae, tero2006physarum}. It also mimics other types of basal cognition, such as macromolecular networks in microbes \cite{westerhoff2014macromolecular}, and signalling in fungi \cite{schmieder2019bidirectional}.

%In a stationary environment the current or flow of particles converges to an efficient transport network on the shortest path. Previous work on current-reinforced models has proven that a shortest path is established in a model where the flow of particles is assumed to be at steady state \cite{ito2011convergence,bonifaci2012physarum}. We have found that if we explicitly model the dynamics of that flow then, while flow is still concentrated on the shortest path, there exist non-hyperbolic stable node configurations. These correspond to particles which, might be said to be `stuck' on the longer path.  Biologically, this implies that while particles are no longer actively contributing to the main flow, they remain in certain regions where they could be 'reactivated' under specific conditions, akin to latent memory traces.  We suggest a connection between these stuck particles and the way in which many organisms with basal cognition have an external memory. By leaving external cues in the environment, they can react when the environment changes \cite{sims2022externalized}. We have established that the stuck particles are associated with a slow manifold. This means we have very slowly changing particle numbers on the longer path spanning the slow manifold, while flow on the network converges to the shortest path. In this way, the model has an external memory: the particles on the nodes corresponding to the longer path `remember' failed solutions, and might be reactivated when conditions change. 

In a stationary environment, the current or flow of particles converges to an efficient transport network along the shortest path. Previous work on current-reinforced models has demonstrated that the shortest path is established in a model where the flow of particles is assumed to be in a steady state \cite{ito2011convergence,bonifaci2012physarum}. We have found that when explicitly modelling the dynamics of that flow that, while the flow still concentrates on the shortest path, non-hyperbolic stable node configurations emerge. Specifically, non-hyperbolic stable nodes represent configurations where particles remain present on the non-optimal path, despite there being no flow in or out of that path. Unlike hyperbolic configurations, which exhibit strong attraction or repulsion, non-hyperbolic configurations feature slow dynamics which mean that particles "linger" or are "stuck" in suboptimal regions. There is a connection between these stuck particles and the concept of external memory in organisms with basal cognition. By leaving external cues in the environment, organisms can adapt their behaviour when environmental conditions change \cite{sims2022externalized}. The particles on nodes corresponding to the longer path 'remember' failed solutions and may be reactivated when conditions change.

The dynamic networks in our simulations produce both efficient flow and long-range oscillatory dynamics. While the nodes oscillate greatly in terms of number of particles at them, the connecting edges of the graph experience only small oscillations (this can be seen by contrasting the scale of the oscillations in conductivity and number of particles in Figure \ref{fig:res-osln-1s1s}). The resulting pattern has low frequency responses to changes in formation over a long range, combined with a rapid flow of information across the system. We see this in Figure \ref{fig:large-hexagon}, which is reminiscent of cross-frequency coupling, i.e. coupling between neurons of different frequencies, which has been proposed as a mechanism for working memory \cite{lisman2013theta, jensen2014temporal}. This is in contrast to the Kuramoto model, which captures synchronisation, but does not capture information transfer in the same way as our model. In our model, information transfer emerges from the interaction of coupled oscillators, which has also been observed experimentally in slime moulds, where the coupling of oscillatory dynamics allows for adaptive responses and efficient decision-making \cite{ray2019information}.
%Information transfer ... coupled oscillators... something something cite \cite{ray2019information}. 

Comparing our model to recent slime mould models created by Alim et al. \cite{alim2017mechanism}, our analysis reveals that current-based reinforcement coupled with oscillators can indeed find the shortest path, as suggested by Alim et al., and similar to what is found by Watanabe et al. \cite{watanabe2014transportation}. However, in contrast to Watanabe et al., the sum of the input and output rates in our model are not set to zero. Yet, after reaching a stable limit cycle, the input and output rates in our model cancel out, suggesting that the oscillators self-organize to have the total flux to/from the system balanced. Our work also aligns with Reid's perspective on slime mould as a connected mass of oscillating units \cite{reid2023thoughts}. It further echoes early experiments by Takamatsu et al., displaying similar oscillatory behaviour in conductivity (i.e. thickness of the plasmodium network) in rings of slime mould oscillators  \cite{takamatsu2000controlling, takamatsu2001spatiotemporal}.

%Comparing our model to recent slime mould models created by Alim et al. \cite{alim2017mechanism}, our analysis reveals that reinforcement coupled with oscillators can indeed find the shortest path, as suggested by these authors, and similar to what is found by Watanabe et al. \cite{watanabe2014transportation}. However, in contrast to Watanabe et al., the sum of the input and output rates in our model are not set to zero. Yet, after reaching a stable limit cycle, the input and output rates in our model cancel out, suggesting that the oscillators self-organize to have the total flux to/from the system balanced. Our work also aligns with Reid's perspective on slime mould as a connected mass of oscillating units \cite{reid2023thoughts} and with early experiments by Matsumoto et al., showing how slime moulds oscillation patterns that showed that coupled oscillators synchronise and propagate signals in a coordinated manner across networks \cite{matsumoto1988reversal, matsumoto1986propagation}. It further echoes early experiments by Takamatsu et al., displaying similar oscillatory behaviour in conductivity (i.e. thickness of the plasmodium network) in rings of slime mould oscillators  \cite{takamatsu2000controlling, takamatsu2001spatiotemporal}.
%As well as even earlier expermients tbat indicate that these oscillating units  ...... \cite{matsumoto1988reversal,matsumoto1986propagation}. 

Expanding our discussion to other species with basal cognition, we can draw parallels between our model and experiments in bacterial colonies  that employ in-phase and anti-phase oscillatory behaviours during resource scarcity \cite{liu2017coupling}. Similarly, the in-phase and out-of-phase oscillatory behaviours in our model are akin to the emergence of in-phase and out-of-phase oscillatory behaviours that is also found in plant shoots. For example, emergent maize plants grow in a group which exhibits synchronized oscillatory motions that may be in-phase or anti-phase \cite{ciszak2016plant}. Drawing connections to fungal mycelia, our model resonates with recent studies highlighting the bidirectional transport of signals and nutrients \cite{schmieder2019bidirectional}. The flexibility of nodes to switch between acting as sources and sinks in our model provides a promising foundation for modelling bidirectional transportation within fungal mycelia \cite{schmieder2019bidirectional}. 

Our analysis shows that not only phase but also amplitude and frequency of oscillations can induce the construction of efficient transport networks. This confirms the hypothesis (in a model) of Boussard et al., who suggested that all three variations could play a role in how slime moulds build networks \cite{boussard2021adaptive}. Phase differences certainly play a relatively more important role, especially when the oscillatory nodes are close to being out of phase. In this case, the model nearly always directs almost all its flow along the shortest path between the nodes that are out of phase. However, any variation in phase, amplitude, frequency, or a combination thereof, is sufficient to produce the basal cognition we observe in our simulations.

We do observe that very big frequency differences inhibits shortest path formation. For example, the oscillating nodes become disconnected when the frequency ratio is 10 (see bottom row of Figure \ref{fig:tab-fp}). This, along with the need for the oscillators to be (somewhat) out of phase, gives some restrictions on the types of oscillations needed in order to create a flow between nodes. Amplitude primarily affects how far the particles can explore in the graph: too small amplitudes do not allow long range communication. Our observations echo earlier studies on slime moulds, which show that frequency mismatches and phase relationships are crucial in coordinating oscillatory behavior and directional movement in response to stimuli. These studies demonstrate that when the frequency of external stimuli deviates significantly from the organism’s internal rhythm, it can disrupt coordinated movement, much like the disconnections observed in our model when frequency ratios are large \cite{matsumoto1986propagation,matsumoto1988reversal}.

Our model does not include feedback between oscillatory nodes, which we know are part of how living organisms assess, engage and adapt to the world around them. Such feedback could also potentially change the phase, amplitude and frequency of the oscillations towards values which better facilitate communications. Our next modelling steps would therefore be to incorporate feedback mechanisms for the oscillators into the model (as suggested by \cite{boussard2021adaptive}). The introduction of feedback could involve the consideration of diverse forms of local information, $lc_i(t)$, such as the potential difference in particles ($\sum_{j\in\Gamma(i)}(N_i(t) - N_j(t))$), the absolute potential difference, or a combination of potential difference and conductivity ($\sum_{j\in\Gamma(i)}(N_i(t) - N_j(t))D_{ij}(t)$). Moreover, an exploration could encompass various strategies for the phase transition, including the option of direct proportionality to local information ($\frac{d\phi_i(t)}{dt} = lc_i(t)$). An initial exploration using smaller networks could be undertaken to discern the impact of network topology on feedback forms, as done by Ma et al. \cite{ma2009defining}. Alternatively, an evolutionary adaptive approach, as advocated by Beer \cite{beer2020lost}, offers another way to learn a feedback scheme.

In conclusion, the model we propose here already has similar properties to many biological systems which exhibit basal (or even more complex forms of) cognition. We see it as a promising starting point for future simulation models of these phenomena.

%There are various possible approaches to starting investigations of these. For example, one can propose potential feedback functions based on those identified by neuroscience \cite{abbott2000synaptic}.  Alternatively, one can learn the feedback scheme with an evolutionary adaptive approach \cite{beer2020lost}, 
\newpage
\bibliography{references_slime}

\begin{thebibliography}{10}

\bibitem{bacsar2000brain}
E.~Ba{\c{s}}ar, C.~Ba{\c{s}}ar-Ero{\u{g}}lu, S.~Karaka{\c{s}}, and
  M.~Sch{\"u}rmann, ``Brain oscillations in perception and memory,'' {\em
  International journal of psychophysiology}, vol.~35, no.~2-3, pp.~95--124,
  2000.

\bibitem{begus2020rhythm}
K.~Begus and E.~Bonawitz, ``The rhythm of learning: Theta oscillations as an
  index of active learning in infancy,'' {\em Developmental Cognitive
  Neuroscience}, vol.~45, p.~100810, 2020.

\bibitem{herweg2020theta}
N.~A. Herweg, E.~A. Solomon, and M.~J. Kahana, ``Theta oscillations in human
  memory,'' {\em Trends in cognitive sciences}, vol.~24, no.~3, pp.~208--227,
  2020.

\bibitem{gnaedinger2019multisensory}
A.~Gnaedinger, H.~Gurden, B.~Gour{\'e}vitch, and C.~Martin, ``Multisensory
  learning between odor and sound enhances beta oscillations,'' {\em Scientific
  Reports}, vol.~9, no.~1, p.~11236, 2019.

\bibitem{hanslmayr2011role}
S.~Hanslmayr, J.~Gross, W.~Klimesch, and K.~L. Shapiro, ``The role of alpha
  oscillations in temporal attention,'' {\em Brain research reviews}, vol.~67,
  no.~1-2, pp.~331--343, 2011.

\bibitem{lisman2013theta}
J.~E. Lisman and O.~Jensen, ``The theta-gamma neural code,'' {\em Neuron},
  vol.~77, no.~6, pp.~1002--1016, 2013.

\bibitem{jensen2014temporal}
O.~Jensen, B.~Gips, T.~O. Bergmann, and M.~Bonnefond, ``Temporal coding
  organized by coupled alpha and gamma oscillations prioritize visual
  processing,'' {\em Trends in neurosciences}, vol.~37, no.~7, pp.~357--369,
  2014.

\bibitem{martin2000synaptic}
S.~J. Martin, P.~D. Grimwood, and R.~G. Morris, ``Synaptic plasticity and
  memory: an evaluation of the hypothesis,'' {\em Annual review of
  neuroscience}, vol.~23, no.~1, pp.~649--711, 2000.

\bibitem{holscher1999synaptic}
C.~H{\"o}lscher, ``Synaptic plasticity and learning and memory: Ltp and
  beyond,'' {\em Journal of neuroscience research}, vol.~58, no.~1, pp.~62--75,
  1999.

\bibitem{fisher2017reinforcement}
S.~D. Fisher, P.~B. Robertson, M.~J. Black, P.~Redgrave, M.~A. Sagar, W.~C.
  Abraham, and J.~N. Reynolds, ``Reinforcement determines the timing dependence
  of corticostriatal synaptic plasticity in vivo,'' {\em Nature
  communications}, vol.~8, no.~1, p.~334, 2017.

\bibitem{shindou2019silent}
T.~Shindou, M.~Shindou, S.~Watanabe, and J.~Wickens, ``A silent eligibility
  trace enables dopamine-dependent synaptic plasticity for reinforcement
  learning in the mouse striatum,'' {\em European Journal of Neuroscience},
  vol.~49, no.~5, pp.~726--736, 2019.

\bibitem{van2006principles}
M.~Van~Duijn, F.~Keijzer, and D.~Franken, ``Principles of minimal cognition:
  Casting cognition as sensorimotor coordination,'' {\em Adaptive Behavior},
  vol.~14, no.~2, pp.~157--170, 2006.

\bibitem{lyon2021reframing}
P.~Lyon, F.~Keijzer, D.~Arendt, and M.~Levin, ``Reframing cognition: getting
  down to biological basics,'' {\em Philosophical Transactions of the Royal
  Society B}, vol.~376, no.~1820, p.~20190750, 2021.

\bibitem{boussard2021adaptive}
A.~Boussard, A.~Fessel, C.~Oettmeier, L.~Briard, H.-G. D{\"o}bereiner, and
  A.~Dussutour, ``Adaptive behaviour and learning in slime moulds: the role of
  oscillations,'' {\em Philosophical Transactions of the Royal Society B},
  vol.~376, no.~1820, p.~20190757, 2021.

\bibitem{behar2010understanding}
M.~Behar and A.~Hoffmann, ``Understanding the temporal codes of intra-cellular
  signals,'' {\em Current opinion in genetics \& development}, vol.~20, no.~6,
  pp.~684--693, 2010.

\bibitem{purvis2013encoding}
J.~E. Purvis and G.~Lahav, ``Encoding and decoding cellular information through
  signaling dynamics,'' {\em Cell}, vol.~152, no.~5, pp.~945--956, 2013.

\bibitem{baluvska2013root}
F.~Balu{\v{s}}ka and S.~Mancuso, ``Root apex transition zone as oscillatory
  zone,'' {\em Frontiers in Plant Science}, vol.~4, p.~354, 2013.

\bibitem{traas2010oscillating}
J.~Traas and T.~Vernoux, ``Oscillating roots,'' {\em Science}, vol.~329,
  no.~5997, pp.~1290--1291, 2010.

\bibitem{oettmeier2017physarum}
C.~Oettmeier, K.~Brix, and H.-G. D{\"o}bereiner, ``Physarum polycephalum—a
  new take on a classic model system,'' {\em Journal of Physics D: Applied
  Physics}, vol.~50, no.~41, p.~413001, 2017.

\bibitem{vallverdu2018slime}
J.~Vallverd{\'u}, O.~Castro, R.~Mayne, M.~Talanov, M.~Levin, F.~Balu{\v{s}}ka,
  Y.~Gunji, A.~Dussutour, H.~Zenil, and A.~Adamatzky, ``Slime mould: the
  fundamental mechanisms of biological cognition,'' {\em Biosystems}, vol.~165,
  pp.~57--70, 2018.

\bibitem{smith2020needs}
J.~Smith-Ferguson and M.~Beekman, ``Who needs a brain? slime moulds,
  behavioural ecology and minimal cognition,'' {\em Adaptive Behavior},
  vol.~28, no.~6, pp.~465--478, 2020.

\bibitem{boisseau2016habituation}
R.~P. Boisseau, D.~Vogel, and A.~Dussutour, ``Habituation in non-neural
  organisms: evidence from slime moulds,'' {\em Proceedings of the Royal
  Society B: Biological Sciences}, vol.~283, no.~1829, p.~20160446, 2016.

\bibitem{saigusa2008amoebae}
T.~Saigusa, A.~Tero, T.~Nakagaki, and Y.~Kuramoto, ``Amoebae anticipate
  periodic events,'' {\em Physical review letters}, vol.~100, no.~1, p.~018101,
  2008.

\bibitem{nakagaki2000maze}
T.~Nakagaki, H.~Yamada, and {\'A}.~T{\'o}th, ``Maze-solving by an amoeboid
  organism,'' {\em Nature}, vol.~407, no.~6803, pp.~470--470, 2000.

\bibitem{beekman2015brainless}
M.~Beekman and T.~Latty, ``Brainless but multi-headed: decision making by the
  acellular slime mould physarum polycephalum,'' {\em Journal of molecular
  biology}, vol.~427, no.~23, pp.~3734--3743, 2015.

\bibitem{tero2010rules}
A.~Tero, S.~Takagi, T.~Saigusa, K.~Ito, D.~P. Bebber, M.~D. Fricker, K.~Yumiki,
  R.~Kobayashi, and T.~Nakagaki, ``Rules for biologically inspired adaptive
  network design,'' {\em Science}, vol.~327, no.~5964, pp.~439--442, 2010.

\bibitem{reid2013amoeboid}
C.~R. Reid, M.~Beekman, T.~Latty, and A.~Dussutour, ``Amoeboid organism uses
  extracellular secretions to make smart foraging decisions,'' {\em Behavioral
  Ecology}, vol.~24, no.~4, pp.~812--818, 2013.

\bibitem{reid2013solving}
C.~R. Reid and M.~Beekman, ``Solving the towers of hanoi--how an amoeboid
  organism efficiently constructs transport networks,'' {\em Journal of
  Experimental Biology}, vol.~216, no.~9, pp.~1546--1551, 2013.

\bibitem{takamatsu2000controlling}
A.~Takamatsu, T.~Fujii, H.~Yokota, K.~Hosokawa, T.~Higuchi, and I.~Endo,
  ``Controlling the geometry and the coupling strength of the oscillator system
  in plasmodium of physarum polycephalum by microfabricated structure,'' {\em
  Protoplasma}, vol.~210, pp.~164--171, 2000.

\bibitem{takamatsu2001spatiotemporal}
A.~Takamatsu, R.~Tanaka, H.~Yamada, T.~Nakagaki, T.~Fujii, and I.~Endo,
  ``Spatiotemporal symmetry in rings of coupled biological oscillators of
  physarum plasmodial slime mold,'' {\em Physical Review Letters}, vol.~87,
  no.~7, p.~078102, 2001.

\bibitem{dussutour2021learning}
A.~Dussutour, ``Learning in single cell organisms,'' {\em Biochemical and
  Biophysical Research Communications}, vol.~564, pp.~92--102, 2021.

\bibitem{beer1992evolving}
R.~D. Beer and J.~C. Gallagher, ``Evolving dynamical neural networks for
  adaptive behavior,'' {\em Adaptive behavior}, vol.~1, no.~1, pp.~91--122,
  1992.

\bibitem{beer1996toward}
R.~D. Beer {\em et~al.}, ``Toward the evolution of dynamical neural networks
  for minimally cognitive behavior,'' {\em From animals to animats}, vol.~4,
  pp.~421--429, 1996.

\bibitem{brancazio2022easy}
N.~Brancazio, ``Easy alliances: The methodology of minimally cognitive behavior
  (mmcb) and basal cognition,'' in {\em PSA 2022: The 28th Biennial Meeting of
  the Philosophy of Science Association}, 2022.

\bibitem{liu2017coupling}
J.~Liu, R.~Martinez-Corral, A.~Prindle, D.-y.~D. Lee, J.~Larkin,
  M.~Gabalda-Sagarra, J.~Garcia-Ojalvo, and G.~M. S{\"u}el {\em Science},
  vol.~356, no.~6338, pp.~638--642, 2017.

\bibitem{moroz2021neural}
L.~L. Moroz, D.~Y. Romanova, and A.~B. Kohn, ``Neural versus alternative
  integrative systems: molecular insights into origins of neurotransmitters,''
  {\em Philosophical Transactions of the Royal Society B}, vol.~376, no.~1821,
  p.~20190762, 2021.

\bibitem{pezzulo2021bistability}
G.~Pezzulo, J.~LaPalme, F.~Durant, and M.~Levin, ``Bistability of somatic
  pattern memories: stochastic outcomes in bioelectric circuits underlying
  regeneration,'' {\em Philosophical Transactions of the Royal Society B},
  vol.~376, no.~1821, p.~20190765, 2021.

\bibitem{beer2020lost}
R.~D. Beer, ``Lost in words,'' {\em Adaptive Behavior}, vol.~28, no.~1,
  pp.~19--21, 2020.

\bibitem{schmieder2019bidirectional}
S.~S. Schmieder, C.~E. Stanley, A.~Rzepiela, D.~van Swaay, J.~Saboti{\v{c}},
  S.~F. N{\o}rrelykke, A.~J. deMello, M.~Aebi, and M.~K{\"u}nzler,
  ``Bidirectional propagation of signals and nutrients in fungal networks via
  specialized hyphae,'' {\em Current Biology}, vol.~29, no.~2, pp.~217--228,
  2019.

\bibitem{westerhoff2014macromolecular}
H.~V. Westerhoff, A.~N. Brooks, E.~Simeonidis, R.~Garc{\'\i}a-Contreras, F.~He,
  F.~C. Boogerd, V.~J. Jackson, V.~Goncharuk, and A.~Kolodkin, ``Macromolecular
  networks and intelligence in microorganisms,'' {\em Frontiers in
  microbiology}, vol.~5, p.~379, 2014.

\bibitem{wan2023active}
K.~Y. Wan, ``Active oscillations in microscale navigation,'' {\em Animal
  Cognition}, pp.~1--14, 2023.

\bibitem{tero2006physarum}
A.~Tero, R.~Kobayashi, and T.~Nakagaki, ``Physarum solver: A biologically
  inspired method of road-network navigation,'' {\em Physica A: Statistical
  Mechanics and its Applications}, vol.~363, no.~1, pp.~115--119, 2006.

\bibitem{ito2011convergence}
K.~Ito, A.~Johansson, T.~Nakagaki, and A.~Tero, ``Convergence properties for
  the physarum solver,'' {\em arXiv preprint arXiv:1101.5249}, 2011.

\bibitem{bonifaci2012physarum}
V.~Bonifaci, K.~Mehlhorn, and G.~Varma, ``Physarum can compute shortest
  paths,'' {\em Journal of theoretical biology}, vol.~309, pp.~121--133, 2012.

\bibitem{ma2013current}
Q.~Ma, A.~Johansson, A.~Tero, T.~Nakagaki, and D.~J. Sumpter,
  ``Current-reinforced random walks for constructing transport networks,'' {\em
  Journal of the Royal Society Interface}, vol.~10, no.~80, p.~20120864, 2013.

\bibitem{kuramoto1975international}
Y.~Kuramoto, ``International symposium on mathematical problems in theoretical
  physics,'' {\em Lecture notes in Physics}, vol.~30, p.~420, 1975.

\bibitem{breakspear2010generative}
M.~Breakspear, S.~Heitmann, and A.~Daffertshofer, ``Generative models of
  cortical oscillations: neurobiological implications of the kuramoto model,''
  {\em Frontiers in human neuroscience}, vol.~4, p.~190, 2010.

\bibitem{cabral2011role}
J.~Cabral, E.~Hugues, O.~Sporns, and G.~Deco, ``Role of local network
  oscillations in resting-state functional connectivity,'' {\em Neuroimage},
  vol.~57, no.~1, pp.~130--139, 2011.

\bibitem{maistrenko2007multistability}
Y.~L. Maistrenko, B.~Lysyansky, C.~Hauptmann, O.~Burylko, and P.~A. Tass,
  ``Multistability in the kuramoto model with synaptic plasticity,'' {\em
  Physical Review E}, vol.~75, no.~6, p.~066207, 2007.

\bibitem{timms2014synchronization}
L.~Timms and L.~Q. English, ``Synchronization in phase-coupled kuramoto
  oscillator networks with axonal delay and synaptic plasticity,'' {\em
  Physical Review E}, vol.~89, no.~3, p.~032906, 2014.

\bibitem{ruangkriengsin2022low}
T.~Ruangkriengsin and M.~A. Porter, ``Low-dimensional analysis of a kuramoto
  model with inertia and hebbian learning,'' {\em arXiv preprint
  arXiv:2203.12090}, 2022.

\bibitem{alim2017mechanism}
K.~Alim, N.~Andrew, A.~Pringle, and M.~P. Brenner, ``Mechanism of signal
  propagation in physarum polycephalum,'' {\em Proceedings of the National
  Academy of Sciences}, vol.~114, no.~20, pp.~5136--5141, 2017.

\bibitem{watanabe2014transportation}
S.~Watanabe and A.~Takamatsu, ``Transportation network with fluctuating
  input/output designed by the bio-inspired physarum algorithm,'' {\em PloS
  one}, vol.~9, no.~2, p.~e89231, 2014.

\bibitem{ben2022structural}
Y.~Ben-Ami, G.~W. Atkinson, J.~M. Pitt-Francis, P.~K. Maini, and H.~M. Byrne,
  ``Structural features of microvascular networks trigger blood flow
  oscillations,'' {\em Bulletin of Mathematical Biology}, vol.~84, no.~8,
  p.~85, 2022.

\bibitem{tero2007mathematical}
A.~Tero, R.~Kobayashi, and T.~Nakagaki, ``A mathematical model for adaptive
  transport network in path finding by true slime mold,'' {\em Journal of
  theoretical biology}, vol.~244, no.~4, pp.~553--564, 2007.

\bibitem{tero2008flow}
A.~Tero, K.~Yumiki, R.~Kobayashi, T.~Saigusa, and T.~Nakagaki, ``Flow-network
  adaptation in physarum amoebae,'' {\em Theory in biosciences}, vol.~127,
  pp.~89--94, 2008.

\bibitem{dirnberger2017characterizing}
M.~Dirnberger and K.~Mehlhorn, ``Characterizing networks formed by p.
  polycephalum,'' {\em Journal of Physics D: Applied Physics}, vol.~50, no.~22,
  p.~224002, 2017.

\bibitem{dussutour2024flow}
A.~Dussutour and C.~Arson, ``Flow-network adaptation and behavior in slime
  molds,'' {\em Fungal Ecology}, vol.~68, p.~101325, 2024.

\bibitem{wiggins1990}
S.~Wiggins, {\em Introduction to Applied Nonlinear Dynamical Systems and
  Chaos}.
\newblock New York: Springer, 1990.

\bibitem{sims2022externalized}
M.~Sims and J.~Kiverstein, ``Externalized memory in slime mould and the
  extended (non-neuronal) mind,'' {\em Cognitive Systems Research}, vol.~73,
  pp.~26--35, 2022.

\bibitem{ray2019information}
S.~K. Ray, G.~Valentini, P.~Shah, A.~Haque, C.~R. Reid, G.~F. Weber, and
  S.~Garnier, ``Information transfer during food choice in the slime mold
  physarum polycephalum,'' {\em Frontiers in Ecology and Evolution}, vol.~7,
  p.~430494, 2019.

\bibitem{reid2023thoughts}
C.~R. Reid, ``Thoughts from the forest floor: a review of cognition in the
  slime mould physarum polycephalum,'' {\em Animal Cognition}, pp.~1--15, 2023.

\bibitem{ciszak2016plant}
M.~Ciszak, E.~Masi, F.~Balu{\v{s}}ka, and S.~Mancuso, ``Plant shoots exhibit
  synchronized oscillatory motions,'' {\em Communicative \& integrative
  biology}, vol.~9, no.~5, p.~e1238117, 2016.

\bibitem{matsumoto1986propagation}
K.~Matsumoto, T.~Ueda, and Y.~Kobatake, ``Propagation of phase wave in relation
  to tactic responses by the plasmodium of physarum polycephalum,'' {\em
  Journal of Theoretical Biology}, vol.~122, no.~3, pp.~339--345, 1986.

\bibitem{matsumoto1988reversal}
K.~Matsumoto, T.~Ueda, and Y.~Kobatake, ``Reversal of thermotaxis with
  oscillatory stimulation in the plasmodium of physarum polycephalum,'' {\em
  Journal of Theoretical Biology}, vol.~131, no.~2, pp.~175--182, 1988.

\bibitem{ma2009defining}
W.~Ma, A.~Trusina, H.~El-Samad, W.~A. Lim, and C.~Tang, ``Defining network
  topologies that can achieve biochemical adaptation,'' {\em Cell}, vol.~138,
  no.~4, pp.~760--773, 2009.

\end{thebibliography}

\newpage
\appendix
\section{Non-oscillatory sources and sinks}\label{sec:app-non-osc}

\subsection{Steady states}
This section of the appendix derives the results for the steady states as presented in Section \ref{sec:res-const}. The equilibrium points of the system are determined by the solutions to the following set of equations:
\begin{align}\label{eq:C_5_eq_steady} 
 & a + \frac{(N_2^*-N_1^*)}{l_{12}}D_{12}^*+\frac{(N_5^*-N_1^*)}{l_{51}}D_{51} = 0, \\
 &  \frac{(N_1^*-N_2^*)}{l_{12}}D_{12}^*+\frac{(N_3^*-N_2^*)}{l_{23}}D_{23}^*=0, \\
  &-bN_3^*+ \frac{(N_2^*-N_3^*)}{l_{23}}D_{23}^*+\frac{(N_4^*-N_3^*)}{l_{34}}D_{34}^* =0,\\
  & \frac{(N_3^*-N_4^*)}{l_{34}}D_{34}^*+\frac{(N_5^*-N_4^*)}{l_{45}}D_{45}^* =0,\\
  & \frac{(N_1^*-N_5^*)}{l_{51}}D_{51}^*+\frac{(N_4^*-N_5^*)}{l_{45}}D_{45}^*=0, \\
 &  q\frac{\abs{N_1^* - N_2^*}}{l_{12}}D_{12}^* - \lambda D_{12}^*=0, \\
 & q\frac{\abs{N_2^* - N_3^*}}{l_{23}}D_{23}^* - \lambda D_{23}^*=0, \\
& q\frac{\abs{N_3^* - N_4^*}}{l_{34}}D_{34}^* - \lambda D_{34}^*=0, \\
 & q\frac{\abs{N_4^* - N_5^*}}{l_{45}}D_{45}^* - \lambda D_{45}^*=0, \\
& q\frac{\abs{N_5^* - N_1^*}}{l_{51}}D_{51}^*- \lambda D_{51}^*  =0.
\end{align}
In general, the steady states for the conductivity are given by the equation:
\begin{align}\label{eq:steady_N}
   q\frac{\abs{N_i^* - N_j^*}}{l_{ij}}D_{ij}^* - \lambda D_{ij}^* = 0.
\end{align}
For this to hold, either $D_{ij}^*=0$ or $\abs{N_i^* - N_j^*}= \frac{\lambda l_{ij}}{q}$. If  $D_{ij}^* \neq 0$, then $\abs{N_i^* - N_j^*}= \frac{\lambda l_{ij}}{q}$, and thus, Equation \eqref{eq:steady_N} can be simplified to 
\begin{equation} 
\pm \frac{\lambda}{q}D_{{i-1},i}^*+ \pm \frac{\lambda}{q}D_{i,{i+1}}^*  = 0.
\end{equation}
As $D_{ij}(t)$ denotes the conductivity, we know that $D_{ij}(t) \geq 0$ and thus, the two addends must have opposite signs. This gives the solution
\begin{equation} \label{eq:same_steady}
    D_{{i-1},i}^*=D_{i,{i+1}}^*.
\end{equation}
This also implies that for a node $i$, which is neither a source nor a sink, $(N_{i-1}^* - N_i^*)$ and $(N_{i+1} - N_i)$ must have opposite signs. We can now consider the following cases: 

\subsubsection*{Case 1:  $D_{51}^*=0 \textmd{ and } D_{12}^*\neq 0$}

If $D_{51}^*$ is zero and $D_{12}$ is non-zero, then $\abs{N_2^* - N_1^*}= \frac{\lambda l_{12}}{q}$.  We thus have that 
\begin{equation} \label{eq:steady_N_source}
a \pm \frac{\lambda l_{12}}{l_{12}q}D_{12}^*  = 0.
\end{equation}
We know that the conductivity must be positive, so
\begin{equation} 
D_{12}^*= \frac{a q}{\lambda},
\end{equation}
as $a$, $q$, and $\lambda$ are positive.  This means that 
\begin{equation}
(N_2^* - N_1^*)=-\frac{\lambda l_{12}}{q}.
\end{equation}
From Equation \eqref{eq:same_steady} we know that $D_{12}^*=D_{23}^*$ and $D_{34}^*=D_{45}^*=D_{51}^*$, so the steady states for the conductivity will in this case be given by $(D_{12}^*,D_{23}^*,D_{34}^*,D_{45}^*,D_{51}^*) = (\frac{a q}{\lambda}, \frac{a q}{\lambda}, 0, 0, 0)$.

Now consider the equation for the change of particles at node with a sink:

\begin{equation}
    \dv{N_3(t)}{t} = -bN_3+ \frac{(N_2-N_3)}{l_{23}}D_{23}+\frac{(N_4-N_3)}{l_{34}}D_{34}.
\end{equation}
The steady states will be given the solution to the equation 
\begin{equation}
-bN_3^*+ \frac{(N_2^*-N_3^*)}{l_{23}}D_{23}^*+\frac{(N_4^*-N_3^*)}{l_{34}}D_{34}^*=0. \\
\end{equation}
Now, if  $(D_{12}^*,D_{23}^*,D_{34}^*,D_{45}^*,D_{51}^*) = (\frac{a q}{\lambda}, \frac{a q}{\lambda}, 0, 0, 0)$, then the equation can be simplified to 

\begin{equation}
-bN_3^*+\frac{(N_2^*-N_3^*)}{l_{23}}D_{23}^*=0, 
\end{equation}

where $N_2^*-N_3^*=\pm \frac{\lambda l_{23}}{q}$ and $D_{23}^*=\frac{a q}{\lambda}$, which gives the following 

\begin{equation}
-bN_3^*\pm \frac{\frac{\lambda l_{23}}{q}}{l_{23}}\frac{a q}{\lambda}=0.
\end{equation}
As $N_3$ denotes the number of particles at node 3, $N_3$ must be positive and we get 

\begin{equation}
N_3^* = \frac{a}{b}.
\end{equation}

This implies that $N_2^*-N_3^*=+ \frac{\lambda l_{23}}{q}$, and thus

\begin{equation}
    N_2^*=\frac{a}{b}+\frac{\lambda l_{23}}{q}.
\end{equation}

We know that $(N_2^* - N_1^*)=-\frac{\lambda l_{12}}{q}$, so 

\begin{equation}
    N_1^*=\frac{a}{b}+ \frac{\lambda l_{23}}{q} + \frac{\lambda l_{12}}{q}.
\end{equation}

There are no exact solutions for $N_4^*$ and $N_5^*$, so for case 1, we have the following steady state: 

\begin{align*}
    (N_1^*,N_2^*,N_3^*,N_4^*, N_5^*, D_{12}^*,D_{23}^*,D_{34}^*,D_{45}^*,D_{51}^*) = \\
    \frac{a}{b}+ \frac{\lambda l_{23}}{q} + \frac{\lambda l_{12}}{q}, \frac{a}{b}+\frac{\lambda l_{23}}{q}, \frac{a}{b}, N_4^*, N_5^*,  \frac{a q}{\lambda}, \frac{a q}{\lambda},0,0,0). 
\end{align*}

\subsubsection*{Case 2: $D_{12}^*=0$ \textmd{and} $D_{51}^* \neq 0$}

If $D_{12}$ is zero  and $D_{51}^*$ is non-zero, then $\abs{N_5^* - N_1^*}= \frac{\lambda l_{51}}{q}$, and thus 
\begin{equation} 
D_{51}^*= \frac{a q}{\lambda},
\end{equation}
and
\begin{equation}
    N_5^*-N_1^*=-\frac{\lambda l_{51}}{q}.
\end{equation}
Again, consider the equation for the change of particles at node with a sink:
\begin{equation}
    \dv{N_3(t)}{t} =  -bN_3+ \frac{(N_2-N_3)}{l_{23}}D_{23}+\frac{(N_4-N_3)}{l_{34}}D_{34}. \\
\end{equation}
The steady states will be given the solution to the equation 
\begin{equation}
-bN_3^*+ \frac{(N_2^+-N_3^+)}{l_{23}}D_{23}^*+\frac{(N_4^*-N_3^*)}{l_{34}}D_{34}^*=0. \\
\end{equation}
Now, if  $(D_{12}^*,D_{23}^*,D_{34}^*,D_{45}^*,D_{51}^*) = (0,0, \frac{a q}{\lambda}, \frac{a q}{\lambda}, \frac{a q}{\lambda})$, then the equation can be simplified to 

\begin{equation}
-bN_3^*+\frac{(N_4^*-N_3^*)}{l_{34}}D_{34}^*=0, 
\end{equation}

where $N_4^*-N_3^*=\pm \frac{\lambda l_{34}}{q}$ and $D_{34}^*=\frac{a q}{\lambda}$, which gives the following 

\begin{equation}
-bN_3^*\pm \frac{\frac{\lambda l_{34}}{q}}{l_{34}}\frac{a q}{\lambda}=0.
\end{equation}
As $N_3$ denotes the number of particles at node 3, $N_3$ must  be positive and thus $N_4^*-N_3^*=+\frac{\lambda l_{34}}{q}$. Hence, we get 

\begin{equation}
N_3^* = \frac{a}{b}.
\end{equation}

We know that $N_4^*-N_3^*=\frac{\lambda l_{34}}{q}$, so
\begin{equation}
N_4^*= \frac{a}{b}+\frac{\lambda l_{34}}{q}.
\end{equation}

As 4 is neither a source, nor sink, we know that $N_4^*-N_5^*$ and $N_4^*-N_3^*$ have opposite signs. As $N_4^*-N_3^*$ is positive, this implies that
\begin{equation}
    N_4^*-N_5^*=-\frac{\lambda l_{45}}{q},
\end{equation}
and thus 
\begin{equation}
N_5^*=\frac{a}{b}+\frac{\lambda l_{34}}{q}+\frac{\lambda l_{45}}{q}.
\end{equation}
As $N_4^*-N_5^*$ is negative, $N_1^*-N_5^*$ must be positive and thus
\begin{equation}
    N_5^*-N_1^*=\frac{\lambda l_{51}}{q},
\end{equation}
which gives
\begin{equation}
   N_1^* - \left(\frac{a}{b} + \frac{\lambda l_{34}}{q} +\frac{\lambda l_{45}}{q}\right)
  = \frac\lambda {l_{51}}{q},
\end{equation}
so
\begin{equation}
   N_1^* = \frac{a}{b} + \frac{\lambda l_{34}}{q} + \frac{\lambda l_{45}}{q} +  \frac{\lambda l_{51}}{q}.
\end{equation}
As for $N_4^*$ and $N_5^*$ in case 1, there are no exact solution for $N_2^*$ in case 2. Thus the steady states are given by
\begin{align*}
    (N_1^*,N_2^*,N_3^*,N_4^*, N_5^*, D_{12}^*,D_{23}^*,D_{34}^*,D_{45}^*,D_{51}^*) = \\
    (\frac{a}{b} + \frac{\lambda l_{34}}{q} + \frac{\lambda l_{45}}{q} +  \frac{\lambda l_{51}}{q}, N_2^*, \frac{a}{b}, \frac{a}{b}+\frac{\lambda l_{34}}{q}, \frac{a}{b} + \frac{\lambda l_{34}}{q} + \frac{\lambda l_{45}}{q}, 0,0, \frac{a q}{\lambda}, \frac{a q}{\lambda}, \frac{a q}{\lambda}). 
\end{align*}

\subsubsection*{Case 3: $D_{12}^* \neq 0$ \textmd{and} $D_{51}^* \neq 0$}
Now, if neither $D_{12}$ nor $D_{51}$ is non-zero we have that
\begin{equation} 
a \pm \frac{\lambda}{q}D_{12}^* \pm \frac{\lambda }{q}D_{51}^*  = 0,
\end{equation}
which gives
\begin{equation} 
\pm D_{12}^* \pm D_{51}^*  = \frac{a q}{\lambda}.
\end{equation}
As the conductivity is the same on the edges going from a non-source and non-sink, we know that
\begin{equation}
    D_{12}^*= D_{23}^*,
\end{equation}
and
\begin{equation}
    D_{34}^*= D_{45}^*= D_{51}^*.
\end{equation}
If we assume that the conductivity is the same on all edges, we get that 
\begin{equation} 
a \pm \frac{\lambda}{q}D_{12}^* \pm \frac{\lambda }{q}D_{12}^*  = 0.
\end{equation}
As the conductivity must be positive, we have 
\begin{equation} 
a - \frac{\lambda}{q}D_{12}^* - \frac{\lambda }{q}D_{12}^*  = 0,
\end{equation}
and thus
\begin{equation} 
D_{12}^* = \frac{aq}{2\lambda}.
\end{equation}
Hence, 
\begin{equation}
    D_{12}^*= D_{23}^*= D_{34}^*= D_{45}^*= D_{51}^*= \frac{aq}{2\lambda}.
\end{equation}
This also means that  
\begin{equation}
(N_2^* - N_1^*)=-\frac{\lambda l_{12}}{q},
\end{equation}
and
\begin{equation}
(N_5^* - N_1^*)=-\frac{\lambda l_{51}}{q}. 
\end{equation}
Now using that for a node, $i$, which is neither a source nor a sink, $(N_{i-1}^* - N_i^*)$ and $(N_{i+1} - N_i)$ must have opposite signs, we know that
\begin{align}
(N_3^* - N_2^*)=-\frac{\lambda l_{23}}{q},\\
(N_4^* - N_5^*)=-\frac{\lambda l_{45}}{q},\\
(N_3^* - N_4^*)=-\frac{\lambda l_{34}}{q}.
\end{align}
To find the steady states of the number particles, we first look at the following equation: 
\begin{equation}
-bN_3^*+ \frac{(N_2^*-N_3^*)}{l_{23}}D_{23}^*+\frac{(N_4^*-N_3^*)}{l_{34}}D_{34}^*=0. 
\end{equation}
Inserting $(N_2^* - N_3^*)=\frac{\lambda l_{23}}{q}$, $(N_4^* - N_3^*)=\frac{\lambda l_{34}}{q}$, and $D_{23}^*=D_{34}^*=\frac{aq}{2\lambda}$, we obtain
\begin{equation}
-bN_3^*+ \frac{a}{2}+\frac{a}{2}=0,
\end{equation}
and thus
\begin{equation}
N_3^*=\frac{a}{b}.
\end{equation}
This means that 
\begin{equation}
    N_4^*=\frac{a}{b}+\frac{\lambda l_{34}}{q},
\end{equation}
\begin{equation}
    N_2^*=\frac{a}{b}+\frac{\lambda l_{23}}{q},
\end{equation}
\begin{equation} \label{eq:N1_equal}
    N_1^*=\frac{a}{b}+\frac{\lambda l_{23}}{q}+\frac{\lambda l_{12}}{q},
\end{equation}
and 
\begin{equation} \label{eq:N5_equal}
    N_5^*=\frac{a}{b}+\frac{\lambda l_{34}}{q}+\frac{\lambda l_{45}}{q}.
\end{equation}
But we also know that $(N_5^* - N_1^*)=-\frac{\lambda l_{51}}{q}$, so combining equations \ref{eq:N1_equal} and \ref{eq:N5_equal} we get
\begin{equation}
    N_5^*- N_1^*=\frac{a}{b}+\frac{\lambda l_{34}}{q}+\frac{\lambda l_{45}}{q}-\frac{a}{b}+\frac{\lambda l_{23}}{q}+\frac{\lambda l_{12}}{q}=-\frac{\lambda l_{51}}{q},
\end{equation}
which implies that 
\begin{equation}
    l_{12}+l_{23}=l_{34}+l_{45}+l_{51}.
\end{equation}
Thus for the steady states to be
\begin{align*}
    (N_1^*,N_2^*,N_3^*,N_4^*, N_5^*, D_{12}^*,D_{23}^*,D_{34}^*,D_{45}^*,D_{51}^*) = \\
    (\frac{a}{b}+\frac{\lambda l_{23}}{q}+\frac{\lambda l_{12}}{q}, \frac{a}{b}+\frac{\lambda l_{23}}{q}, \frac{a}{b}, \frac{a}{b}+\frac{\lambda l_{34}}{q}, \frac{a}{b} + \frac{\lambda l_{34}}{q} + \frac{\lambda l_{45}}{q}, \frac{aq}{2\lambda},\frac{aq}{2\lambda}, \frac{a q}{2\lambda}, \frac{a q}{2\lambda}, \frac{a q}{2\lambda}),
\end{align*}
the different paths between the source and the sink must have the same total length. 

\subsection{Stability}

As an initial step to investigate the stability of the system described by Equation \eqref{eq:C_5_eq}, we analyze the Jacobian matrix, denoted as $\mathbf{J}$. The expression for $\mathbf{J}$ is given by the following extensive matrix:

\begin{equation}
 \mathbf{J}= \resizebox{0.85\textwidth}{!}{$
    \begin{pmatrix}
-\frac{D_{12}}{l_{12}}-\frac{D_{51}}{l_{51}} & \frac{D_{12}}{l_{12}} & 0 & 0 & \frac{D_{51}}{l_{51}} & \frac{N_2-N_1}{l_{12}} & 0 & 0 & 0& \frac{N_5-N_1}{l_{51}}   \\
\frac{D_{12}}{l_{12}} & -\frac{D_{12}}{l_{12}}-\frac{D_{23}}{l_{23}}  & \frac{D_{23}}{l_{23}} & 0 & 0 & \frac{N_2-N_1}{l_{12}} &\frac{N_3-N_2}{l_{23}} & 0 & 0& 0  \\
0 & \frac{D_{23}}{l_{23}}  & -b-\frac{D_{23}}{l_{23}}-\frac{D_{34}}{l_{34}} & \frac{D_{34}}{l_{34}} & 0 & 0 &\frac{N_3-N_2}{l_{23}} & \frac{N_4-N_3}{l_{34}}& 0& 0  \\
0 & 0  & \frac{D_{34}}{l_{34}} & -\frac{D_{34}}{l_{34}}-\frac{D_{45}}{l_{45}} & \frac{D_{45}}{l_{45}} & 0 & 0 & \frac{N_3-N_4}{l_{34}} & \frac{N_5-N_4}{l_{45}} & 0 \\
\frac{D_{51}}{l_{51}} & 0  & 0  & \frac{D_{45}}{l_{45}} & -\frac{D_{45}}{l_{45}} - \frac{D_{51}}{l_{51}} & 0 & 0 & 0 & \frac{N_4-N_5}{l_{45}} & \frac{N_1-N_5}{l_{51}} \\

\frac{q \sign{(N_1-N_2)}D_{12}}{l_{12}} & \frac{q \sign{(N_1-N_2)}D_{12}}{l_{12}} & 0 & 0 & 0 & \frac{q |N_1-N_2|}{l_{12}}-\lambda & 0 & 0 & 0 & 0 \\
0 & \frac{q \sign{(N_2-N_3)}D_{23}}{l_{23}} & \frac{q \sign{(N_2-N_3)}D_{23}}{l_{23}}  & 0 & 0 &0  &  \frac{q |N_2-N_3|}{l_{23}}-\lambda & 0 & 0 & 0 \\
0 & 0 & \frac{q \sign{(N_3-N_4)}D_{34}}{l_{34}}  & \frac{q \sign{(N_3-N_4)}D_{34}}{l_{34}}  & 0 &0  & 0 & \frac{q |N_3-N_4|}{l_{34}}-\lambda & 0 & 0 \\
0 & 0 & 0  & \frac{q \sign{(N_4-N_5)}D_{45}}{l_{45}}  & \frac{q \sign{(N_4-N_5)}D_{45}}{l_{45}}  &0  & 0 & 0 & \frac{q |N_4-N_5|}{l_{45}}-\lambda & 0 \\
\frac{q \sign{(N_5-N_1)}D_{51}}{l_{51}}  & 0 & 0  & 0 & \frac{q \sign{(N_5-N_1)}D_{51}}{l_{51}}  &0  & 0 & 0 & 0 & \frac{q |N_5-N_1|}{l_{51}}-\lambda 
\end{pmatrix}$}.
\end{equation}
Upon evaluating $\mathbf{J}$ at equilibrium $\mathcal{E}_1$, the resulting matrix $\mathbf{J}_{\mathcal{E}_1}$ is given by:
\begin{equation}
 \mathbf{J}_{\mathcal{E}_1}=  \resizebox{0.85\textwidth}{!}{$
        \left(\begin{array}{cccccccccc} -\frac{a\,q}{l_{12}\,\lambda } & \frac{a\,q}{l_{12}\,\lambda } & 0 & 0 & 0 & -\frac{\lambda }{q} & 0 & 0 & 0 & -\frac{\frac{a}{b}-N_{5}+\frac{l_{12}\,\lambda }{q}+\frac{l_{23}\,\lambda }{q}}{l_{51}}\\ \frac{a\,q}{l_{12}\,\lambda } & -\frac{a\,l_{23}\,q}{\lambda }-\frac{a\,q}{l_{12}\,\lambda } & \frac{a\,l_{23}\,q}{\lambda } & 0 & 0 & \frac{\lambda }{q} & -\frac{{l_{23}}^2\,\lambda }{q} & 0 & 0 & 0\\ 0 & \frac{a\,q}{l_{23}\,\lambda } & -b-\frac{a\,q}{l_{23}\,\lambda } & 0 & 0 & 0 & \frac{\lambda }{q} & \frac{N_{4}-\frac{a}{b}}{l_{34}} & 0 & 0\\ 0 & 0 & 0 & 0 & 0 & 0 & 0 & -\frac{N_{4}-\frac{a}{b}}{l_{34}} & -l_{45}\,\left(N_{4}-N_{5}\right) & 0\\ 0 & 0 & 0 & 0 & 0 & 0 & 0 & 0 & l_{45}\,\left(N_{4}-N_{5}\right) & \frac{\frac{a}{b}-N_{5}+\frac{l_{12}\,\lambda }{q}+\frac{l_{23}\,\lambda }{q}}{l_{51}}\\ \frac{a\,q^2\,}{l_{12}\,\lambda } & -\frac{a\,q^2\,}{l_{12}\,\lambda } & 0 & 0 & 0 & 0 & 0 & 0 & 0 & 0\\ 0 & \frac{a\,q^2\,}{l_{23}\,\lambda } & -\frac{a\,q^2\,}{l_{23}\,\lambda } & 0 & 0 & 0 & 0 & 0 & 0 & 0\\ 0 & 0 & 0 & 0 & 0 & 0 & 0 & \frac{q\,\left|N_{4}-\frac{a}{b}\right|}{l_{34}}-\lambda  & 0 & 0\\ 0 & 0 & 0 & 0 & 0 & 0 & 0 & 0 & \frac{q\,\left|N_{4}-N_{5}\right|}{l_{45}}-\lambda  & 0\\ 0 & 0 & 0 & 0 & 0 & 0 & 0 & 0 & 0 & \frac{q\,\left|\frac{a}{b}-N_{5}+\frac{l_{12}\,\lambda }{q}+\frac{l_{23}\,\lambda }{q}\right|}{l_{51}}-\lambda  \end{array}\right)
        $}.
\end{equation}
As we have two zero columns, there will be two zero-eigenvalues, and thus the determinant will also be zero, meaning that the equilibrium is non-hyperbolic. The two eigenvectors corresponding to the two zero eigenvalues will be $(0,0,0,N_4^*,0,0,0,0,0,0)$ and $(0,0,0,0,N_5^*,0,0,0,0,0)$. 

Evaluating $\mathbf{J}$ at $\mathcal{E}_2$ we obtain:
\begin{equation}
 \mathbf{J}_{\mathcal{E}_2}=  \resizebox{0.85\textwidth}{!}{$
           \left(\begin{array}{cccccccccc} -\frac{a\,q}{l_{51}\,\lambda } & 0 & 0 & 0 & \frac{a\,q}{l_{51}\,\lambda } & -\frac{\frac{a}{b}-N_{2}+\frac{l_{34}\,\lambda }{q}+\frac{l_{45}\,\lambda }{q}+\frac{l_{51}\,\lambda }{q}}{l_{12}} & 0 & 0 & 0 & -\frac{\lambda }{q}\\ 0 & 0 & 0 & 0 & 0 & \frac{\frac{a}{b}-N_{2}+\frac{l_{34}\,\lambda }{q}+\frac{l_{45}\,\lambda }{q}+\frac{l_{51}\,\lambda }{q}}{l_{12}} & -l_{23}\,\left(N_{2}-\frac{a}{b}\right) & 0 & 0 & 0\\ 0 & 0 & -b-\frac{a\,q}{l_{34}\,\lambda } & \frac{a\,q}{l_{34}\,\lambda } & 0 & 0 & \frac{N_{2}-\frac{a}{b}}{l_{23}} & \frac{\lambda }{q} & 0 & 0\\ 0 & 0 & \frac{a\,q}{l_{34}\,\lambda } & -\frac{a\,l_{45}\,q}{\lambda }-\frac{a\,q}{l_{34}\,\lambda } & \frac{a\,l_{45}\,q}{\lambda } & 0 & 0 & -\frac{\lambda }{q} & \frac{{l_{45}}^2\,\lambda }{q} & 0\\ \frac{a\,q}{l_{51}\,\lambda } & 0 & 0 & \frac{a\,l_{45}\,q}{\lambda } & -\frac{a\,l_{45}\,q}{\lambda }-\frac{a\,q}{l_{51}\,\lambda } & 0 & 0 & 0 & -\frac{{l_{45}}^2\,\lambda }{q} & \frac{\lambda }{q}\\ 0 & 0 & 0 & 0 & 0 & \frac{q\,\left|\frac{a}{b}-N_{2}+\frac{l_{34}\,\lambda }{q}+\frac{l_{45}\,\lambda }{q}+\frac{l_{51}\,\lambda }{q}\right|}{l_{12}}-\lambda  & 0 & 0 & 0 & 0\\ 0 & 0 & 0 & 0 & 0 & 0 & \frac{q\,\left|N_{2}-\frac{a}{b}\right|}{l_{23}}-\lambda  & 0 & 0 & 0\\ 0 & 0 & -\frac{a\,q^2\,\mathrm{sign}\left(\frac{l_{34}\,\lambda }{q}\right)}{l_{34}\,\lambda } & \frac{a\,q^2\,\mathrm{sign}\left(\frac{l_{34}\,\lambda }{q}\right)}{l_{34}\,\lambda } & 0 & 0 & 0 & 0 & 0 & 0\\ 0 & 0 & 0 & -\frac{a\,q^2\,\mathrm{sign}\left(\frac{l_{45}\,\lambda }{q}\right)}{l_{45}\,\lambda } & \frac{a\,q^2\,\mathrm{sign}\left(\frac{l_{45}\,\lambda }{q}\right)}{l_{45}\,\lambda } & 0 & 0 & 0 & 0 & 0\\ \frac{a\,q^2\,\mathrm{sign}\left(\frac{l_{51}\,\lambda }{q}\right)}{l_{51}\,\lambda } & 0 & 0 & 0 & -\frac{a\,q^2\,\mathrm{sign}\left(\frac{l_{51}\,\lambda }{q}\right)}{l_{51}\,\lambda } & 0 & 0 & 0 & 0 & 0 \end{array}\right)$}.
\end{equation}

As there is a column of zeros, there will be at least one zero eigenvalue, meaning that this equilibrium is also non-hyperbolic. The eigenvector corresponding to the zero eigenvalue will be $(0,N_2^*,0,N_4^*,0,0,0,0,0,0)$. 

Upon evaluating the Jacobian at the steady state when $l_{12} + l_{23} = l_{34} + l_{45} + l_{51}$, $\mathbf{J}_{\mathcal{E}_3}$ the expression is given by:

\begin{equation}
 \mathbf{J}_{\mathcal{E}_3}=  \resizebox{0.85\textwidth}{!}{$
\left(\begin{array}{cccccccccc} -\frac{a\,q}{2\,l_{12}\,\lambda }-\frac{a\,q}{2\,l_{51}\,\lambda } & \frac{a\,q}{2\,l_{12}\,\lambda } & 0 & 0 & \frac{a\,q}{2\,l_{51}\,\lambda } & -\frac{\lambda }{q} & 0 & 0 & 0 & -\frac{\frac{l_{12}\,\lambda }{q}+\frac{l_{23}\,\lambda }{q}-\frac{l_{34}\,\lambda }{q}-\frac{l_{45}\,\lambda }{q}}{l_{51}}\\ \frac{a\,q}{2\,l_{12}\,\lambda } & -\frac{a\,l_{23}\,q}{2\,\lambda }-\frac{a\,q}{2\,l_{12}\,\lambda } & \frac{a\,l_{23}\,q}{2\,\lambda } & 0 & 0 & \frac{\lambda }{q} & -\frac{{l_{23}}^2\,\lambda }{q} & 0 & 0 & 0\\ 0 & \frac{a\,q}{2\,l_{23}\,\lambda } & -b-\frac{a\,q}{2\,l_{23}\,\lambda }-\frac{a\,q}{2\,l_{34}\,\lambda } & \frac{a\,q}{2\,l_{34}\,\lambda } & 0 & 0 & \frac{\lambda }{q} & \frac{\lambda }{q} & 0 & 0\\ 0 & 0 & \frac{a\,q}{2\,l_{34}\,\lambda } & -\frac{a\,l_{45}\,q}{2\,\lambda }-\frac{a\,q}{2\,l_{34}\,\lambda } & \frac{a\,l_{45}\,q}{2\,\lambda } & 0 & 0 & -\frac{\lambda }{q} & \frac{{l_{45}}^2\,\lambda }{q} & 0\\ \frac{a\,q}{2\,l_{51}\,\lambda } & 0 & 0 & \frac{a\,l_{45}\,q}{2\,\lambda } & -\frac{a\,l_{45}\,q}{2\,\lambda }-\frac{a\,q}{2\,l_{51}\,\lambda } & 0 & 0 & 0 & -\frac{{l_{45}}^2\,\lambda }{q} & \frac{\frac{l_{12}\,\lambda }{q}+\frac{l_{23}\,\lambda }{q}-\frac{l_{34}\,\lambda }{q}-\frac{l_{45}\,\lambda }{q}}{l_{51}}\\ \frac{a\,q^2\,\mathrm{sign}\left(\frac{l_{12}\,\lambda }{q}\right)}{2\,l_{12}\,\lambda } & -\frac{a\,q^2\,\mathrm{sign}\left(\frac{l_{12}\,\lambda }{q}\right)}{2\,l_{12}\,\lambda } & 0 & 0 & 0 & 0 & 0 & 0 & 0 & 0\\ 0 & \frac{a\,q^2\,\mathrm{sign}\left(\frac{l_{23}\,\lambda }{q}\right)}{2\,l_{23}\,\lambda } & -\frac{a\,q^2\,\mathrm{sign}\left(\frac{l_{23}\,\lambda }{q}\right)}{2\,l_{23}\,\lambda } & 0 & 0 & 0 & 0 & 0 & 0 & 0\\ 0 & 0 & -\frac{a\,q^2\,\mathrm{sign}\left(\frac{l_{34}\,\lambda }{q}\right)}{2\,l_{34}\,\lambda } & \frac{a\,q^2\,\mathrm{sign}\left(\frac{l_{34}\,\lambda }{q}\right)}{2\,l_{34}\,\lambda } & 0 & 0 & 0 & 0 & 0 & 0\\ 0 & 0 & 0 & -\frac{a\,q^2\,\mathrm{sign}\left(\frac{l_{45}\,\lambda }{q}\right)}{2\,l_{45}\,\lambda } & \frac{a\,q^2\,\mathrm{sign}\left(\frac{l_{45}\,\lambda }{q}\right)}{2\,l_{45}\,\lambda } & 0 & 0 & 0 & 0 & 0\\ \frac{a\,q^2\,\mathrm{sign}\left(\frac{l_{12}\,\lambda }{q}+\frac{l_{23}\,\lambda }{q}-\frac{l_{34}\,\lambda }{q}-\frac{l_{45}\,\lambda }{q}\right)}{2\,l_{51}\,\lambda } & 0 & 0 & 0 & -\frac{a\,q^2\,\mathrm{sign}\left(\frac{l_{12}\,\lambda }{q}+\frac{l_{23}\,\lambda }{q}-\frac{l_{34}\,\lambda }{q}-\frac{l_{45}\,\lambda }{q}\right)}{2\,l_{51}\,\lambda } & 0 & 0 & 0 & 0 & \frac{q\,\left|\frac{l_{12}\,\lambda }{q}+\frac{l_{23}\,\lambda }{q}-\frac{l_{34}\,\lambda }{q}-\frac{l_{45}\,\lambda }{q}\right|}{l_{51}}-\lambda  \end{array}\right)$}.
\end{equation}

This expression can be further simplified to:

\begin{equation}
 \mathbf{J}_{\mathcal{E}_3}=  \resizebox{0.85\textwidth}{!}{$
\left(\begin{array}{cccccccccc} -\frac{a\,q}{2\,l_{12}\,\lambda }-\frac{a\,q}{2\,l_{51}\,\lambda } & \frac{a\,q}{2\,l_{12}\,\lambda } & 0 & 0 & \frac{a\,q}{2\,l_{51}\,\lambda } & -\frac{\lambda }{q} & 0 & 0 & 0 & - \frac{\lambda}{q}\\ 
\frac{a\,q}{2\,l_{12}\,\lambda } & -\frac{a\,l_{23}\,q}{2\,\lambda }-\frac{a\,q}{2\,l_{12}\,\lambda } & \frac{a\,l_{23}\,q}{2\,\lambda } & 0 & 0 & \frac{\lambda }{q} & -\frac{{l_{23}}^2\,\lambda }{q} & 0 & 0 & 0\\ 
0 & \frac{a\,q}{2\,l_{23}\,\lambda } & -b-\frac{a\,q}{2\,l_{23}\,\lambda }-\frac{a\,q}{2\,l_{34}\,\lambda } & \frac{a\,q}{2\,l_{34}\,\lambda } & 0 & 0 & \frac{\lambda }{q} & \frac{\lambda }{q} & 0 & 0\\ 
0 & 0 & \frac{a\,q}{2\,l_{34}\,\lambda } & -\frac{a\,l_{45}\,q}{2\,\lambda }-\frac{a\,q}{2\,l_{34}\,\lambda } & \frac{a\,l_{45}\,q}{2\,\lambda } & 0 & 0 & -\frac{\lambda }{q} & \frac{{l_{45}}^2\,\lambda }{q} & 0\\ 
\frac{a\,q}{2\,l_{51}\,\lambda } & 0 & 0 & \frac{a\,l_{45}\,q}{2\,\lambda } & -\frac{a\,l_{45}\,q}{2\,\lambda }-\frac{a\,q}{2\,l_{51}\,\lambda } & 0 & 0 & 0 & -\frac{{l_{45}}^2\,\lambda }{q} & \frac{\lambda}{q}\\ 
\frac{a\,q^2\,}{2\,l_{12}\,\lambda } & -\frac{a\,q^2\,}{2\,l_{12}\,\lambda } & 0 & 0 & 0 & 0 & 0 & 0 & 0 & 0\\ 
0 & \frac{a\,q^2\,}{2\,l_{23}\,\lambda } & -\frac{a\,q^2\,}{2\,l_{23}\,\lambda } & 0 & 0 & 0 & 0 & 0 & 0 & 0\\ 
0 & 0 & -\frac{a\,q^2\,}{2\,l_{34}\,\lambda } & \frac{a\,q^2\,}{2\,l_{34}\,\lambda } & 0 & 0 & 0 & 0 & 0 & 0\\ 
0 & 0 & 0 & -\frac{a\,q^2\,}{2\,l_{45}\,\lambda } & \frac{a\,q^2\,}{2\,l_{45}\,\lambda } & 0 & 0 & 0 & 0 & 0\\ 
\frac{a\,q^2\,}{2\,l_{51}\,\lambda } & 0 & 0 & 0 & -\frac{a\,q^2\,}{2\,l_{51}\,\lambda } & 0 & 0 & 0 & 0 & 0  \end{array}\right)$}.
\end{equation}
The determinant of $\mathbf{J}_{\mathcal{E}_3}$ is given by:
\begin{equation} \label{eq:Det_E3}
       \det\left(\mathbf{J}_{\mathcal{E}_3}\right)= -\frac{a^4\,q^3\left(b\,l_{23}\,l_{45}\,q\,\left|l_{12}\,\lambda +l_{23}\,\lambda -l_{34}\,\lambda -l_{45}\,\lambda \right|-b\,l_{23}\,l_{45}\,l_{51}\,\lambda \,q\right)}{16\,l_{12}\,l_{34}\,l_{51}}.
\end{equation}
At the equilibrium point $\mathcal{E}_3$, we now that both paths possess equal lengths, implying that: 
\begin{equation}
    \left|l_{12},\lambda +l_{23},\lambda -l_{34},\lambda -l_{45},\lambda \right| = l_{51}. \label{eq:equal_len}
\end{equation} 
Substituting equality \eqref{eq:equal_len} into \eqref{eq:Det_E3}, we find that the determinant becomes zero, indicating that $\mathbf{J}_{\mathcal{E}_3}$ also  corresponds to a non-hyperbolic equilibrium.

\section{Further results}
\begin{figure}[H]
    \centering
    \begin{tabular}{cc}
        \includegraphics[width=.4\textwidth]{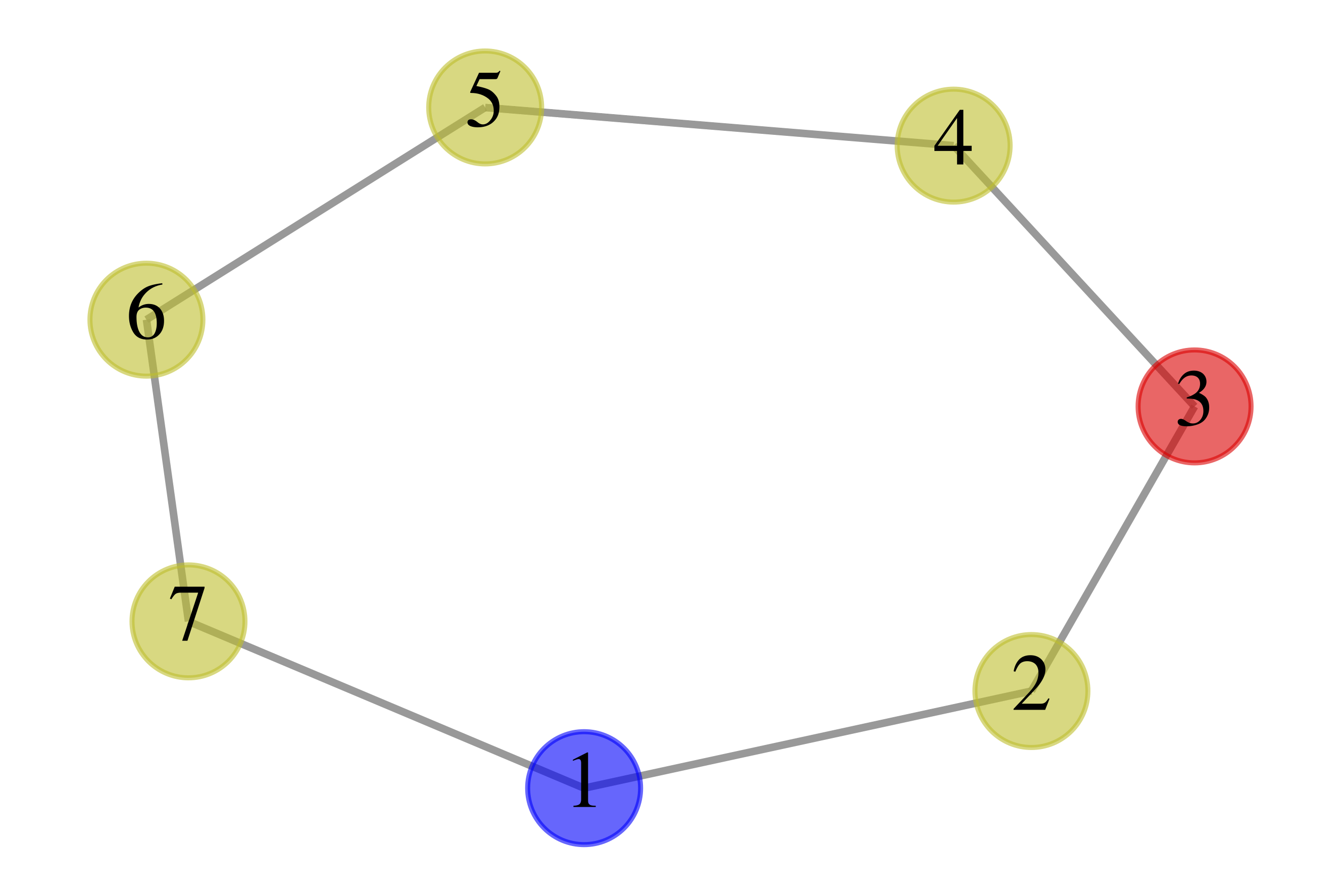} & \includegraphics[width=.4\textwidth]{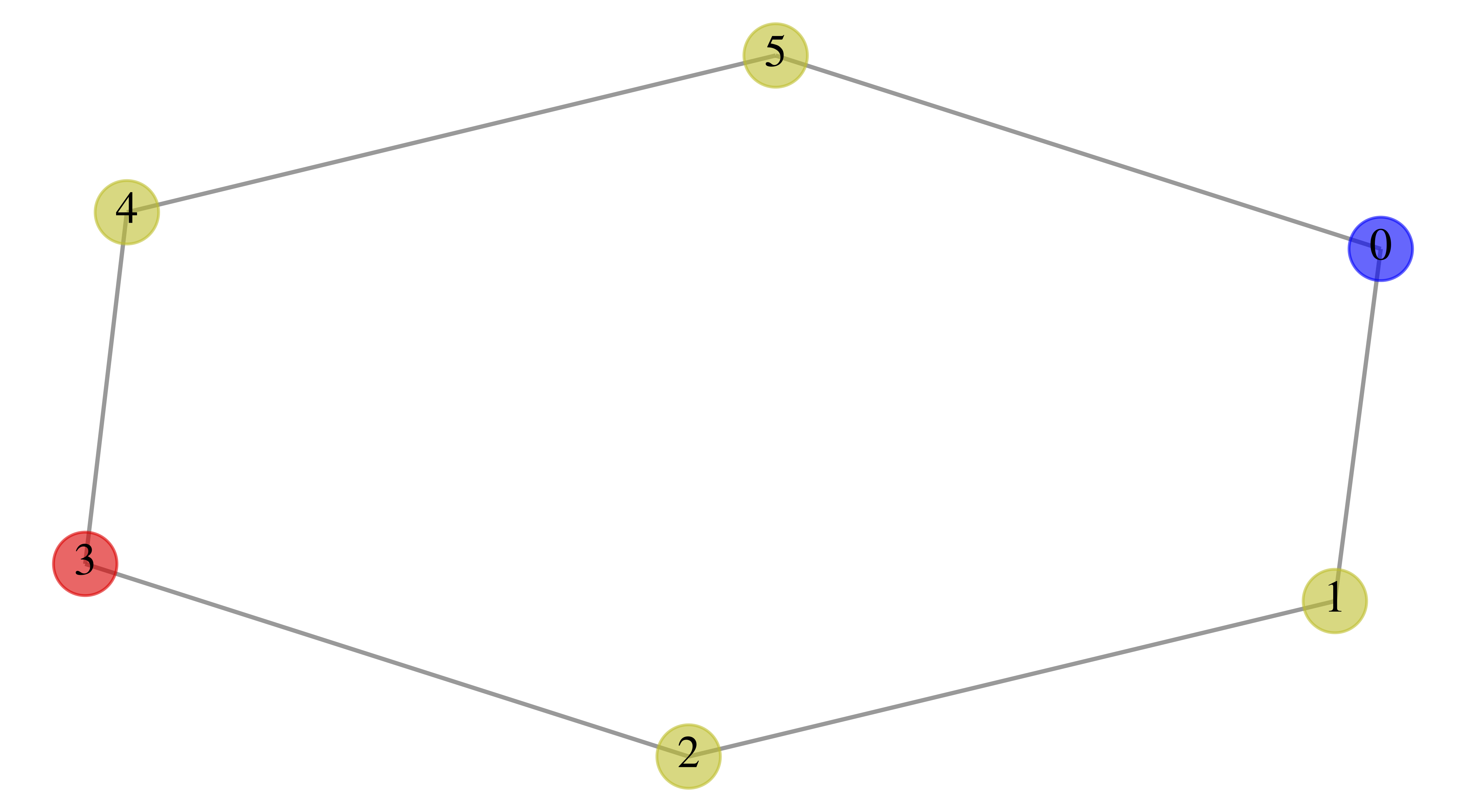}
    \end{tabular}
    \caption{A $C_7$ graph (left) and a $C_6$ graph plus an extra edge between nodes $4$ and $6$ (right).}
    \label{fig:7cycle+}
\end{figure}
\subsection{Non-oscillatory sources and sinks}\label{sec:app-nonoscillatory}
\begin{figure}[H]
    \centering
    \includegraphics[width=.96\textwidth]{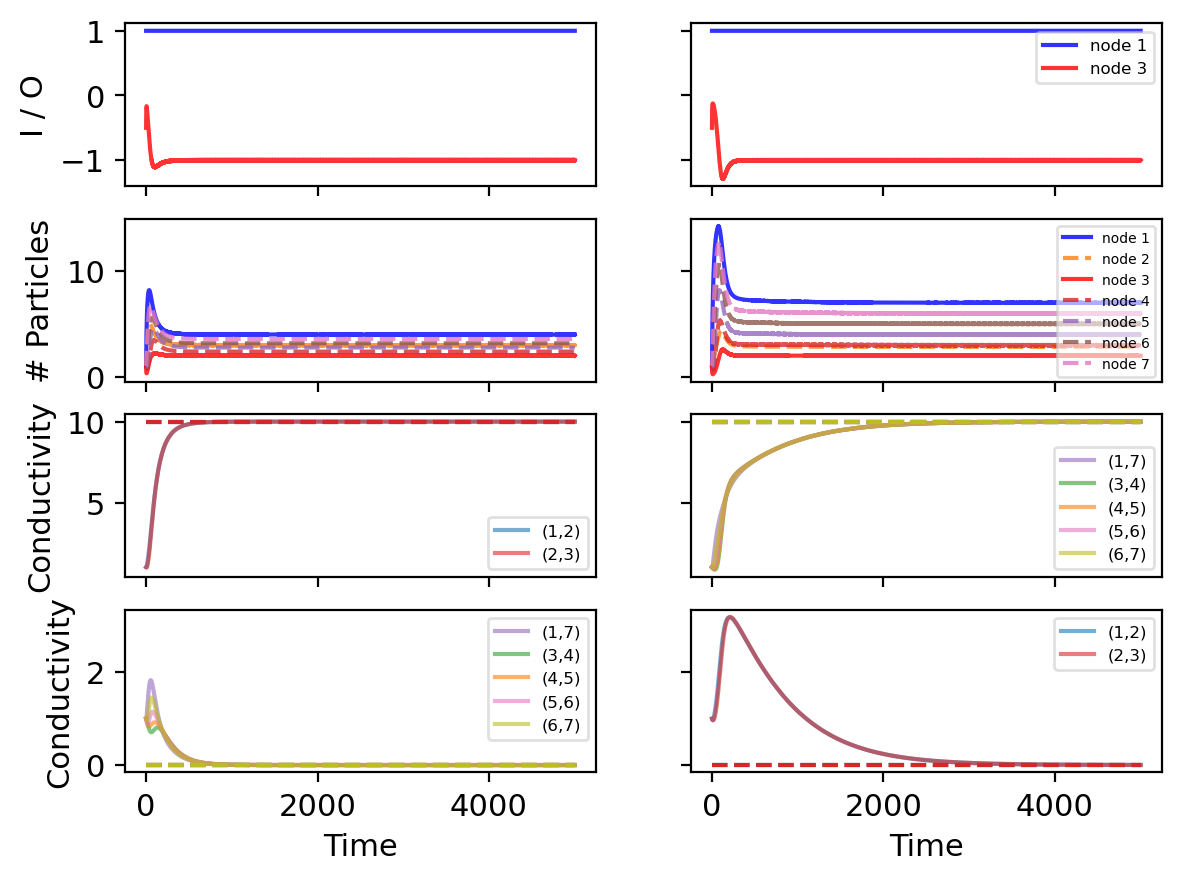}
    \caption{Results from the model on a $C_7$ graph, with node $1$ as a source, node $3$ as a sink, an inflow at rate $a=1$, an outflow at rate $b=0.5$, and the lengths of all edges are $l_{ij} = 10$ apart from $l_{12}$. On the left, we set $l_{12} = 10$, thus the path going through 1-2-3, is the shortest one. On the right, we set $l_{12} = 50$, thus the path going through 1-7-6-5-4-3 is the shortest one. The first row shows the input at node $1$ and output at node $3$, the second row plots the number of particles on each node, $N_i(t)$, the third row indicates the conductivity of the edges in the shortest path, and the last row indicates the conductivity of the remaining edges in the graph, where the dashed lines in the corresponding colour show the analytical results of the steady state. }
    \label{fig:res-1s1s-n7}
\end{figure}

\begin{figure}[H]
    \centering
    \includegraphics[width=.96\textwidth]{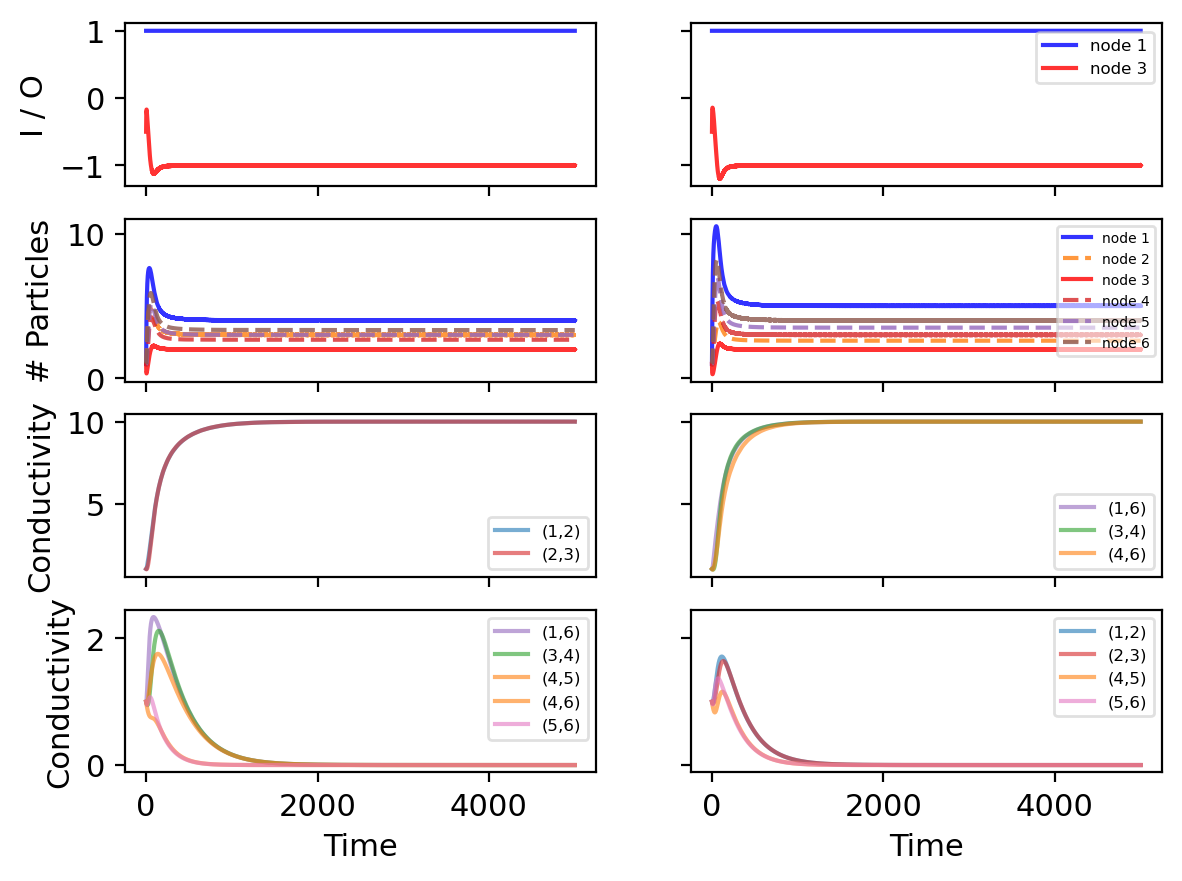}
    \caption{Results from the model on a $C_6$ graph plus an extra edge between nodes 4 and 6 (see the rightmost graph in Figure \ref{fig:7cycle+}), with node $1$ as a source, node $3$ as a sink, an inflow at rate $a=1$, an outflow at rate $b=0.5$, and the lengths of all edges are $l_{ij} = 10$ apart from $l_{12}$. On the left, we set $l_{12} = 10$, thus the path going through 1-2-3, is the shortest one. On the right, we set $l_{12} = 40$, thus the path going through 1-6-4-3 is the shortest one. The first row shows the input at node $1$ and output at node $3$, the second row plots the number of particles on each node, $N_i(t)$, the third row indicates the conductivity of the edges in the shortest path, and the last row indicates the conductivity of the remaining edges in the graph.}
    \label{fig:res-1s1s-n6}
\end{figure}

In this section, we include more results from the simulations on cycle graphs of larger sizes and more general graph structure; see Figures \ref{fig:res-1s1s-n7} and \ref{fig:res-1s1s-n6}. We observe that the slime mould model can still find the shortest path in both cases.


\subsection{Oscillating nodes}\label{sec:app-oscil}
\begin{figure}[H]
    \centering
    \includegraphics[width=.96\textwidth]{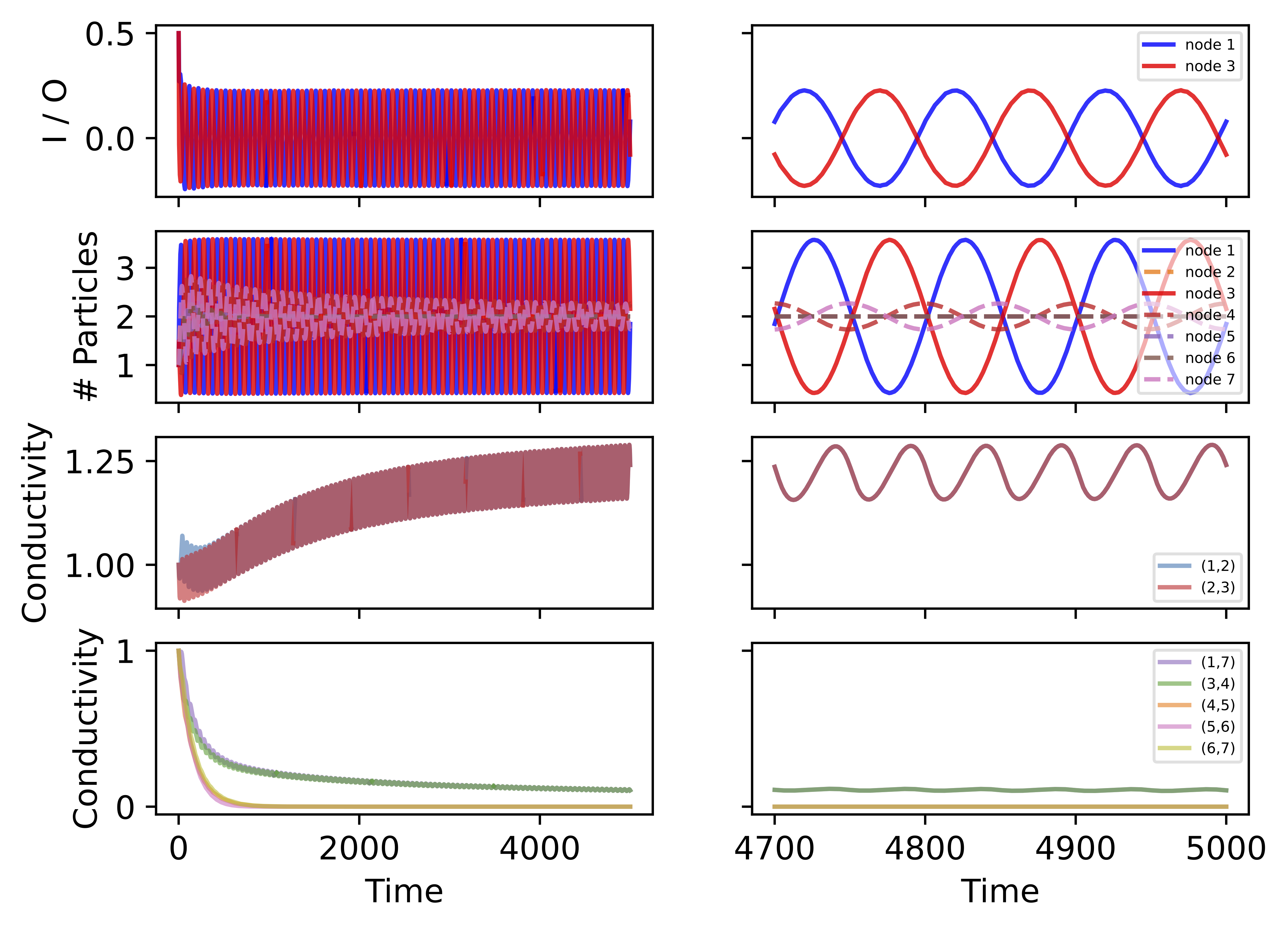}
    \caption{Results from the model on a $C_7$ graph, with nodes 1 and 3 as the oscillating nodes, rate $b=0.5$, amplitude $A_1 = A_3 = A=1$, frequency $\theta_1 = \theta_3 =\theta=0.01$, and phases $\phi_1 = 0$ and $\phi_3 = \pi$, with the same reinforcement parameter $q=0.1$ and decay parameter $\lambda=0.01$ as in Figure \ref{fig:res-1s1s}. The first row plots the actual input/output to the two nodes, the second row shows the number of particles in each node, the third row indicates the conductivity of the edges in the shortest path and the last row indicates the conductivity of the remaining edges in the graph, with the full profile on the left and the change in the long term on the right.}
    \label{fig:res-osln-1s1s-n7}
\end{figure}

\begin{figure}[H]
    \centering
    \includegraphics[width=.96\textwidth]{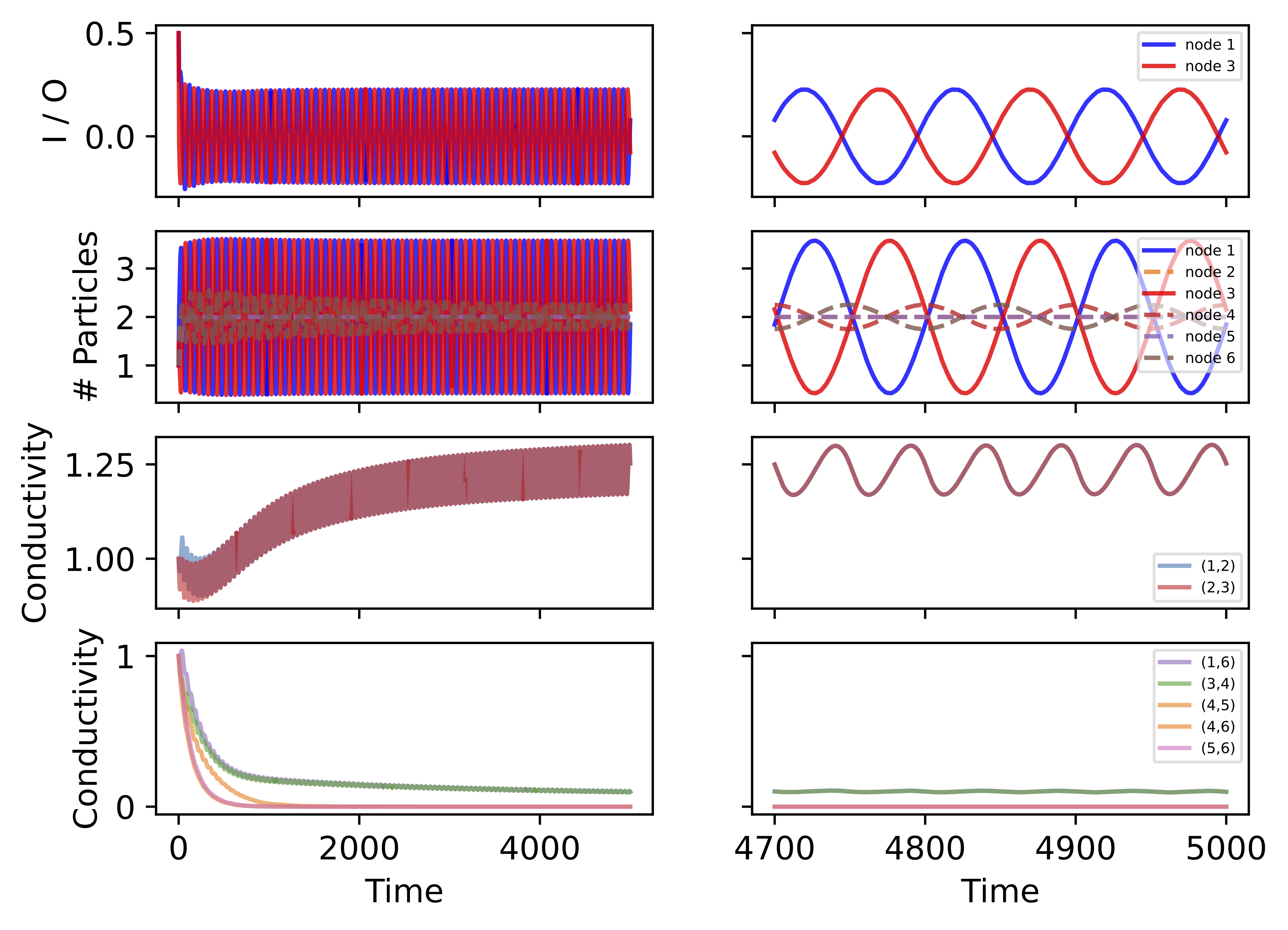}
    \caption{Results from the model on a $C_6$ graph plus an extra edge between nodes 4 and 6 and the right in Figure \ref{fig:7cycle+}, with nodes 1 and 3 as the oscillating nodes, rate $b=0.5$, amplitude $A_1 = A_3 = A=1$, frequency $\theta_1 = \theta_3 =\theta=0.01$, and phases $\phi_1 = 0$ and $\phi_3 = \pi$, with the same reinforcement parameter $q=0.1$ and decay parameter $\lambda=0.01$ as in Figure \ref{fig:res-1s1s}. The first row plots the actual input/output to the two nodes, the second row shows the number of particles in each node, the third row indicates the conductivity of the edges in the shortest path and the last row indicates the conductivity of the remaining edges in the graph, with the full profile on the left and the change in the long term on the right.}
    \label{fig:res-osln-1s1s-n6}
\end{figure}

\end{document}